\documentclass[onecolumn]{emulateapj}
\begin{document}
%\sloppy
\slugcomment{Accepted to Ap.J. Suppl.}

\title{Atomic and Molecular Opacities for Brown Dwarf and Giant Planet
Atmospheres}

\author{C.M. Sharp\altaffilmark{1},
A. Burrows\altaffilmark{1}}

\altaffiltext{1}{Department of Astronomy and Steward Observatory, The
University of Arizona,
Tucson, AZ \ 85721; csharp@as.arizona.edu, burrows@zenith.as.arizona.edu}

\begin{abstract}

We present a comprehensive description of the theory and practice of 
opacity calculations from the infrared to the ultraviolet needed to generate models of the 
atmospheres of brown dwarfs and extrasolar giant planets.  Methods for using 
existing line lists and spectroscopic databases in disparate formats
are presented and plots of the resulting absorptive opacities versus
wavelength for the most important molecules and atoms at representative 
temperature/pressure points are provided.  Electronic, ro-vibrational, 
bound-free, bound-bound, free-free, and collision-induced
transitions and monochromatic opacities are derived, discussed, and 
analyzed.  The species addressed include the alkali metals, iron, heavy metal
oxides, metal hydrides, $H_2$, $H_2O$, $CH_4$, $CO$, $NH_3$, $H_2S$, $PH_3$,
and representative 
grains. Once monochromatic absorption cross sections for all consistuents 
have been derived, chemical abundances have to be obtained before 
the resulting product can be summed to obtain total opacities. Hence, we 
include a review of the thermochemistry, techniques, and databases needed to 
derive equilibrium abundances and provide some sample results.

\end{abstract}


\section{INTRODUCTION}
\label{intro}

The discoveries of extrasolar giant planets and brown dwarfs in 1995 have
opened up exciting new fields in astrophysics.  These enable model atmospheres
to be directly tested against observations for atmospheric temperatures
and pressures at which a large number of molecules that had not been
considered before in stellar astrophysics reside.  Moreover, many of
the extrasolar planets, including the first to be discovered, are much
closer to their parent star than Mercury is to the Sun
so the atmospheric conditions of these planets are totally unlike
any previously investigated.  In such situations, many
of these molecules are subjected to a large ultraviolet flux from the
parent star.

The absorption of radiation by molecules is generally much more
complicated than by atoms.  Since many of the molecules in the 
atmospheres of brown dwarfs and extrasolar giant planets had not been studied in
detail in an astrophysical context before 1995, their properties are still
poorly known.  Even those that had been investigated, had not been
for the temperatures (100 - 3000 K) and pressures (10$^{-6}$ - 100
atmospheres) found in the atmospheres of such objects.  Moreover, an
additional factor of
considerable importance is the formation of grains.  In this paper, we
address the spectroscopy and opacities of the molecules and atoms central
to an understanding of substellar objects.  Such opacities are required
for models of their spectra and evolution.

Another consequence of the emergence of this young field is the extension
of the spectral classification system beyond late-type M-dwarf stars to
include two new spectroscopic classes at lower temperatures, L and T dwarfs.
The criteria for these new classes depend upon the strengths of atomic and
molecular spectral lines, which in turn are governed by absorption cross
sections and the abundances of various species.

In this paper, we discuss the methods used to calculate opacities from
extant line lists or by directly calculating the energy levels.  The
absorption due to spectral lines is the most important source of opacity,
as well as the most difficult to calculate, and this is considered in
\S\ref{lines}.  After dealing with the background theory for calculating
line strengths and broadening, and how to handle a frequency grid that
samples the lines, we review the chemical species used for calculating
substellar atmospheres.  These are grouped together when they share
important similarities, such as being metal hydrides or absorbing in the
same general part of the spectrum.  Table 1 lists the species considered
in this work and gives the sections where the species are discussed in
detail, including several condensates addressed with the Mie theory.

In \S\ref{hitran}, we discuss the species with data obtained from the
HITRAN (Rothman et al. 2003, 2005) and GEISA (Jacquinet-Husson et al. 1999,
2003, 2005) databases, together with other sources.  The spectral
range covered is the visible and infrared.  The molecules considered in
detail are $H_2O$ (Partridge \& Schwenke 1997; Barber et al. 2006),
$CH_4$ (Strong et al.
1993; Karkoscka 1994; Borysow et al. (2003); Brown et al. 1997),
$NH_3$ (Brown el al. 2000), $CO$ (Goorvitch, 1994).  Of these, $CH_4$ and
$NH_3$ are important at the lower temperatures of interest, $CO$ is important
at the upper temperatures of interest, and $H_2O$ is very important over
the whole temperature range.

In \S\ref{oxides}, we discuss the two heavy metal oxides, $TiO$ (Allard,
Hauschildt \& Schwenke 2000) and $VO$ (Plez 1998).  Of the heavy metal oxides,
these are the most important sources of opacity at the higher end of the
temperatures of interest.  Both have very similar properties.  This is
followed by \S\ref{hydrides}, where we discuss the metal hydrides $TiH$
(Burrows et al. 2005), $CrH$ (Burrows et al. 2002), $FeH$
(Dulick et al. 2003), $MgH$ (Skory et al. 2003; Weck et al. 2003a),
and $CaH$ (Leininger \& Jeung 1995), which all have similar
properties.

As much as possible, the most extensive and up-to-date spectroscopic data
available are considered.  In the case of $H_2S$ and $PH_3$, tables of
precomputed opacities can be used, so calculations are restricted to
interpolating in the ranges of temperature and pressure in the tables, but
since these species have low abundances, they are relatively unimportant.

Ultraviolet opacities are important when a planet is irradiated by the
star it orbits, and in \S\ref{uvopac} we discuss these opacities for
atomic hydrogen (Menzel 1969; Bethe \& Salpeter 1957; Carson 1988a), atomic
iron (Kurucz 1995), and the molecules $H_2$, $CO$, $SiO$, $H_2O$
(Kurucz 1993), and $H_2S$ (Lee, Wang \& Suto 1987).  The available data vary
considerably, and this dictates the methods used to calculate the opacities.

The absorption of the alkali elements in monatomic form: $Li$, $Na$,
$K$, $Rb$, and $Cs$ using the Vienna Atomic Line Data, (VALD - Piskunov
1994) is considered in \S\ref{vald}.  The line strengths and widths are all
calculated in a uniform way, except for the wings of the resonance lines
of $Na$ and $K$ (Burrows \&  Volobuyev 2003; Allard et al. 2003;
Zhu, Babb, \& Dalgarno 2006), for which a separate treatment should be applied.

In addition to lines, the absorption due to the underlying continuum is
calculated, as discussed in \S\ref{continuous}.  Rosseland and other
harmonic mean opacities are divergent if the absorption drops to zero at
any point in the frequency region considered, and a realistic truncation
of the wings of spectra lines is required.  The four processes discussed
are free-free, bound-free, collision-induced absorption in the gas
phase, and grain absorption by condensed phases.

For the temperatures and pressures of interest, free-free opacity
sources are the least important, so are only briefly covered in
\S\ref{freefree}, since they can make at most a minor contribution at
the upper end of the temperature range of interest.  We discuss a
number of free-free opacity sources, but in practice the only ones of
even minor importance are $H_{ff}$ (Carson 1988a), $H^-_{ff}$
(Wishart 1979), $He^-_{ff}$ (Bell, Berrington \& Croskery 1982),
and $H_{2ff}^-$ (Bell 1980).  We adopt the convention of including the
free electron in the net charge of the whole system undergoing photon
absorption, so here $H_{ff}$ refers to an electron moving in the field of
a proton, and $H^-_{ff}$ refers to an electron moving in the field of a
neutral hydrogen atom, and likewise for the other species.
 
Of considerably greater importance is bound-free absorption, as
discussed in \S\ref{boundfree}.  We calculate the contribution due to
$H$ with Gaunt factors from Carson (1988a), its negative ion $H^-$
(Wishart 1979; Bell \& Berrington 1987), and the atomic species $Na$
(Cunto \& Mendoza 1992; Cunto et al. 1993), $K$ (Verner \& Yakovlev 1995),
and $Fe$ (Kurucz 1995).  The ion $H^-$ is important only when there is a
supply of free electrons, and this is the case only when the species with
the lowest ionization potentials, namely the alkali elements, become
ionized.

At high pressures, collision-induced absorption, as discussed in
\S\ref{ciaopac}, can be very important in the infrared.  In an
astrophysical mixture for the temperatures of interest to us here,
the two most important contributions are due to $H_2-H_2$
(Borysow et al. 1985; Zheng \& Borysow 1995a; Zheng \& Borysow 1995b;
Borysow \& Frommhold 1990; Lenzuni, Chernoff, \& Salpeter 1991;
Guillot et al. 1992), and $H_2-He$ collisions (Borysow, Frommhold,
\& Birnbaum 1988; Borysow, Frommhold, \& Moraldi 1989; Borysow
\& Frommhold 1989).  Although the contribution to the opacity due to
$H_2-CH_4$ collisions is much weaker, $CH_4$ is an important molecule
and the data are available.

The detailed discussion of opacity calculations is completed in
\S\ref{grainopac} by calculating Mie scattering due to grains formed
by the condensation of species out of the gas phase (Van de Hulst 1957;
Sudarsky 2002).  Both the real and imaginary components of the refractive
indices are needed and used.

Once all the individual opacity sources have been calculated, they must be
weighted by abundances and combined into monochromatic opacities or
suitable mean opacities, as discussed in \S\ref{total}.  For the
abundances calculated in \S\ref{atmosabun}, suitable thermodynamic data
are required, and methods of handling them are discussed.  Once the
abundances are used to derive the total monochromatic opacity, 
Rosseland and Planck mean opacities can be calculated, if desired.

\section{LINE OPACITIES}
\label{lines}

The sources of opacities of greatest interest are bound-bound transitions
of atoms and molecules, and are the main focus of this paper.  These
involve the greatest computational efforts, and often influence most strongly
the computation of model atmospheres of brown dwarfs and giant planets.
Such transitions involve the absorption of a photon by an atom or molecule
in the ground state, or a discrete initial excited state, to a more excited
state, giving rise to a set of spectral lines, whose frequencies, strengths,
and widths have to be calculated.

For some atoms, notably the alkali elements, there are only a few strong lines
and a relatively modest number of weak lines that need to be considered.
For other elements, such as iron, there are a large number of lines.  For
molecules the number of lines that need to be considered can be very large
indeed, often numbered in the millions, hundreds of millions, or billions,
which require a significant amount of computing in order to calculate the
opacity.  All the gaseous species down to $CaH$ in Table 1 contribute line
absorption, with the exception of the $H^-$ ion, which has no lines,
and $He$ which has no lines of importance in the spectral regions of interest.

The calculation of the line strengths for each line of each species depends on
the data available for the species being considered, so different methods have
to be used, as discussed in \S\ref{hitran}, \S\ref{oxides},
\S\ref{hydrides}, \S\ref{uvopac}, and \S\ref{vald}.  In order to reduce the
chances of errors with input data in different forms, it is recommended to
convert, if necessary, all the line strengths into the same uniform system,
with the best being integrated line strengths in cm$^2$s$^{-1}$ species$^{-1}$.
These depend only on the temperature.  The lines should then be broadened
to a suitable profile which is dependent on the pressure, then the
monochromatic absorption in cm$^2$species$^{-1}$ across the profile
should be computed, with the contributions from any overlapping profiles
being summed.  The monochromatic absorption for each species so obtained
depends on the temperature and pressure.  The total monochromatic opacity
of the gas is obtained by summing up the individual contributions weighted
by the abundances as number densities in cm$^{-3}$ for each of the species,
yielding the total monochromatic volume opacity in cm$^2$cm$^{-3}$, i.e.
cm$^{-1}$; however, the total monochromatic mass opacity in cm$^2$g$^{-1}$
is usually the required result, and is obtained by dividing the volume
opacity by the gas mass density.  This is discussed later in \S\ref{total}.

In its most general LTE (Local Thermodynamic Equilibrium) form, 
the integrated strength $S$ of a spectral line in
cm$^2$ s$^{-1}$ species$^{-1}$ is obtained from

\begin{equation}
S = \frac{\pi e^2 g_if_{ij}}{m_ec} \frac{e^{-hcF_i/kT}}{Q(T)}
\left[1 - e^{-hc(F_j-F_i)/kT}\right]\, ,
\label{oscil}
\end{equation}

\noindent where $g_i$ and $f_{ij}$ are, respectively, the statistical weight of
the $i^{th}$ energy level and the oscillator strength for a transition from
that level to a higher level $j$, $F_i$, and $F_j$ are, respectively, the term
values (excitation energies) in cm$^{-1}$ of the $i^{th}$ and $j^{th}$ levels
participating in the transition, and $Q(T)$ is the partition function of the
species at some temperature $T$.  The other symbols have their usual
meanings.  Note that the first term in eq. (\ref{oscil}) gives the line
strength in cm$^2$ s$^{-1}$ absorber$^{-1}$, the next term with the
Boltzmann factor and the partition function converts this to the required
line strength, and the last term is the stimulated emission correction factor,
where $F_j-F_i$ is the transition frequency in wavenumbers, i.e.
$\bar{\nu}$ in cm$^{-1}$.  Although monochromatic opacities are displayed
later as functions of wavelength, it is recommended that all opacity
calculations be performed internally in wavenumbers, even if some
of the input data are given in wavelengths, as most molecular spectroscopic
constants and energy levels are expressed in cm$^{-1}$, and adopting a
uniform system of units reduces the chances of error.

However, much of the data available are not expressed in the form of
oscillator strengths and statistical weights, collectively given as
$gf$-values, but in other forms that must be converted to the required
lines strengths (see \S\ref{hitran}).

The general expression for calculating the partition function is given by

\begin{equation}
Q(T) = \sum_{i=1}^n g_ie^{-hcF_i/kT}\, ,
\label{partfun}
\end{equation}

\noindent where the summation is performed over the first $n$ levels, whose
contributions are required at the highest temperatures of interest.  The
term value of the lowest level $F_1$, i.e. the gound state, is zero by
construction.

For atoms, eq. (\ref{partfun}) can be evaluated directly with the data from
Moore (1949) up to term values of about 20,000 cm$^{-1}$, above which the
contribution to the partition function is less than 1\% at all relevant 
temperatures.  Formally the partition function is divergent, but
for the pressures of interest, the higher levels that can cause this problem
are not populated.  However, because molecules contain a large number of
energy levels, directly summing the levels is usually impractical and other
methods should be used, depending on the data.  For diatomic molecules, a good
approximation is to assume that the rotational, vibrational, and
electronic contributions to the partition function can be separately
evaluated for each electronic state, then combined by summing their
products weighted by the Boltzmann factor of each electronic state.
This gives

\begin{equation}
Q(T) = \frac{1}{\sigma} \sum_{i=1}^n Q_{{e_i}} Q_{v_{i}} Q_{r_{i}}
e^{-hcT_i/kT}\, ,
\label{molpart}
\end{equation}

\noindent where $\sigma$ is the symmetry number, equal to 1 or 2
for heteronuclear and homonoclear molecules, respectively, $Q_{e_{i}}$,
$Q_{v_{i}}$, and $Q_{r_{i}}$ are the electronic, vibrational, and rotational
contributions, respectively, to the partition function of the $i^{th}$
electronic state, and $T_i$ is the term value in cm$^{-1}$ of the $i^{th}$
state, with, as before, the ground state energy having zero energy by
definition.  As with the atoms, levels above about 20,000 cm$^{-1}$ contribute
less than 1\% to the partition function and, therefore, can be neglected.

When present, multiplet spin splitting of an electronic state is small
compared to $kT/hc$, and the electronic partition function can be
considered a statistical weight without any temperature dependence and
is given by (on dropping the $i$ subscript for clarity)

\begin{equation}
Q_e = (2S+1)(2-\delta_{\Lambda,0})\, ,
\label{qepart}
\end{equation}

\noindent where $S$ is the electron spin quantum number, $2S+1$ is the
multiplicity, $\Lambda$ is the electronic orbital quantum number: 0, 1,
2, 3... for $\Sigma$, $\Pi$, $\Delta$, $\Phi$... electronic states,
respectively, and $\delta$ is the Kronecker delta.  All states other
than $\Sigma$ states contribute an additional factor of two.  If the
spin splitting is large compared to $kT/hc$, then the individual
multiplet states can be treated as separate substates, and
weighted by the Boltzmann factor of that substate.  The vibrational and
rotational contributions to the partition function for each electronic state
in eq. (\ref{molpart}) can be calculated using asymptotic formulae derived
from Kassel (1933a, 1933b).

In some cases where partition functions of diatomic molecules are already
available in the literature as tables, as is the case with $TiO$, it may
be more convientient to use polynomial approximations, rather than evaluating
eq. (\ref{molpart}) using the methods outlined above.  Expressions for
calculating the partition functions of polyatomic molecules are in general
very much more complicated, but with the available data in \S\ref{hitran},
only the partition functions of $CH_4$, $H_2O$, and $NH_3$ are required.
Of these, that of $CH_4$ is easily calculated by evaluating a formula,
as given in \S\ref{hitran}.  Unlike the other molecules considered, $CH_4$
is a spherical top molecule with three equal principle moments of inertia.
The partition functions of the other molecules can be obtained
from polynomial fits.

In the case of diatomic molecules, each electronic state is split up into
vibrational and rotational levels, and transitions between two electronic
states, or even within the same (usually the ground state), make up a band
system.  Transitions between two given vibrational levels make up a band,
which in turn is resolved into individual rotational lines that have to be
calculated separately.  If the $gf$ values of these rotational lines are
available, then eq. (\ref{oscil}) can be used as above, but if they are not,
the line strengths can still be calculated, provided an oscillator strength
is available for either the band system, or the particular band being
calculated.  In the latter case, eq. (\ref{oscil}) can be replaced by

\begin{equation}
S = \frac{\pi e^2}{m_ec} f_{v'v''} \frac{\bar\nu}{\bar\nu_{v'v''}}
S^{\Lambda'\Lambda''}_{J'J''} \frac{e^{-hcF_i/kT}}{Q(T)}
\left[1-e^{-hc(F_j-F_i)/kT}\right]\, ,
\label{stren}
\end{equation}

\noindent where $f_{v'v''}$ is the band oscillator strength for the
$v' \leftarrow v''$ transition (by convention, in molecular spectroscopic
notation the lower state is double-primed, and is given after the upper
state), $\bar\nu$ is the transition frequency of the band origin (the
hypothetical transition with $J'=J''=0$), and $S^{\Lambda'\Lambda''}_{J'J''}$
is the H\"onl-London factor (Herzberg 1950; Kov\'{a}cs 1969)
for the rotational transition $J' \leftarrow J''$
between the electronic states with the corresponding values $\Lambda'$ and
$\Lambda''$.  Note that the H\"onl-London factors express the contributions
to the total line strengths of the rotational wavefunctions of the
participating levels, and should be normalized such that they satisfy eq.
(\ref{qepart}), which accounts for the statistical weight.

In cases where individual band oscillator strengths are not given, the line
strengths can be calculated from the electronic strength of the band system,
from which the band oscillator strengths can be calculated using

\begin{equation}
f_{v'v''} = q_{v'v''}f_{el}(\bar\nu_{v'v''})\, ,
\label{franckcond}
\end{equation}

\noindent where $q_{v'v''}$ is the Franck-Condon factor and
$f_{el}(\bar\nu_{v'v''})$ is the electronic oscillator strength of the system.
Note that the Franck-Condon factors are defined only for transitions between
different electronic states, express the relative contributions to the
band strengths due to the vibrational wavefunctions of the participating
levels, and are normalized such that they sum to unity from a given $v''$ to
all possible values of $v'$, (likewise, for the reverse transitions,
assuming that the electronic oscillator strength is a constant over the
system).  In general, this will be a function of the frequencies of the band
origins, and can be accounted for by using r-centroids, as first
introduced by Fraser (1954), and subsequently used, by, for example,
Nicholls \& Jarmain (1956), Nicholls (1965), and Sharp (1984). 
%<<<
When such data are not readily available, and the variation of the 
electronic component of the oscillator strength over the band system is
unknown, it could only be obtained by experiment or detailed quantum
mechanical calculations.  Both of these are beyond the scope of our work,
and the simplest assumption is to assume it is constant.  This is justified
by Schadee (1967), who provides a number of figures and expressions that show
the variation of the electronic component of the band oscillor strength as a
function of wavelength for one or more band systems of each of the
molecules $C_2$, $CN$, $N_2$, $N_2^+$, $CO^+$, $NO$, and $O_2$.  With
the exception of the red system of $CN$ ($A^2\Pi_i - X^2\Sigma^+$) and
possibly the Schumann-Runge system of $O_2$ ($B^3\Sigma^-_u - X^3\Sigma^-_g$),
these associated oscillator strengths vary by not much more than a factor of
three at worst across the bands; even the exceptions still vary by less than
an order of magnitude.
%>>>
For polyatomic
molecules a similar, though considerably more complicated, theory is in
general involved, but data in forms suitable for direct opacity calculations
are often available, as discussed in \S\ref{hitran}.

Many elements have several isotopes which can have a significant abundance.
For example, the dominant isotope of titanium, $^{48}Ti$, makes up about
74\% of all isotopes of titanium, with four other isotopes making up the
balance.  The isotopic shifts of atomic lines are negligible, but are
significant for molecules, and have to be considered for calculating line 
positions, though not for calculating partition functions (see below).  
Many datasets included line strengths and positions of isotopically substituted
molecules, so no additional provisions have to be made.  However, for those
diatomic molecules which involve elements with significant fractions of
minor isotopic species, and for which only data on the most abundant
isotopic form are available, the isotopically shifted lines can be
calculated.  This can only be performed if the data for each transition
include the vibrational and rotational quantum numbers of the participating
states.

If $\mu$ and $\mu^i$ are, respectively, the reduced masses of the main
isotopic version and an isotopically substituted version of the molecule,
then given from Herzberg (1950)

\begin{equation}
\rho = \sqrt{\frac{\mu}{\mu^i}}\, ,
\label{rho}
\end{equation}

\noindent the isotopic shifts of the harmonic, anharmonic and second-order
anharmonic vibrational constants are, respectively,
$\omega_e^i = \rho\omega_e$, $\omega_e^ix_e^i = \rho^2\omega_ex_e$ and
$\omega_e^iy_e^i = \rho^3\omega_ey_e$, the rigid and non-rigid rotational
constants are, respectively, $B_e^i = \rho^2B_e$ and $D_e^i = \rho^4D_e$,
and the first- and second-order vibration-rotation coupling constants are,
respectively, $\alpha_e^i = \rho^3\alpha_e$ and $\beta_e^i = \rho^5\beta_e$.
%<<<
The shifts of higher order constants, even when available, are neglected.
This is because they make only a small correction to the energy levels.
For multiplet states, various spin-orbit and spin-spin coupling constants,
again if they are available, are also neglected, and it is assumed that the
levels are shifted by only the effects already mentioned.  This is a higher
order effect beyond the scope of this work, and is expected to have a small
influence.
%>>>
With the isotopically
shifted constants, together with the vibrational and rotational quantum
number, the positions of the shifted energy levels can be calculated, and,
thus, also the positions of the lines in the spectrum.  The line strength
are obtained by just multiplying the strength from eq.(\ref{stren}) by
the fractional abundance of the isotope of the element.  This is applied
also to the main isotopic version of the molecule, if more than one are
considered. 
%<<<
%new paragraph

In nearly all cases the main isotopic form of a molecule is more than an order
of magnitude more abundant than the next most abundant isotopic form, so
abundance effect dominates in the line strengths over other isotopic effects.
The effects due to isotopically-shifted partition functions are generally only
a few percent, and so are neglected.  The line strengths will also be
influenced by the changed wavefunctions due to the shifted energy levels of the
participating states, but this is also expected to be small, as the isotopic
shifts of energy levels are small compared to the depth of a typical
internuclear potential well, whose shape is to a high order independent of the
nuclear masses.  Even if the necessary data are available, calculating such
wavefunctions is beyond the scope of this work and would provide at best a
very marginal improvement in accuracy.  In the case of homonuclear diatomic
molecules, the line strengths also depend on the nuclear spins, and alternate
between levels with even and odd parities (Herzberg 1950).  For nuclei with
zero spin, alternate levels are missing.  When one of the atoms is replaced by
an isotope, the molecule becomes heteronuclear and this symmetry is lost.
If both atoms are replaced by another isotope causing the molecule to be
homonuclear again, the line strengths alternate with parity depending on the
nuclear spins.  Thus, in the case of homonuclear molecules these effects have
to be accounted for.  The only homonuclear molecule considered in the opacity
calculations in this work is $H_2$, but deuterium is not discussed due to its
low abundance.  In the case of polyatomic molecules, the shifts are
considerably more complicated to calculate, and most of the data on these
molecules already include the isotopes.  In situations where no isotopic data
are given, no data on the levels involved in the transitions are given either.
%>>>

Once the integrated line strength for a particular transition being
considered has been obtained, including any isotopic versions of a
molecule, its contribution to the monochromatic absorption has to be
calculated.  This requires the line to be broadened into a suitable
profile, and the calculation of the absorption across each
profile is the most intensive part of any calculation.

With the exception of the resonance lines of sodium and potassium, a
Lorentzian profile should be used for all the broadening calculations.
Because of the generally high pressures and low temperatures, Lorentzian
broadening dominates over Gaussian Doppler broadening, so not employing
general Voigt profiles is justified, particularly when there are
large uncertainties with broadening.  Accordingly, the monochromatic
absorption in cm$^2$species$^{-1}$ can be calculated using

\begin{equation}
\sigma(\bar\nu) = \left(\frac{Sb}{c}\right) \frac{\Delta\bar\nu/2\pi}
{(\Delta\bar\nu/2)^2 + (\bar\nu-\bar\nu_o)^2}\, ,
\label{loren}
\end{equation}

\noindent where as before $S$ is the integrated line strength, $b$ is the
normalization correction factor, $c$ is the velocity of light,
$\Delta\bar\nu$ is the Lorentzian full-width at half-maximum in cm$^{-1}$,
and $\bar\nu_o$ is the frequency of the line center in cm$^{-1}$.  If the
profile is calculated out to large distances in the wings from the line
center, then the normalization correction factor is unity.  However, a
major problem in calculating profiles is to know how to treat the far wings.
If the wings are simply extrapolated out to large values of
$|\bar\nu-\bar\nu_o|$, the absorption is unrealistically large, and an
excessive amount of computing time is used.  As the absorption generally
drops off in the far wings at a rate larger than a simple Lorentzian profile,
a suitable treatment in the absence of a detailed theory would be to
cut off the profile at some distance $d$ from the line center, given by
a simple presciption of the form $d=\min(25P,100)$ cm$^{-1}$,
where $P$ is the total gas pressure in atmospheres.  The $\min()$ function
ensures that the lines do not become unrealistically broad.

Because the profile is truncated at a distance $d$ from $\bar\nu_o$, in
order to ensure that the total strength of the profile is conserved, the
normalization correction factor $b$ in eq. (\ref{loren}) is greater than
unity, and is given by

\begin{equation}
b = \left(\frac{2}{\pi}\right) \arctan\left(\frac{2d}{\Delta\bar\nu}\right)\, .
\label{corrsmall}
\end{equation}

However, another problem occurs when the profiles are much narrower than the
grid intervals, i.e. when $\Delta\bar\nu \ll w$, where $w$ is the grid
interval in cm$^{-1}$, which is likely to happen at low pressures when the
lines are very narrow.  On average, most profiles will fall between grid
points and only the far wings will be sampled, or even missed completely
because of the truncation discussed above, so such profiles will be
undersampled or not sampled at all.  In these cases, the line centers should be
moved to the nearest grid point, but this will overcompensate and cause the
lines to be overestimated in strength.  This is because the profile is
effectively replaced by a triangle whose apex is the profile's peak, which
has a larger area than the original profile.  In this case, the integrated
strength causes the profile to appear to be too strong, so the normalization
factor must reduce the total strength.  After taking account of the small
absorption in the two grid points immediately adjacent to the grid point at
the center of the profile, the normalization factor used in eq. (\ref{loren})
is given by

\begin{equation}
b = \left(\frac{\pi}{4}\right) \left(\frac{\Delta\bar\nu}{w}\right)
\left[\frac{4+(\Delta\bar\nu/w)^2}{2+(\Delta\bar\nu/w)^2}\right]\, ,
\label{corrbig}
\end{equation}

\noindent where in this case $b$ is less than unity and the criterion that $\Delta\bar\nu < w^2/2$ is
satisfied.  In these cases, truncating the wings is done simply to avoid
unnecessary computation, and the truncation is performed sufficiently far out
in the wings so as not to affect eq. (\ref{corrbig}), and if necessary may
exceed the distance $d$ using the criterion based on the total gas pressure
mentioned above.  In the limit of very small widths, the contributions
made by the two grid points adjacent to the profile's center become
negligible, and the right hand side of eq. (\ref{corrbig}) reduces to
$\pi\Delta\bar\nu/2w$.

The isotopic shifts of the energy levels of atoms are negligibly 
small (typically, one part in ten thousand), and
can be ignored.  For the molecules, the isotopic shifts can be significant,
and can affect both the partition function and the spectral lines.  However,
the effect on the partition function is relatively small for
%<<<
the molecules of interest, as previously mentioned, and most molecules are
%>>>
dominated by one major isotope, so
calculations can be simplified considerably by using only the partition
function of the most abundant isotope for all isotopic forms.  In the case
of spectral lines the isotope effects need to be considered, and separate
sets of shifted lines should be calculated for each isotopic version.
Many sources of line data already incorporate the isotopically shifted
lines with their fractional abundances included with the line strengths,
but when such data are not available, the isotopic shifts have to be
specifically calculated (see eq. \ref{rho}).

In calculating models of brown dwarfs and extrasolar giant planets,
the monochromatic opacities produced by the species known to be important
sources of absorption are required at the temperature and pressure at
each level in the model.  For an object relatively remote from its parent
star where stellar irradiation is unimportant, the majority of the radiation
is in the infrared, where molecules such as $CH_4$ and $H_2O$ have strong
vibration-rotation (and in the case of the latter also pure rotation)
absorption, but are relatively transparent in the visible and the near
ultraviolet parts of the spectrum.  This is because these molecules have
stable closed electron shells, and only highly excited electronic states
exist above the ground state.  At the temperatures of interest the
population of these levels is negligibly small, and transitions from the
ground state to these excited states lie further in the ultraviolet,
where the flux is neglibibly small, so ultraviolet opacities are unimportant.
In the case of a substellar object orbiting close to its parent star,
it can be irradiated at short wavelengths, which can affect its modeling,
so ultraviolet opacities are required.

For those opacity sources that do not require much processing time, they can
be calculated ``on the fly" as required.  This includes all calculations of
processes that produce continuous absorption, such as bound-free processes,
those where the data can be obtained in the form of a smoothed absorption of a
band system, and where a relatively small number of spectral lines have to be
individually computed.

%<<<
For those atomic and molecular species for which data are available
containing millions of spectral lines that have to be individually
computed, the processing time per thermodynamic point can be too long,
of the order of 30 minutes on a modern computer with a processor speed
of about 2 GHz, for them to be calculated on the fly.
%>>>
In these cases a good option is to generate a large number of
files containing pre-computed monochromatic opacities for each species for a
range of temperatures and pressures.  Then, interpolation can be performed at
the time the opacities are required.  A good choice would be to consider 50
logarithmic steps in temperature from 50 K to 3000 K, and 50 logarithmic steps
in pressure from $1.00\times 10^{-7}$ to $3.16\times 10^2$ atmospheres. 
Each table would consist of monochromatic opacities in
cm$^2$ species$^{-1}$ over a range of wavenumbers.  For those opacities
important in the infrared, a suitable tabulation could be from
100 cm$^{-1}$ to 30,000 cm$^{-1}$ in steps of 1 cm$^{-1}$
(100 $\mu$m to 0.33 $\mu$m, respectively), which would include the visible
part of the spectrum when data are available.  For those that are important
sources in the visible and ultraviolet parts of the spectrum, but not the
infrared, the tabulation would cover a suitable range appropriate for the
species (see the discussion in section \S\ref{uvopac}), but starting from at
least 10,000 cm$^{-1}$ (1 $\mu$m) in steps of 2 cm$^{-1}$.  The suggested
tabulation intervals in temperature, pressure, and wavenumber represent
realistic compromises to render the data as accurately as possible with
a manageable amount of processing time and
%<<<
memory, and interpolations in temperature and pressure should be accurate
to one part in 10$^3$ or better.
%>>>

Because opacities are usually confined to specific spectral regions, often
with large sections of zero absorption on either side in the spectral range
of interest, in many cases considerable disk space and processing time can be
saved by noting the blank regions and not storing the zero data in these
regions.  This is particularly the case between the last non-zero grid point
and 30,000 cm$^{-1}$ for the relevant infrared opacities.  Further savings
can be made by not calculating and storing opacities of refractory species,
such as $TiO$, below 1000 K, as their abundances would be too small to be of
any relevance.  In such cases very short dummy files can be generated, and
when the opacities are obtained by interpolation, zero values are handled
very quickly without having to read in any data.

\subsection{HITRAN or GEISA-like Data}
\label{hitran}

Several of the molecules that are important sources of opacity in
brown dwarf and extrasolar giant planet atmospheres in the
infrared have their data expressed in HITRAN (Rothman et al. 2003, 2005)
or GEISA-like (Jacquinet-Husson et al. 1999, 2003, 2005) format.  For the
molecules of interest here, more extensive data are available from sources
other than HITRAN or GEISA.

All HITRAN and GEISA-like line strengths are expressed in units of cm
species$^{-1}$, which can be converted to the required strengths in
cm$^2$ s$^{-1}$ species$^{-1}$ by multiplying by the velocity of light
and various factors containing the Boltzmann factor and the partition
function.  Specifically, this can be written as

\begin{equation}
S = cS_h \frac{Q(T_o)}{Q(T)} e^{F(hc/kT_o-hc/kT)}
\frac{\left(1-e^{-hc\bar\nu/kT}\right)}{\left(1-e^{-hc\bar\nu/kT_o}\right)}\, ,
\label{hitstren}
\end{equation}

\noindent where $S_h$ is the tabulated line strength in cm species$^{-1}$ in
HITRAN or GEISA format, and $T_o$ is the HITRAN or GEISA reference
temperature of 296 K.  Thus,
$Q(T_o)$ is the partition function at that temperature.  The term after
the partition functions is the Boltzmann factor, after taking account of the
Boltzmann factor at 296 K, and the last term is the stimulated emission
correction factor, which has to allow for the correction factor at 296 K.
Isotopes, when relevant, are accounted for by their fractional abunadances
already being included in $S_h$, so no separate treatment is required.
Some of the datasets give the transition moment or the Einstein A coefficient
in addition to $S_h$, but this is unnecessary, except as a
check on the data.

The line widths are calculated from broadening parameters using

\begin{equation}
\Delta\nu = \alpha \left(\frac{T_o}{T}\right)^{\beta} \frac{kT}{A_o} N_t\, ,
\label{hitwid}
\end{equation}

\noindent where $\Delta\nu$ is the full-width at half-maximum in cm$^{-1}$,
as given in eq.(\ref{loren}), $\alpha$ and $\beta$ are two widths parameters,
as given in the HITRAN or GEISA databases, $A_o$ is the standard atmosphere in
dyne cm$^{-2}$, and
$N_t$ is the total number density of particles in cm$^{-3}$.  In practice for
most of the temperatures and pressures of interest, most of the gas is
in the form of $H_2$ and $He$.  So, it is convenient to assume that the
broadening is the weighted average of these two gases.  In the case of
$CO$, the individual contributions due to $H_2$ and $He$ can be calculated,
and the combined width is found from the weighted fractional abundances by
number of these two gases.  The individual values of $\alpha$ are given,
and $\beta$ can be assumed to be 0.6 and 0.4, respectively, for $H_2$ and $He$.
With the widths calculated, the monochromatic absorption in cm$^{-2}$
species$^{-1}$ across the spectrum can be calculated, as discussed in
\S\ref{lines}.

All the data in HITRAN or GEISA format cover the infrared part of the spectrum
up to 10,000 cm$^{-1}$, and in some cases to up larger wavenumbers, subject
to the available data, with the following molecules being discussed here:
$H_2O$, $CH_4$, $NH_3$, and $CO$.  Due to the availablility of the
data and some specific restrictions, several molecules require special
treatment as discussed here.

$H_2O$ - Together with $CH_4$, this molecule is one of the most important
sources of opacity in the infrared.  Partridge and Schwenke (1997) and
Barber et al. (2006) have very extensive databases,
%<<<
in addition to those in the latest versions of HITRAN (Rothman et al. 2005)
and GEISA (Jacquinet-Husson et al. 2005),
%>>>
but it is likely that there are too many weak
lines for calculations to be completed in a realistic period of time with
typical computing resources available.  In that case the lines can be
binned in decades of line strength $S_h$, and only those lines stronger
than a threshold value need be used in opacity calculations.  A good
compromise between realistic opacity calculations and processing time is
with $S_h \ge 1\times 10^{-40}$ cm molecule$^{-1}$.  This results in a
considerably reduced database, but still contains over fourty million lines.

The partition function can be obtained from a 6$^{th}$-order polynomial
of the form

\begin{equation}
\ln(Q) = a + bx + cx^2 + dx^3 + ex^4 + fx^5 + gx^6\, ,
\label{fith2o}
\end{equation}

\noindent where $x = \ln(T)$, and $a$ to $g$ are two sets of fitted
coefficients for the ranges $T \le$ 500 and $T >$ 500 K.

\begin{figure}
\epsscale{1.00}
\centerline{\includegraphics[angle=-90,width=19cm]{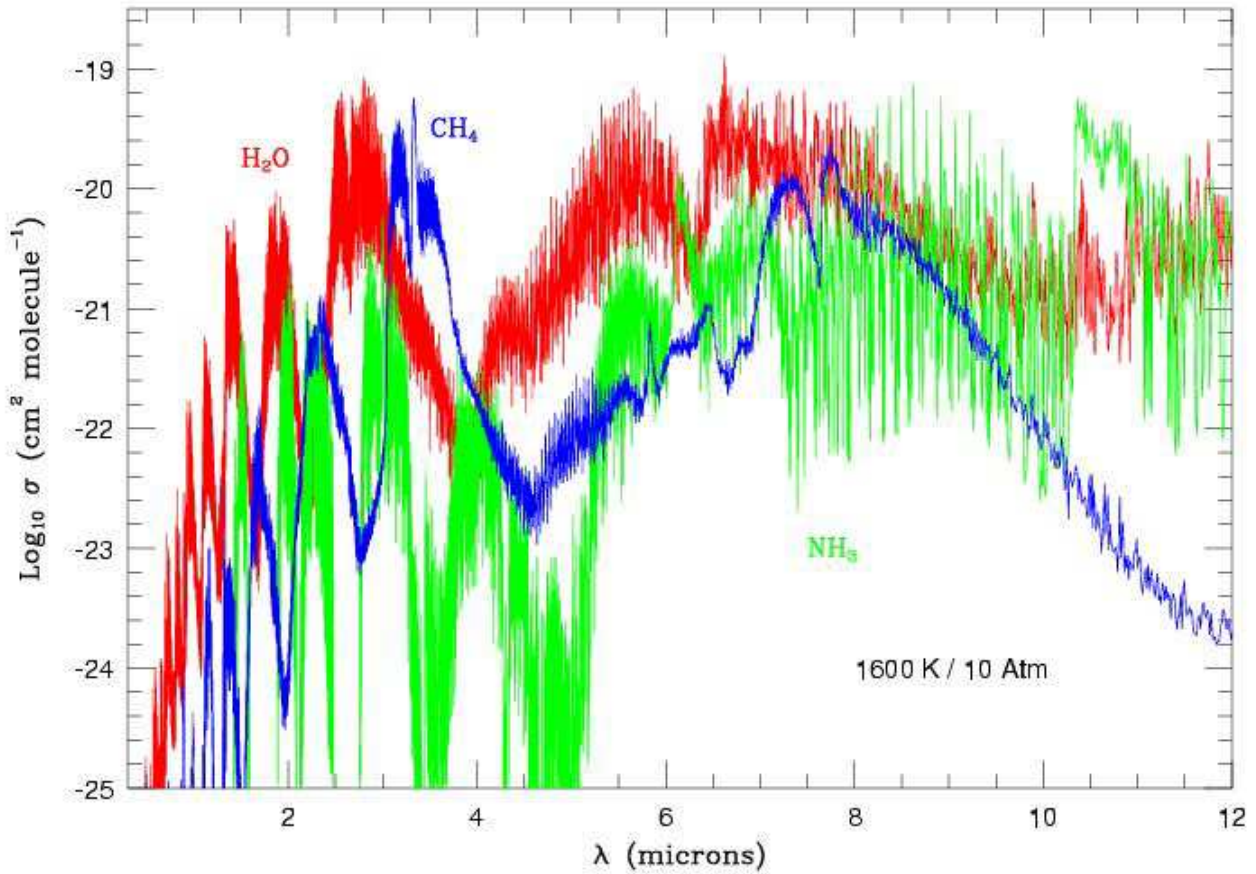}}
\caption{The log (base 10) of the monochromatic absorption $\sigma$ in
cm$^2$molecule$^{-1}$ as a function wavelength $\lambda$ in
$\mu$m in the infrared at a temperature of 1600 K and a pressure of 10
atmospheres for the vibration-rotation transitions of $H_2O$, $NH_3$, and
$CH_4$, as indicated by the red, green, and blue curves, respectively.
The contribution due to different isotopes is included here and in all
other figures, when data are available and relevant.  The temperature and
pressure chosen here are reasonably representative, in particular the lines
are sufficiently broadened at 10 atmospheres that the very rapid
fluctuations in absorption over short wavelength intervals are suppressed
so that the main band features can be more easily seen.  At significantly
lower pressures the broadening of the lines is much smaller and the
absorption can change so rapidly in short wavelength intervals that the
main features do not show up so clearly.  As can be seen here, $H_2O$ has
a strong absorption feature just shortward of 3 $\mu$m, and $CH_4$ has a
strong peak near of 3.3 $\mu$m.  In the region of 8 $\mu$m to 9 $\mu$m
all three molecules absorb strongly; however, between about 10.5 $\mu$m
and 11 $\mu$m $NH_3$ has absorption which is distinctly higher than that
of the other two molecules.  When the combined opacity is calculated, the
individual absorptions must be weighted by the abundances.  (See text for
discussion.)}
\label{fig:1}
\end{figure}

\begin{figure}
\epsscale{1.00}
\centerline{\includegraphics[angle=-90,width=19cm]{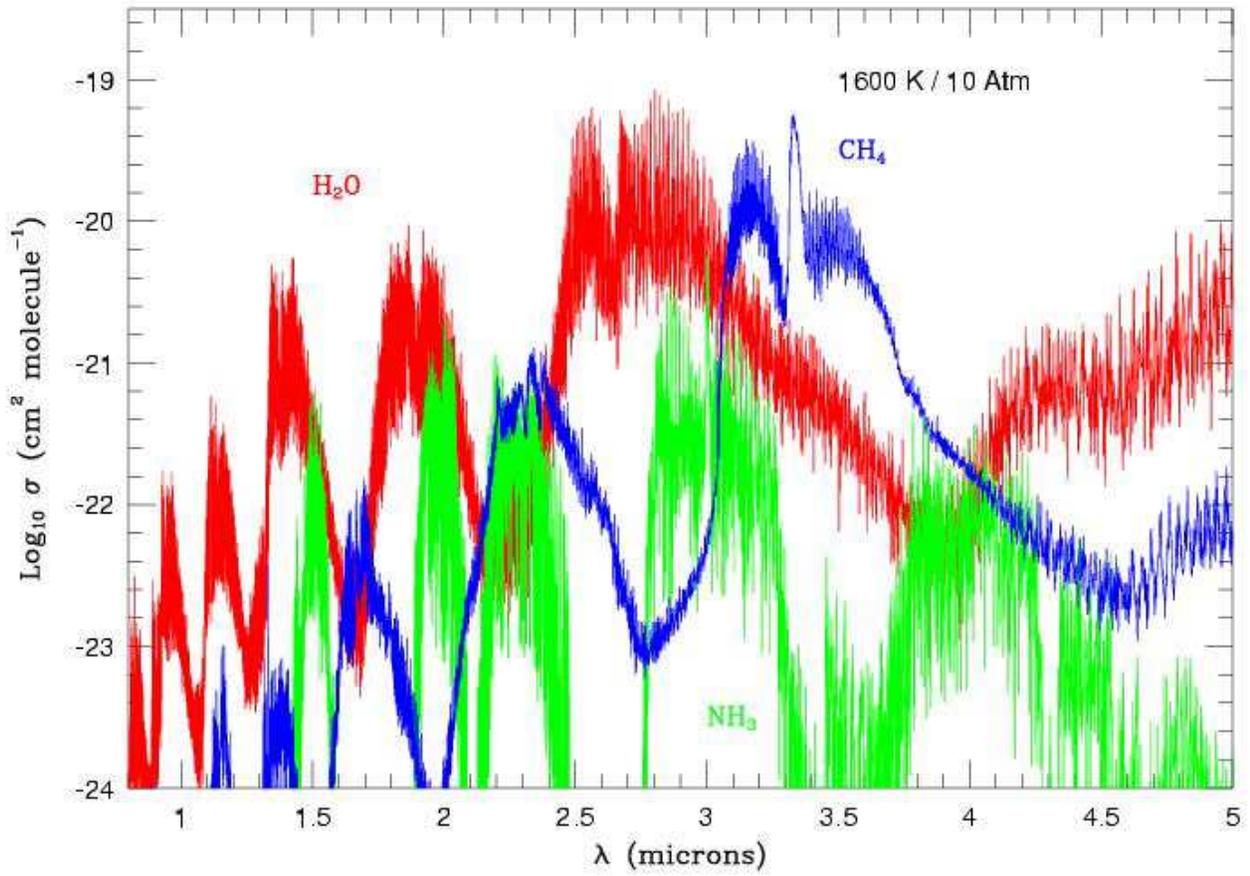}}
\caption{This is the same as the previous figure, but with the shorter
wavelength region much expanded, together with a small increase in scale
of the ordinate, and shows in greater detail the absorption at shorter
infrared wavelengths.  Of particular interest are the positions and
structures of the bands due to $H_2O$ and $CH_4$.}
\label{fig:2}
\end{figure}

Figure \ref{fig:1} shows a plot of the monochromatic absorption of $H_2O$
in the near infrared, together with $CH_4$ and $NH_3$, as discussed later
in this section, for a temperature of 1600 K and a total gas pressure of
10 atmospheres.  It is seen that $H_2O$ has strong absorption over this
region, with a strong peak centered at about 2.7 $\mu$m, and another broader
peak centered at about 6.4 $\mu$m.  The double peak structure of some of the
bands with a trough between them, is due to the gap between the $P$- and
$R$-branches.  Over most of the wavelength range plotted, $H_2O$ is the
strongest intrinsic absorber, and moreover, after $H_2$, which has no
permanent dipole-allowed infrared absorption, $H_2O$ is either the second
or third most abundant molecular species over most of the temperatures and
pressures of interest.  Abundances are discussed in greater detail in
\S\ref{atmosabun}.  Figure \ref{fig:2} is a repeat of Fig. \ref{fig:1},
but plotted on a larger wavelength scale and a slightly larger opacity
scale, for the short wavelength side of Fig. \ref{fig:1}.

$CH_4$ - This is the other significant source of opacity in the infrared, along
with $H_2O$, for much of the temperatures and pressures of interest.  As
no single source of data covers the spectral region of interest, which
extends into the visible part of the spectrum, the best approach is to
merge data from three separate sources having four data files, taking
account of overlapping frequency ranges in order to avoid duplicate
lines, and incompatibilities with the data.

The main dataset covering lower frequencies up to 6825 cm$^{-1}$ is
available from Borysow et al. (2003), as used in our earlier work on
$CH_4$ (Burrows, Marley, \& Sharp 2000), and contains over 17 millions lines.
Brown et al. (1997) contains two files for the frequency coverage
4800 to 7698 and 8002 to 9199 cm$^{-1}$.
%<<<
This is in addition to those in the latest versions of HITRAN
(Rothman et al. 2005) and GEISA (Jacquinet-Husson et al. 2005).
%>>>
Laboratory studies of $CH_4$
are reported in Robert et al. (2000) and Brown, Dulick \& Devi (2001).
In superimposing a plot of the lower range of the data from Brown et al.
(1997) with that from Borysow et al. (2003), it is seen that the coverage of
the lower range overlaps with the data from Borysow et al. (2003), but
includes a large gap in coverage centered at about 5850 cm$^{-1}$,
corresponding to a peak in the opacity in the data from Borysow et al. (2003).
It can also be seen that as the the opacity from the Borysow et al. (2003) data
tails off, it is overlapped by increasing opacity from the Brown et al. (1997)
data, and corresponds to a band system that was not covered by the Borysow
et al. (2003) data.
This can be confirmed by looking at the opacity from Strong et al.
(1993), which we used in our earlier work (Burrows, Marley, \& Sharp 2000)
to extend the opacity to higher frequencies.  By a careful examination of
the data, it can be seen that there is a small additional gap in coverage
from Brown et al. (1997) below 6356 cm$^{-1}$.  Where the data from Brown et al.
(1997) has coverage below this point, it appears to duplicate the data from
Borysow et al. (2003).  However, the overlapping coverage at and above
6356 cm$^{-1}$ appears to cover a different band system, so the data from
Brown et al. (1997) below 6356 cm$^{-1}$ should be discarded.
%<<<
A certain amount of judgment is required in deciding which lines
to include from overlapping databases, and which to discard.  This is
decided by looking at plots of absorption at high resolution in overlapping
regions.
%>>>

The second file from Brown et al. (1997) with the higher frequency
coverage presents no problem with duplication of any other files,
and extends the opacity coverage of $CH_4$ up to 9199 cm$^{-1}$.  In
superimposing data from Strong et al. (1993), the peaks in absorption
covered by the two sets of the data from Brown et al. (1997) are seen to match
well, and the gap in coverage between the two files presents no problem,
as it corresponds to a minimum in the opacity as seen in the Strong et al.
data.  As all three files are in the same format, and match each other
together with the data from Strong et al., where there is overlap,
the same code could be used for the whole range up to 9199 cm$^{-1}$
without any special treatment, with the data from Borysow et al. (2003)
and Brown et al. (1997) joined together.  A good place to join is at 6356 cm$^{-1}$.

However, the upper frequency limit of 9199 cm$^{-1}$ is still
too low because $CH_4$ opacities are also important at shorter wavelengths.
In our earlier work, when we did not have available data from Brown et al.
(1997), the  $CH_4$ opacities could be extended continuously up to 9495
cm$^{-1}$ using data from Strong et al. (1993).  This was followed by an
unavolidable gap in coverage, until data from Karkoschka (1994) starting
at 10,000 cm$^{-1}$ could be used.  The data from Strong et al. and
Karkoschka give only monochromatic opacities as a function of wavenumber
or wavelength.  The data from Strong et al. give a simple formula for
calculating the temperature dependence of the opacity, but no pressure
dependence, and the data from Karkoschka give no dependences at all.
Consequently, various means of scaling could be devised to match these
opacities with the data from Borysow et al. (2003).

However, the small range covered by Strong et al. (1993) beyond the upper
limit from Brown et al. (1997) is probably not worth the extra work in including
the former for a marginal gain.  Additionally, without other information,
there is little to be gained by attempting to fill in the gap in
opacities using interpolation.

%<<<
The data from Karkoschka (1994) cover the range from 1 $\mu$m to 0.3 $\mu$m,
i.e. from about 10,000 cm$^{-1}$ to about 33,300 cm$^{-1}$.  However,
shortward of 0.4 $\mu$m (i.e. 25,000 cm$^{-1}$), the opacity includes gaps
and in our judgement does not seem very reliable.  Since its contribution to
the opacity decreases with decreasing wavelength, whilst other processes
become more important, the amount of processing and storage necessary to
handle this section of data was not considered worthwhile.
%>>>

As the data from Karkoschka (1994) have no pressure or temperature dependency,
a simple method can be devised to match it to the rest of the data.  As the
monochromatic absorption over a number of wavelength points is given without
any information on the lines producing this absorption, there is no way
any pressure dependency can be determined. However, it is possible to adopt a
simple scaling factor to adjust the absorption for different temperatures.
The data provided by Karkoschka are applicable for 160 K; which is somewhat
higher than the effective temperature of Jupiter.  The strongest line in the
upper of the two frequency ranges listed by Brown et al. (1997), as
discussed above (in fact at 8644.2322 cm$^{-1}$), can have its strength
calculated at 160 K, then when opacities are calculated at some required
temperature $T$, the ratio of the strengths at $T$ and 160 K can be used to
scale the Karkoschka opacities.  This very simple scaling will at least allow
the Karkoschka data to follow the trend of the other data with changing
temperatures.

Following Herzberg (1968), the partition function for $CH_4$ can be
calculated from

\begin{equation}
Q(T) = \frac{4}{3}\sqrt{\pi} e^{hcB/4kT}
\left(\frac{kT}{hcB}\right)^{\frac{3}{2}}
\left(1 + \frac{15}{4}\frac{hc}{kT}\frac{D}{B^2}\right)
\prod_{i=1}^4 \frac{1}{\left(1 - e^{-hc\omega_i/kT}\right)^{d_i}}\, ,
\label{fitch4}
\end{equation}

\noindent where $B$ and $D$ are the rigid
rotational and the non-rigid correction constants in cm$^{-1}$ for the
spherical top model of $CH_4$, $\omega_i$ is the vibrational harmonic
constant of the $i^{th}$ vibrational mode, of which there are four, and
$d_i$ is the corresponding degeneracy of that mode, such that for the four
modes the degeneracies are 1, 2, 3, and 3, respectively.  This formula
neglects higher-order effects, such as anharmonicity and vibration-rotation
coupling, and only the partition function of $^{12}CH_4$ is calculated,
with the differences for $^{13}CH_4$ being neglected due to the low abundance 
of $^{13}C$ (approximately one part in a hundred).  $CH_4$ is the best
known example of a spherical top molecule, where the three principle
moments of inertia are the same, as are, hence the rigid rotational
constants.  The rotational and vibrational constants are well known, and can
be obtained from many sources, such as Herzberg (1966) or the JANAF tables
(Chase et al. 1985).

\begin{figure}
\epsscale{1.00}
\centerline{\includegraphics[angle=-90,width=19cm]{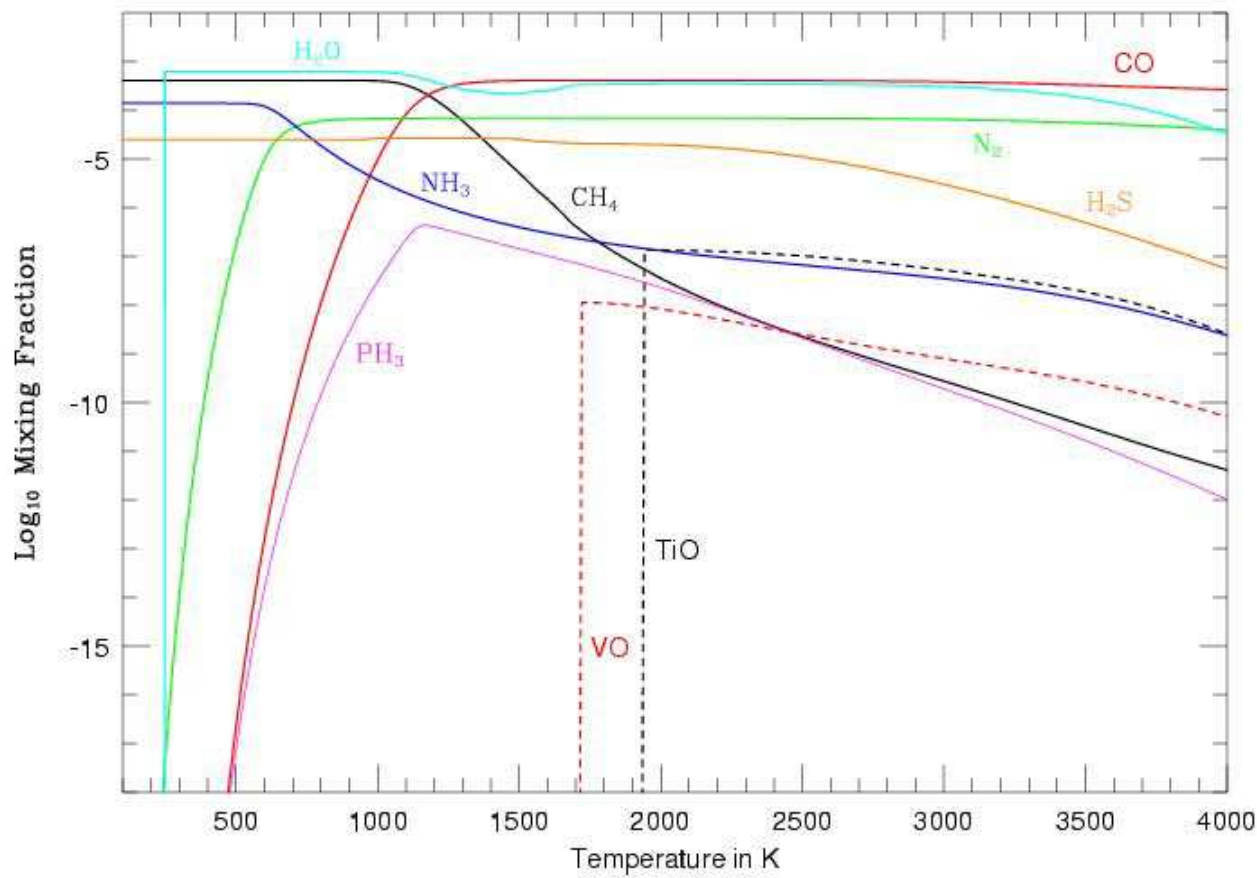}}
\caption{The log (base 10) of the mixing fraction as a function of
temperature at a total gas pressure of 1 atmosphere for the seven molecules
shown with solid curves $CH_4$ (black), $CO$ (red), $N_2$
(green), $NH_3$ (blue), $H_2O$ (cyan), $H_2S$ (orange), and
$PH_3$ (magenta), and the two molecules shown with dashed curves
$TiO$ (black) and $VO$ (red).  At 4000 K,
$CO$ and $N_2$ are the most stable species, containing nearly all
the carbon and nitrogen, respectively.  With decreasing temperature,
$CO$ reacts with $H_2$, forming $CH_4$, which becomes the dominant
carbon-bearing species at low temperatures, and $N_2$ reacts with
$H_2$, forming $NH_3$, which likewise becomes the dominant
nitrogen-bearing species at low temperatures.  Except above about 3000 K,
$H_2O$ is fully associated containing nearly all the available oxygen
that is not bound in $CO$.  Below about 1600 K, its abundance
temporarily falls slightly due to the condensation of silicates which
reduce the available oxygen; however, the mixing fraction of $H_2O$
then rises again when $CO$ is converted to $CH_4$, which releases
the oxygen tied up in $CO$.  Finally, at 273 K $H_2O$ drops
effectively to zero due to the condensation of ice.
With decreasing temperature, both
$TiO$ and $VO$ rise as they associate, then sharply drop to
effectively zero when condensates involving $Ti$ and $V$ form.}
\label{fig:17}
\end{figure}

An example of joining together the data from different sources is the
monochromatic absorption of $CH_4$, shown in Figs. \ref{fig:1} and
\ref{fig:2}; in several places, it fills in windows in the
absorption of $H_2O$.  The strongest band in the wavelength range shown
is centered just above 3.3 $\mu$m, where the central peak between the
$P$- and $R$-branches is due to the $Q$-branch.  As discussed in more
detail in \S\ref{atmosabun}, Fig. \ref{fig:17} shows that
$CH_4$ is only slightly less abundant that $H_2O$ at low temperatures, due
to the lower abundance of carbon compared to oxygen.

$NH_3$ - This is the third most important molecule at low temperatures, after
$H_2O$ and $CH_4$.  Data are available from Brown et al. (2000) listing over
34,000 lines from below 1 cm$^{-1}$ to over 7000 cm$^{-1}$.  Unlike the
other two molecules, no special treatment is required, and a set of
precomputed files for a suitable set of temperatures and pressures,
as discussed above, can be generated relatively rapidly.  Even though
$NH_3$ is important in giant planet atmospheres, suitable sources of data
for a spectral coverage to shorter wavelengths do not appear to be
readily available.

Like $H_2O$, the partition function can conveniently be calculated from
a polynomial fit in the same form as eq. (\ref{fith2o}) for a
6-$^{th}$ order polynomial, using available numerical values in the range
50 $\le T \le$ 3000 K for the isotope $^{14}NH_3$.  This covers the whole
temperature range that is likely to be of interest.  As before, other
isotopic versions of nitrogen can be neglected, with the next most abundant isotope,
$^{15}N$, comprising only $\sim$0.4\% of total nitrogen.  

As with $H_2O$ and $CH_4$, the monochromatic absorption of $NH_3$ is
plotted on Figs. \ref{fig:1} and \ref{fig:2}.  For most of the
wavelengths shown, $NH_3$ is an intrinsically weaker opacity source
than $H_2O$ and $CH_4$ over most wavelengths, and additionally it is less
abundant than the other two molecules because nitrogen is less abundant
than carbon or oxygen.  Nevertheless, it is still a relatively important
opacity source in the infrared, and has a strong absorption peak at about
10.5 $\mu$m, where its absorption exceeds $H_2O$ and $CH_4$ by a large
amount.  As with $H_2O$ and $CH_4$, the abundance of $NH_3$ is plotted on
Fig. \ref{fig:17}, which is discussed in more detail in \S\ref{atmosabun}.

$CO$ - Although in principle $CO$ is a non-refractory molecule like the
three discussed above, below about 1200 K, depending on the pressure, it is
superseded in abundance by $CH_4$; its abundance becomes very small below
about 800 K, if equilibrium is assumed.  Nevertheless, a full set of opacities
for temperatures down to 50 K should still be computed if non-equilibrium
calculations are required.  The data are available from HITRAN
(Rothman et al. 2003, 2005) and GEISA (Jacquinet-Husson et al. 1999, 2003,
2005), which can be used to calculate the absorption from 3 to nearly
8500 cm$^{-1}$ in the infrared for about 99,000 transitions covering the
pure rotational, and the fundamental (first harmonic), second, third and
fourth harmonic vibration-rotation bands.  Unlike the polyatomic molecules
mentioned above, there are relatively few lines of $CO$ due to the fact
that the ground electronic state is a $^1\Sigma$ state, and each band
consists of a single $P$- and $R$-branch.  The other isotopic forms, the
most important of which is $^{13}C^{16}O$, make relatively small
contributions to the opacity when compared with the main isotopic form
$^{12}C^{16}O$.  There are large gaps between the bands with negligibly
small opacities.

The partition function can be obtained from eq. (\ref{molpart}) and asymptotic
formulae for the vibrational and rotational contributions.  Because only
the ground electronic state contributes any significant fraction to
the partition function, and because it is just a singlet $\Sigma$ state,
there is no summation in eq. (\ref{molpart}), which reduces to just
the product of the vibrational and rotational contributions of the ground
state.  The spectroscopic constants used to calculate the partition function
can be obtained from Huber \& Herzberg (1979).

\begin{figure}
\epsscale{1.00}
\centerline{\includegraphics[angle=-90,width=19cm]{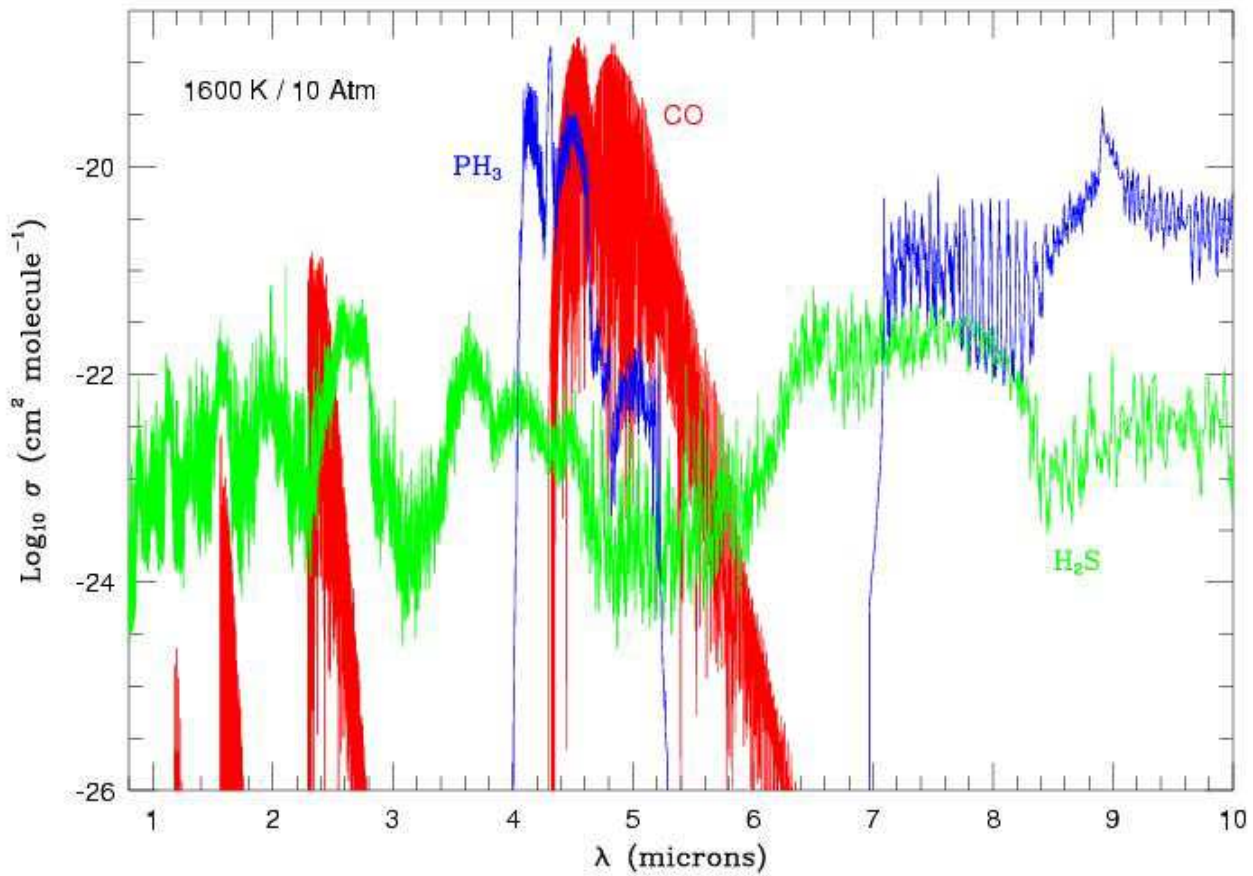}}
\caption{The log (base 10) of the monochromatic absorption
$\sigma$ in cm$^2$molecule$^{-1}$ as a function wavelength $\lambda$ in
$\mu$m in the infrared at the same temperature and pressure of the
previous figures for vibration-rotation transitions of $CO$, $H_2S$, and
$PH_3$, as indicated by the red, green, and blue curves, respectively.
Since $CO$ is a diatomic molecule with no electron spin or orbital angular
momentum in its ground electronic state, the spectra are particularly
simple with bands caused by only $P$- and $R$-branches, together with
isotopic versions.  The very strong absorption at $\sim$5 $\mu$m
is due to the fundamental (first harmonic) vibration-rotation transition
of $CO$.  The progressively weaker bands at progressively shorter
wavelengths are due to the second, third and fourth harmonics.  Note the
strongest peak in $PH_3$ absorption partially overlaps the CO first
harmonic absorption, but a second peak at 9 $\mu$m is clear from $CO$
absorption.  However, both $H_2S$ and $PH_3$ have low abundances, and
have to compete with much more abundant $H_2O$, $NH_3$, and $CO$
or $CH_4$.  Note that for a given pressure the region in temperature
over which $CO$ and $CH_4$ are of comparable abundances is
relatively narrow; otherwise, most carbon is either in $CO$ or
$CH_4$.}
\label{fig:3}
\end{figure}

The monochromatic absorption of $CO$ at 1600 K and a total gas pressure
of 10 atmospheres is shown in Fig. \ref{fig:3}, together with $PH_3$ and
$H_2S$, which are briefly discussed in \S\ref{precomp}.  $CO$ is the only
diatomic molecule shown here and in the previous figures, and has a
particularly simple spectrum, with only simple $P$- and $R$-branches, which
are easily seen in the strong absorption feature centered at about
4.67 $\mu$m.  This corresponds to the fundamental vibration-rotation band
and the associated ``hot" bands, and is the strongest feature in its
infrared absorption spectrum.  The weaker bands at shorter wavelengths
correspond to the second, third, and fourth harmonics with progressively
decreasing wavelengths.  Although $CO$ has some strong absorption,
particularly around the $\sim$4.5--5.5-$\mu$m region where it exceeds the
intrinsic
absorption of the molecules discussed so far, the absorption is confined to
relatively narrow spectral regions, with very low absorption in other
places; it is not a particularly important opacity source in the
infrared.  Figure \ref{fig:17} shows that it replaces $CH_4$ at higher
temperatures, as discussed in more detail in \S\ref{atmosabun}.

\subsection{Heavy Metal Oxides}
\label{oxides}

The oxides of transition metals, where ''metals" are used here in a
chemical, rather than an astrophysical, sense, form a well defined group
with many similar physical, chemical, and spectroscopic properties.
Of these, $TiO$ and $VO$ are by far the most important sources of opacity
over the upper regions of the temperatures of interest.

$TiO$ - Because compounds involving titanium are refractory, gas-phase
species of titanium do not exist at low temperatures.  Thus, both processing
time and memory can be saved by not calculating opacities below 1000 K.
Unlike many volatile gaseous species like $H_2O$, where vibration-rotation
transitions are the main source of absorption in the infrared, the opacity
of $TiO$ in the infrared, as well as in the visible, is due to transitions
between different electronic states.  Two important sets of data of $TiO$
can be found in Allard, Hauschildt, \& Schwenke (2000) and Plez (1998),
with the data for the former being in a HITRAN or GEISA-like format, similar
to the data for the molecules discussed in \S\ref{hitran}.

The dataset from Allard, Hauschildt, \& Schwenke (2000) lists about 12 million
lines in one large file from 3 to nearly 30,000 cm$^{-1}$, covering the
whole infrared and visible parts of the spectrum for the transitions of the
most abundant isotope $^{48}TiO$, together with $^{46}TiO$, $^{47}TiO$,
$^{49}TiO$, and $^{50}TiO$, for the seven different electronic band systems 
$\alpha$($C^3\Delta - X^3\Delta$), $\beta$($c^1\Phi - a^1\Delta$),
$\gamma$($A^3\Phi - X^3\Delta$), $\gamma'$($B^3\Pi - X^3\Delta$),
$\delta$($b^1\Pi - a^1\Delta$), $\epsilon$($E^3\Pi - X^3\Delta$), and
$\phi$($b^1\Pi - d^1\Sigma$).  Because all the data are contained in one
file, no provision is required to handle the isotopes or the electronic
transitions separately.  Also, because the data are given in a HITRAN or
GEISA-like format, the line strengths and widths are calculated with the same
methods discussed in \S\ref{hitran}.

The data from Plez (1998) are contained in five files for each of the
isotopes for nine electronic systems, i.e. those listed immediately above,
plus two additional systems.  The data for $^{48}TiO$ has about 3 million
lines covering the range from about 2800 cm$^{-1}$ to about 29,000 cm$^{-1}$,
and is more restricted for the other isotopes.  The line strengths are
expressed as $gf$ values, and, given the excitation energy of the lower state
and transition frequency, the line strength can be calculated directly from
eq. (\ref{oscil}).  However, no information is available for line
broadening, so a simple formula based on the rotational quantum number
is used, namely

\begin{equation}
\Delta\nu = \left[w_0 - \min(J'',30)w_1\right] kTN_t/A_o\, ,
\label{simplewid}
\end{equation}

\noindent where $w_0$ is the full-width at half-maximum of a transition at
1 atmosphere when the lower rotational quantum number, $J''$, is zero,
and $w_1$ is a coefficient that governs the dependancy of the line
width on $J''$, with the broadening at $J''=30$ being used for larger
values of $J''$.  Numerical values of 0.1 and 0.002 are adopted for $w_0$
and $w_1$, respectively, and in the absence of a suitable theory, appear
to give realistic broadening for a number of molecules whose line
broadening parameters are not available.  This somewhat \textit{ad hoc}
approach is in general agreement with experimental observations, where
broadening increases with $J''$ up to some limit (see, for example,
Varanasi \& Chudamani 1989).

If the opacities of $TiO$ using the two sources of data are compared, a
significant and consistent discrepancy between the two sources of data and
methods employed is found, with the absorption from Allard, Hauschildt,
\& Schwenke (2000) being about two orders of magnitude higher than Plez
(1998) for a number of test calculations at different temperatures and
pressures.  It was discovered that an additional factor of $2J''+1$ had been
included in the line strengths from the former, and when this
factor in the tabulated values is divided out before calculating the
absorption, the agreement between the two sources is good.  (The fact that
the highest tabulated values of $J$ were about 100, means that the average
value of $2J''+1$ is also about 100, which explains this discrepancy.)
Because Allard, Hauschildt, \& Schwenke provide more extensive coverage,
as well as information on line broadening, it is recommended that one use
these data, after correcting for the $2J''+1$ factor, even though the data
cover seven rather than nine electronic systems.

The partition function of $TiO$ can be obtained by interpolating the
table given by J\o{rgensen} (1997), which tabulates it for a range of
temperatures between 200 and 8000 K based on 7 electronic states.  The
coefficients $a$, $b$, $c$, $d$, $e$, and $f$ can be fitted to the
expression

\begin{equation}
\log_{10}Q(T) = aT^{-1} + b + cT + dT^2 + eT^3 +fT\ln(T)\, ,
\label{fittio}
\end{equation}

\noindent for each tabulated value in the range 200 $\le T \le$ 8000 K,
then any partition function for a temperature in this range can
be obtained.  Alternatively, Schwenke (1998) gives some very detailed
spectroscopic data for a large number of electronic states which can be
used to obtain the partition function.

\begin{figure}
\epsscale{1.00}
\centerline{\includegraphics[angle=-90,width=19cm]{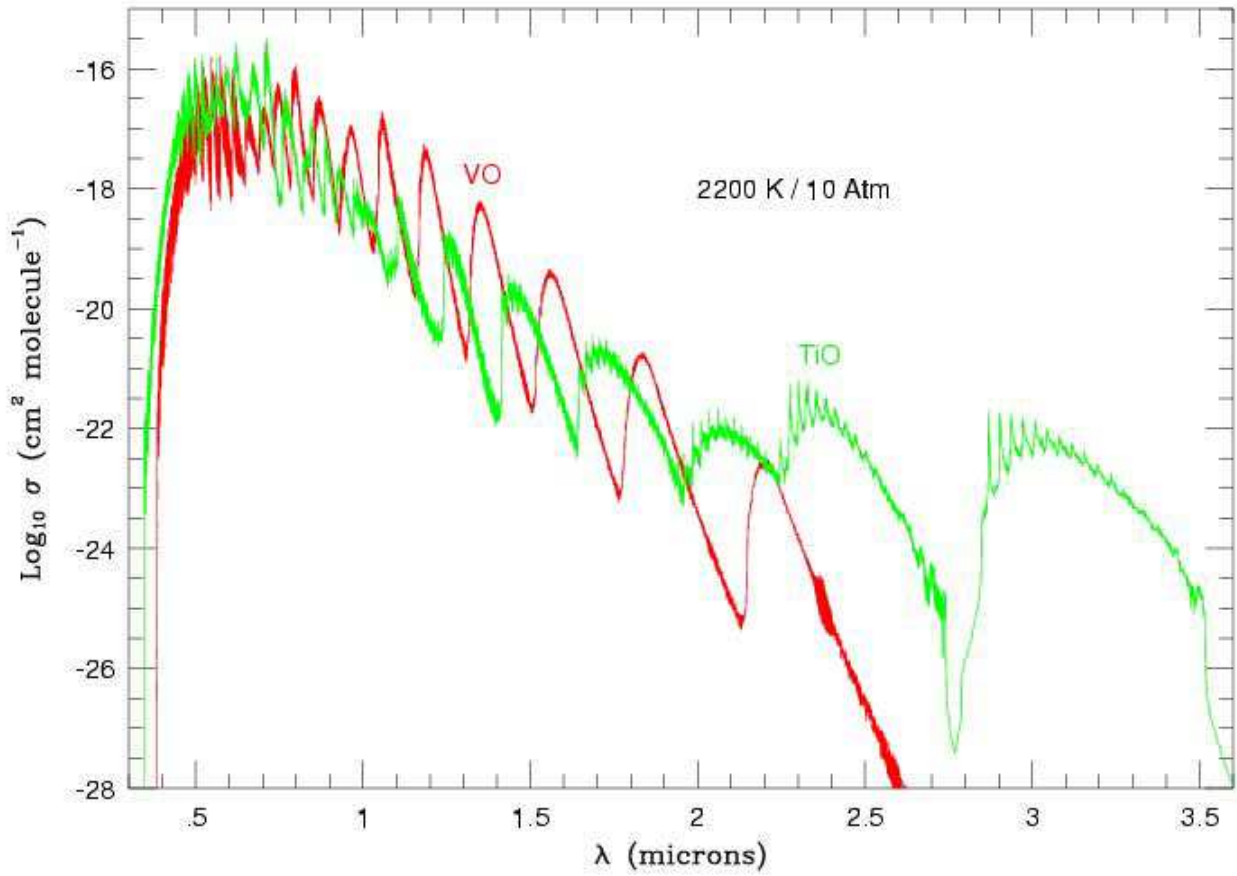}}
\caption{The log (base 10) of the monochromatic absorption $\sigma$ in
cm$^2$molecule$^{-1}$ as a function wavelength $\lambda$ in $\mu$m in
the infrared and visible at a temperature of 2200 K and a pressure of 10
atmospheres for $VO$ (red curve) and $TiO$ (green curve).
Unlike the previous figures, the absorption here is due to electronic
transitions, with three systems being calculated for $VO$ and
seven for $TiO$.  Note that the two molecules are very similar, but
band strengths of $TiO$ drop off more slowly with increasing
wavelengths.}
\label{fig:4}
\end{figure}

\begin{figure}
\epsscale{1.00}
\centerline{\includegraphics[angle=-90,width=19cm]{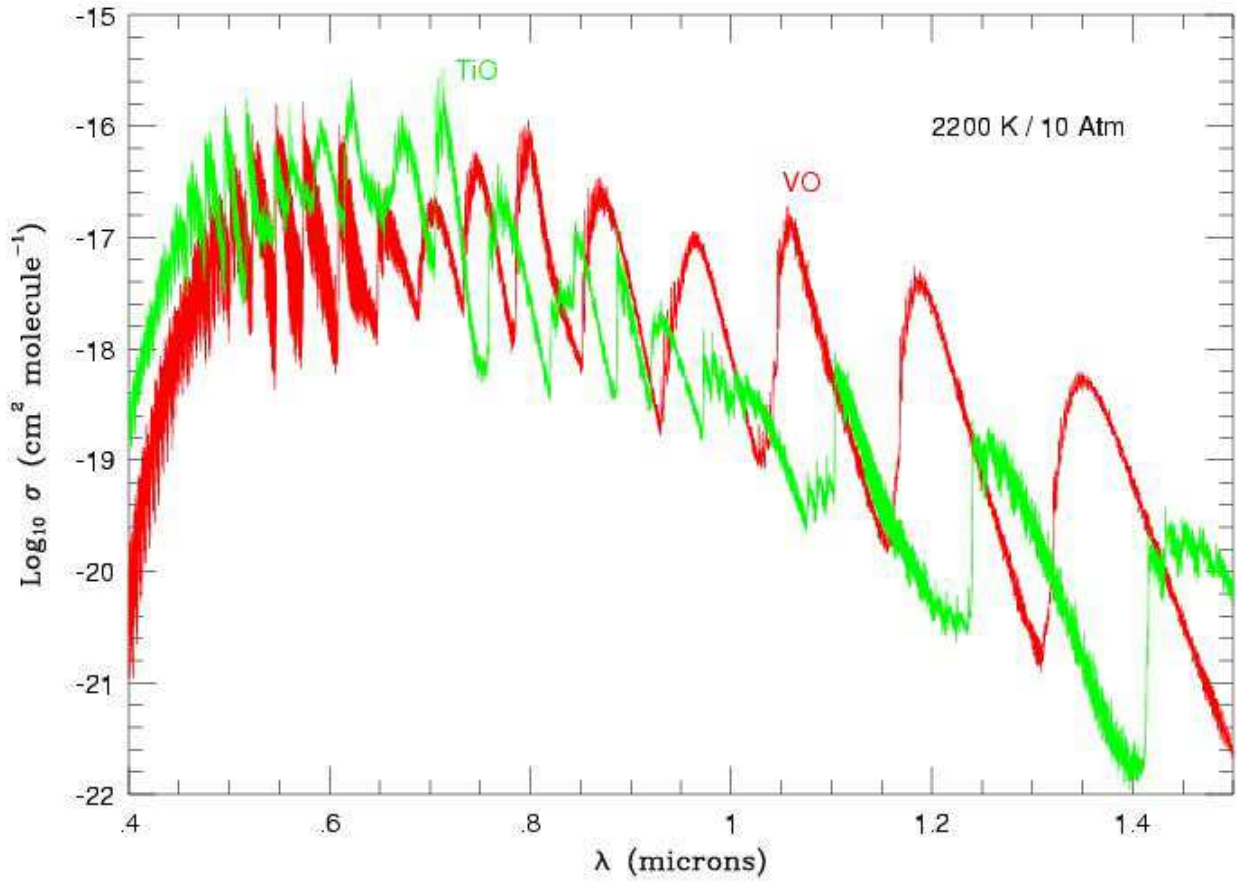}}
\caption{The same as the previous figure, but with the shorter wavelength
region being represented by expanded wavelength and absorption scales, in
order to show the differences between the $VO$ and $TiO$
absorptions.}
\label{fig:5}
\end{figure}

\begin{figure}
\epsscale{1.00}
\centerline{\includegraphics[angle=-90,width=19cm]{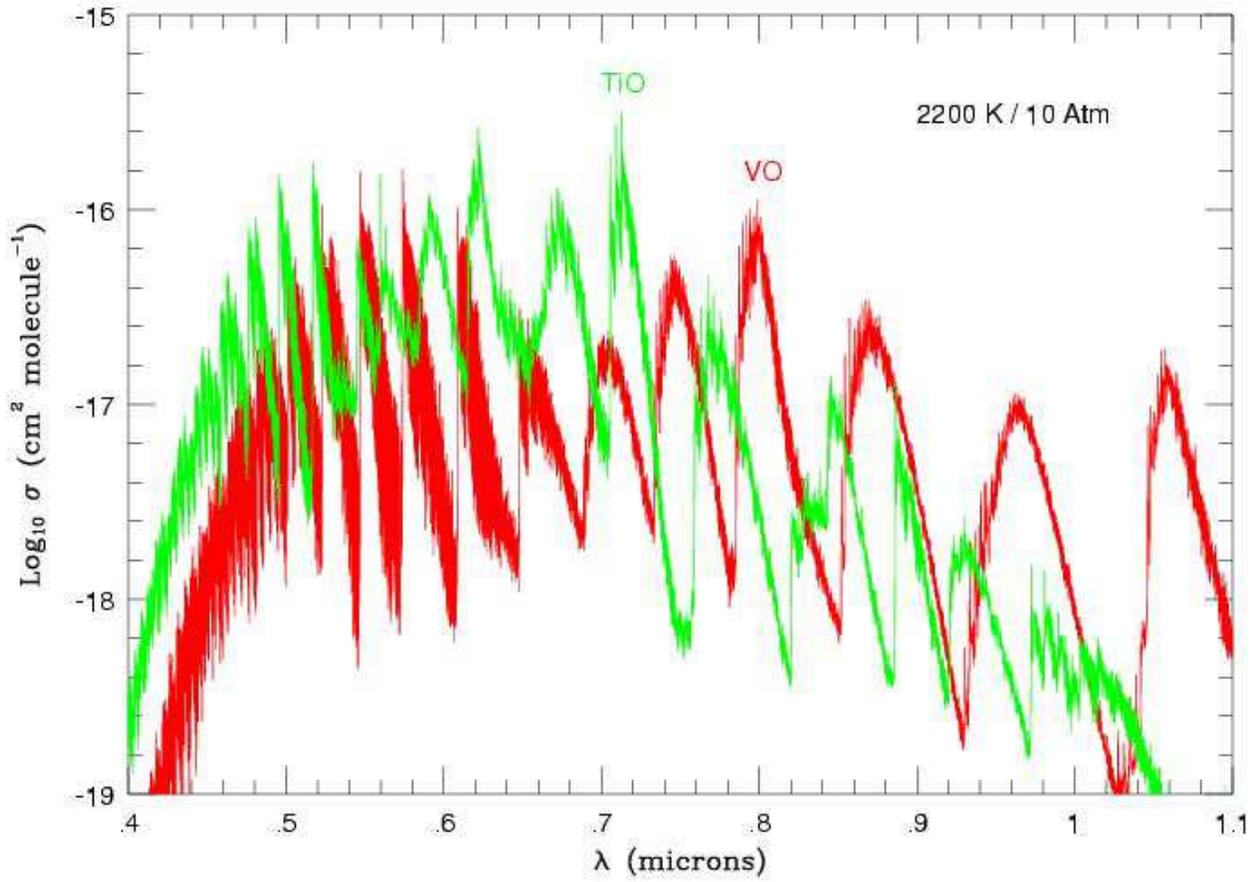}}
\caption{A further expansion in scales of the short wavelength region of
Fig. \ref{fig:4} to show the differences between the $VO$ and $TiO$
absorptions.  As can be seen, the strongest $TiO$ absorption at
about 0.71 $\mu$m occurs where $VO$ absorption is relatively low,
and there are several other places where peaks in $TiO$ absorption
take place where there are troughs in $VO$ absorption, and visa
versa; this is not so clear at smaller scales.}
\label{fig:6}
\end{figure}

Figures \ref{fig:4}, \ref{fig:5}, and \ref{fig:6} show the
monochromatic absorption of $TiO$ and $VO$ at progressively larger
scales at a temperature of 2200 K and a total pressure of 10 atmospheres.
As discussed in more detail in \S\ref{atmosabun}, Fig. \ref{fig:17}
shows the abundances of $TiO$ and $VO$ at a total
pressure of 1 atmosphere.  The sudden decrease with falling temperature
is due to the formation of condensates (Burrows \& Sharp 1999).

$VO$ - Data on this molecules can be found in Plez (1998), where they are
given in the same format as $TiO$ in that reference.

As with the other molecules, a set of opacity files can be precomputed,
but like $TiO$ it is a refractory species, so temperatures below 1000 K
need not be considered.  Also like $TiO$, the opacity is produced by
electronic transitions, and covers a large region of the infrared, as well
as parts of the visible spectrum.  The line data consist of over 3 million
lines from nearly 3800 cm$^{-1}$ to nearly 26,000 cm$^{-1}$ with the $gf$
values of the lines being given, together with the other necessary data,
including the value of $J''$ for each transition for the
$A^4\Pi - X^4\Sigma^-$, $B^4\Pi_r - X^4\Sigma^-$, and
$C^4\Sigma^- X^4\Sigma^-$ band systems.  The line strengths can be
calculated using eq. (\ref{oscil}).  No information on line broadening is
available, so, as with $TiO$ when the Plez (1998) data are used,
eq. (\ref{simplewid}) can be employed to calculate the broadening.
The isotopic version $^{51}VO$ is by far the most abundant, so other
isotopes need not be considered.  The partition function can be calculated
using eq. (\ref{molpart}), with the spectrosopic constants for the ground
and first few excited electronic states taken from Huber \& Herzberg (1979).

\begin{figure}
\epsscale{1.00}
\centerline{\includegraphics[angle=-90,width=19cm]{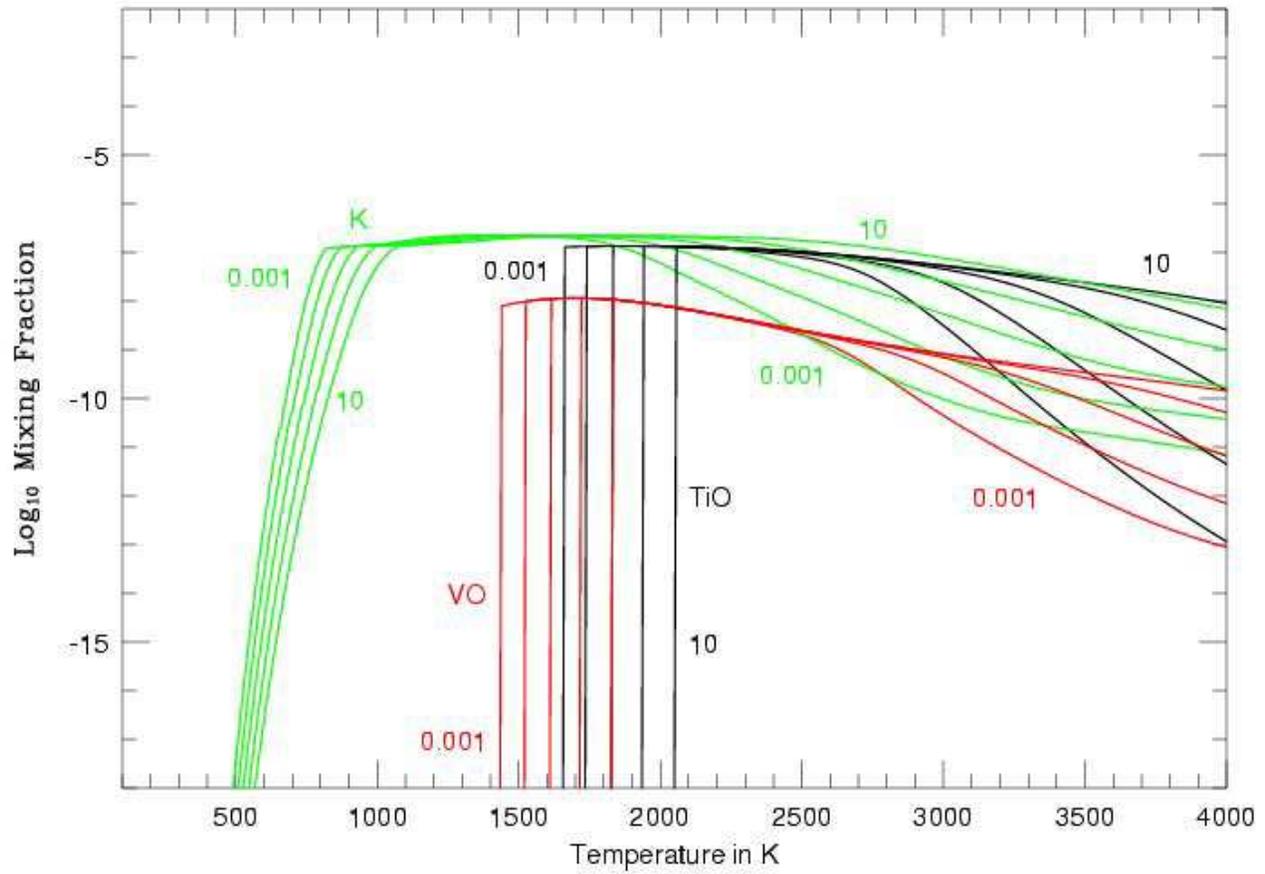}}
\caption{The log (base 10) of the mixing fraction as a function of
temperature for pressures of 0.001, 0.01, 0.1, 1, and 10 atmospheres
for $TiO$ (black curves), $VO$ (red curves), and
$K$ (green curves), with the same scales as the other abundance plots.
As before, only the curves corresponding to 0.001 and 10 atmospheres are
labeled, but in the case of $VO$, due to overlap with curves for
$K$ and $TiO$, the label for the curve corresponding to 10
atmospheres is omitted.  As with $FeH$ and $CrH$ in the previous
figure, the condensation temperatures at which these species are removed
from the gas phase increases with pressure, but when these species are
well above the condensation temperatures, the abundances increase with
pressure.}
\label{fig:22}
\end{figure}

Because the behavior of $VO$ is so similar to $TiO$, it is included on the
same opacity plots in Figs. \ref{fig:4}, \ref{fig:5}, and \ref{fig:6} and
abundance plots in Figs. \ref{fig:17} and \ref{fig:22}.  The opacity figures show
that it has the same general trend as $TiO$, though the band positions are at
different wavelengths.

\subsection{Metal Hydrides}
\label{hydrides}

A major source of molecular opacity
is the metal hydrides, where as in \S\ref{oxides} ``metals" are used in a
chemical, rather than an astrophysical, sense.  The role played by metal
oxides such as $TiO$ is well known, and although the absorption of metal
hydrides is often less than the oxides for a solar composition, to zeroth
order the abundances of the hydrides at a given temperature and pressure
(thus gravity) are approximately linearly dependent on $Z$, the metallicity
(in the astrophysical sense), whereas the abundance of the oxides can very
approximately scale as the square of $Z$, because two elements other than
hydrogen and helium are involved.  The consequence of this is that for
mixtures with a Population II value of $Z$, the hydrides are relatively more
important.

\begin{figure}
\epsscale{1.00}
\centerline{\includegraphics[angle=-90,width=19cm]{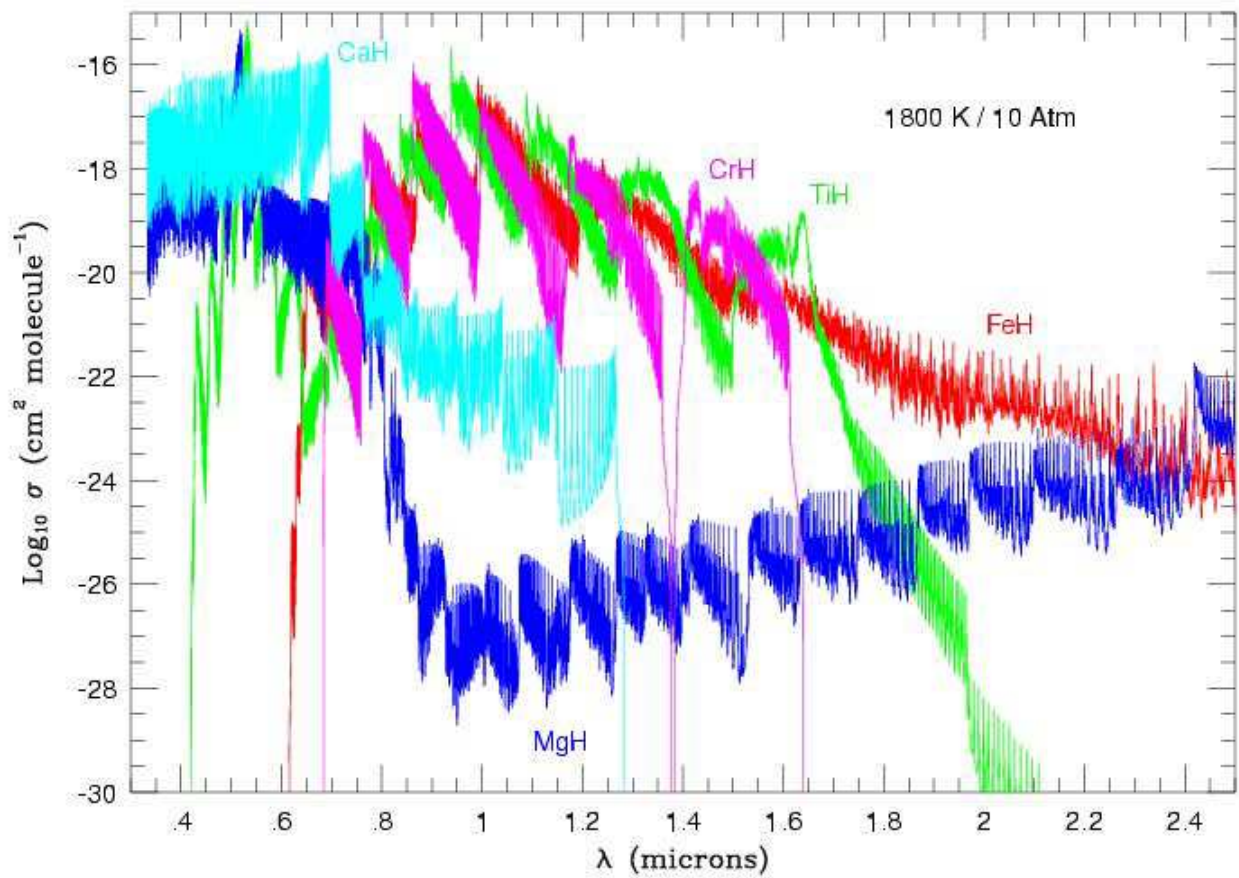}}
\caption{The log (base 10) of the monochromatic absorption $\sigma$ in
cm$^2$molecule$^{-1}$ as a function wavelength $\lambda$ in $\mu$m in
the infrared and visible at a temperature of 1800 K and a pressure of 10
atmospheres for the five metal hydrides $FeH$, $TiH$,
$MgH$, $CrH$, and $CaH$, represented, respectively,
in red, green, blue, magenta, and cyan.  All the bands are due to electronic
transitions, except that for $MgH$, which includes vibration-rotation
transitions in the ground $X$ electronic state, in addition to the $A-X$
and $B'-X$ electronic systems.}
\label{fig:7}
\end{figure}

\begin{figure}
\epsscale{1.00}
\centerline{\includegraphics[angle=-90,width=19cm]{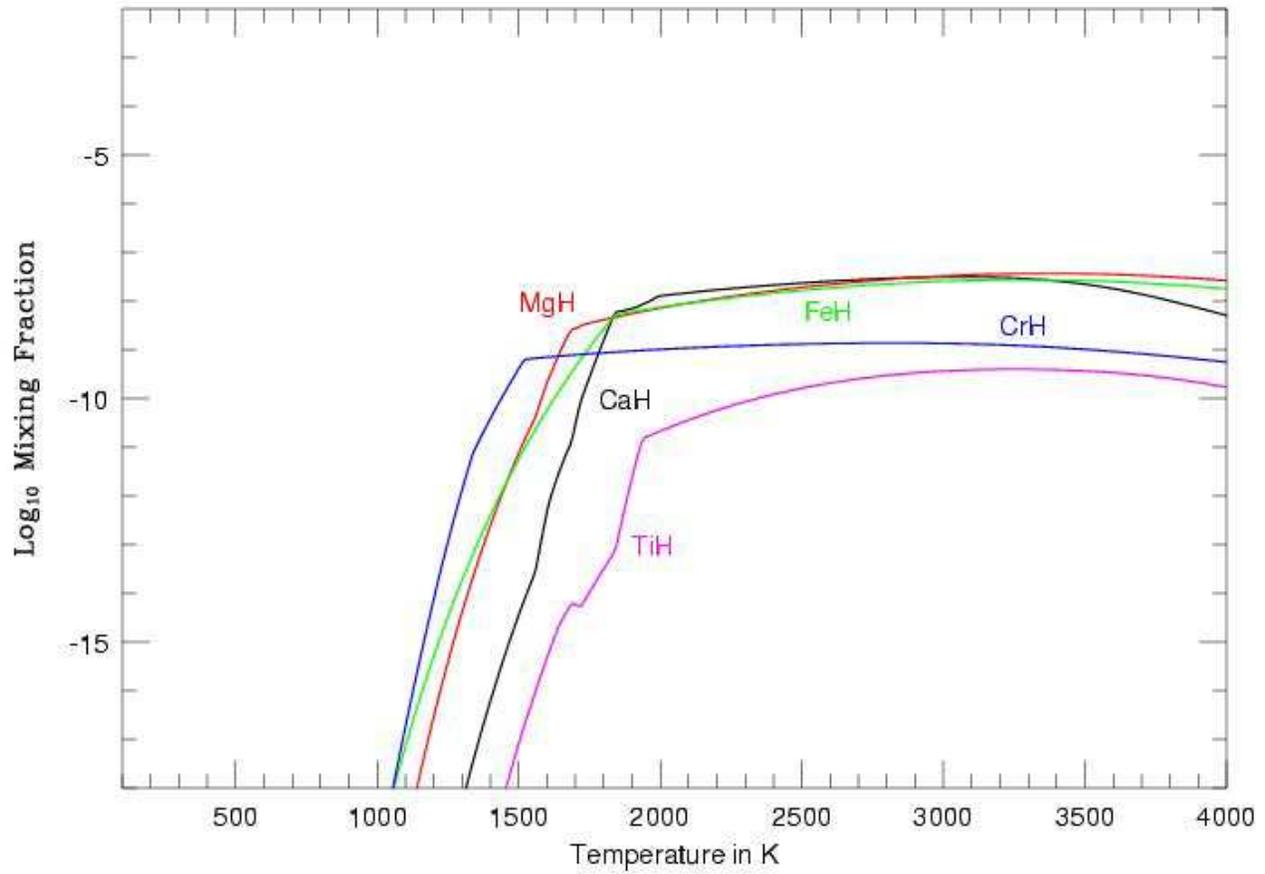}}
\caption{The log (base 10) of the mixing fraction as a function of
temperature at a total gas pressure of 1 atmosphere for the metal hydrides
$CaH$ (black), $MgH$ (red), $FeH$ (green), $CrH$ (blue), and $TiH$
(magenta), shown with the same scale as the previous figure.  The drop in
abundance with decreasing temperature for all these molecules below about
2000 K, is due to the formation of condensates involving the
corresponding metal.  The kinks in the curves, particularly for $TiH$,
are due to the replacement of one dominant condensed phase involving that
element by another.}
\label{fig:18}
\end{figure}

The opacities of the hydrides $TiH$, $CrH$, $FeH$, $MgH$, and $CaH$ can be
calculated using various methods, depending on the available data.  The
characteristics of these molecules are broadly similar, and all involve
electronic transitions in the visible and infrared parts of the spectrum.
As with $TiO$ and $VO$, because of the large number of lines involved, it
is recommended, as before, to precompute sets of files for a grid of the
same temperatures and pressures.  For comparison, opacities of all
five metal hydrides are plotted in Fig. \ref{fig:7} at 1800 K and a
total gas pressure of 10 atmospheres.  Because they have low moments of
inertia, and thus large rotational constants, the rotational fine structure
is visible in several of the bands at the scale used here.  For
comparison, the abundances of the molecules are plotted together on
Fig. \ref{fig:18} for a total pressure of 1 atmosphere, as discussed in
\S\ref{atmosabun}.

$TiH$ - The line frequencies and strengths of the $A^4\Phi - X^4\Phi$ and
$B^4\Gamma  - X^4\Phi$ electronic systems of $TiH$ are available from
detailed quantum mechanical calculations, as described by
Burrows et al. (2005), and include the spectroscopic data and the
H\"onl-London factors for the two quartet electronic transitions.
The calculations there are based on the data on about 91,000 and 101,000
lines for the $A^4\Phi - X^4\Phi$ and $B^4\Gamma  - X^4\Phi$ systems,
respectively.  The lines for the $A - X$ system are based on 20 bands with
$v'$ = 0, 1, 2, 3, 4 and $v''$ = 0, 1, 2, 3, and 24 bands for the $B - X$
system with $v'$ = 0, 1, 2, 3 and $v''$ = 0, 1, 2, 3, 4, 5.  Bands with the
same value of $\Delta v=v'-v''$ generally strongly overlap giving, the
appearance of one band, unless the resolution is high enough
to resolve the individual rotational components, so the general appearance
of a spectrum is given by different sequences of $\Delta v$.  In the case
of the $A - X$ system, 8 sequencies are calculated with $\Delta v$ taking on
values in the range of -3 to 4, and for the $B - X$ system 9 sequencies are
calculated with $\Delta v$ taking on values in the range of -5 to 3.  For
both systems in any transition, the value of $\min(J',J'') \le 50\frac{1}{2}$
is satisfied.  The two systems overlap, giving continuous coverage from
about 5000 cm$^{-1}$ to 24,000 cm$^{-1}$ (about 2 $\mu$m to 0.42 $\mu$m).
The line strengths are given in terms of the Einstein $A$ coefficients,
which can be converted to the required integrated line strength using

\begin{equation}
S = \frac{A}{8 \pi \bar{\nu}^2}(2J'+1) \frac{e^{-hcF''/kT}}{Q(T)}
\left[1-e^{-hc(F'-F'')/kT}\right]\, ,
\label{einsteina}
\end{equation}

\noindent where $J'$ is the rotational angular momentum quantum number of the
upper state.  The Einstein $A$ coefficient for each line in the input data
is given by

\begin{equation}
A = A_{v'v''} S_{J'J''}/(2J'+1)\, ,
\label{avv}
\end{equation}

\noindent where $A_{v'v''}$ is the Einstein $A$ coefficient for the whole
$v'-v''$ band, and $S_{J'J''}$ is the H\"onl-London for the $J'-J''$
rotational transition, including the spin substates.  A number of values of
$A_{v'v''}$ for both electronic transitions are listed in Table 2 of Burrows
et al. (2005).  Unlike $TiO$, where data on several isotopic versions are
available, for $TiH$, only data on the main isotopic version $^{48}TiH$
are available, so the isotopic shifts for the forms containing $^{46}Ti$,
$^{47}Ti$, $^{49}Ti$, and $^{50}Ti$, have to be calculated, using eq.
(\ref{rho}), and the methods briefly discussed in \S\ref{lines}.  The
quantum numbers of the levels involved are known, together with the
vibrational and rotational constants from Burrows et al. (2005) for the
three electronic states, of which the pair $X$ and $A$ or $X$ and $B$
participate in a given transition.  Although the shift of the lines of
$TiD$ is large, even when deuterium has not been destroyed by nuclear
burning, at best, it has an abundance of about five orders of magnitude
below normal hydrogen, so its isotopic shift, and equivalently, those of
the other hydrides, is not required.

The terrestrial isotopic fraction of $Ti$ in the form of $^{48}Ti$ is 0.738,
so this factor should be applied to eq. (\ref{einsteina}) for the main
isotopic version.  For the other isotopes, $^{46}Ti$, $^{47}Ti$, $^{49}Ti$,
and $^{50}Ti$, the corresponding factors of 0.08, 0.073, 0.055, and 0.054,
respectively, can be applied to eq. (\ref{einsteina}), together with the
shifted values of $\bar{\nu}$, $F''$, and
%<<<
$F'$.  However, no account is taken of the slight changes in the Einstein $A$
coefficient due to the shifts of the levels, as discussed in \S\ref{lines}.
Likewise, the isotopically shifted partition functions are not considered.
%>>>
The lines widths are calculated
using eq. (\ref{simplewid}) with the same coefficients.

\begin{figure}
\epsscale{1.00}
\centerline{\includegraphics[angle=-90,width=19cm]{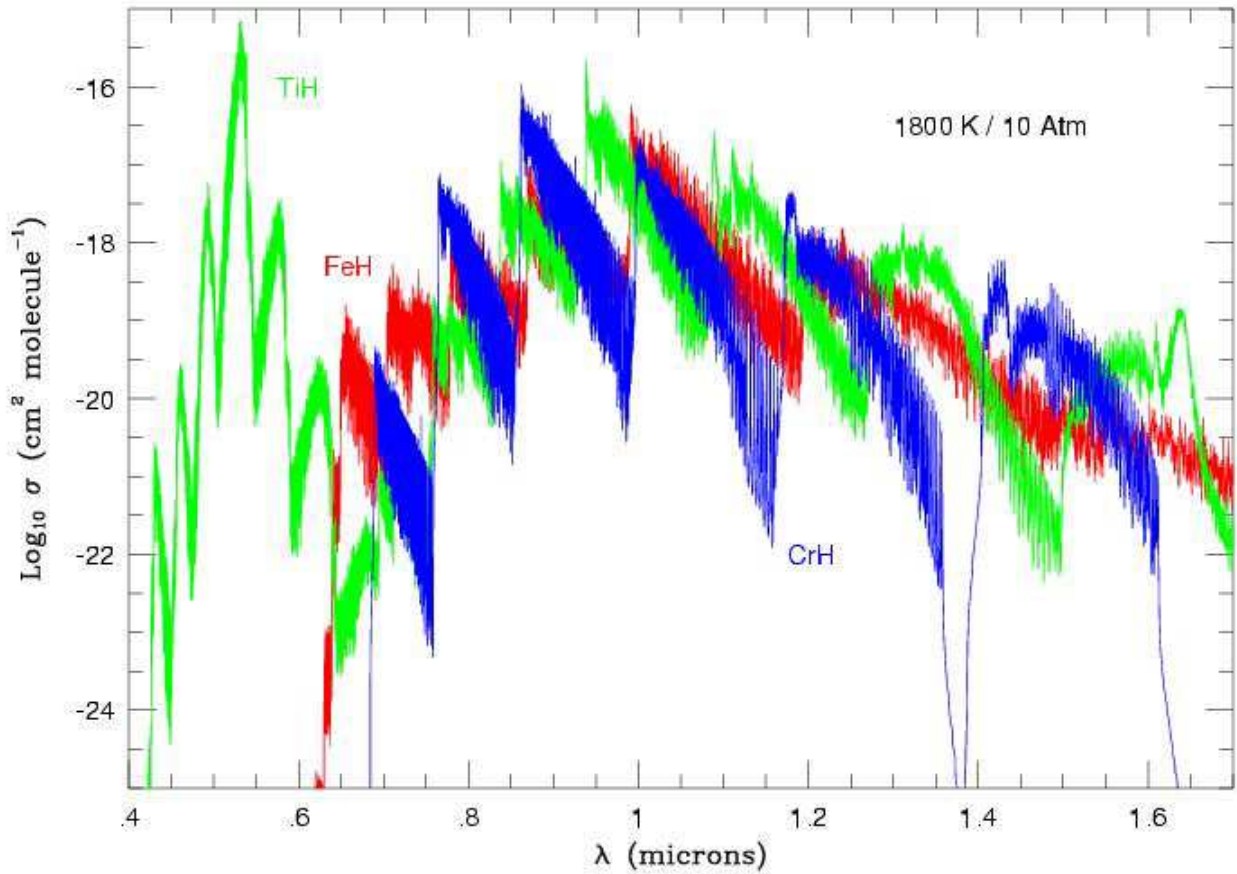}}
\caption{A subset of $FeH$ (red), $TiH$ (green), and
$CrH$ (blue), taken from the previous figure and plotted on a larger
scale to show more detail.  The absorption due to the $F-X$ electronic
system of $FeH$, the $A-X$ system of $TiH$, and the $A-X$
system of $CrH$ all very strongly overlap longward of about 0.65 $\mu$m.
In addition, we have the data for the $B-X$ system of $TiH$, which stands
out on its own here at the shortest wavelengths.}
\label{fig:8}
\end{figure}

In addition to Fig. \ref{fig:7}, the opacities of $TiH$ are plotted on a
larger scale in Fig. \ref{fig:8}, together with $FeH$ and $CrH$.  The
$A-X$ system covers an extensive region in the near infrared longward
of about 0.65 $\mu$m, where it follows a trend very similar to $FeH$ and
$CrH$, which strongly overlaps it.  There are several strong peaks, of
which the strongest is at about 10,500 cm$^{-1}$ or $\sim$0.94 $\mu$m.
This corresponds to the overlapping of many $\Delta v$=0 bands, i.e.
the 0-0, 1-1, 2-2 etc. transitions.  The next strongest bands at a
shorter wavelength correspond to the overlapping of many $\Delta v$=1
bands, i.e. the 1-0, 2-1, 3-2 etc. transitions, and likewise the next
strongest bands at a still longer wavelength are due to the overlapping
of the $\Delta v$=-1 bands, i.e. the 0-1, 1-2, 2-3 etc. transitions.
Most of the band heads are on the $R$-branches because $B' < B''$, as
can be seen by the absorption dropping off more rapidly shortward of the
peak than at longer wavelengths; however, for the peak at the longest
wavelength here corresponding to the $\Delta v$=-3 sequence the band
heads are on the $P$-branches.  This band is particularly interesting.
When $TiH$ is present, $H_2O$ is also present and ``competes" with
$TiH$ absorption; most of the peaks of the bands in the $A-X$ system
coincide with maxima in the the bands of $H_2O$.  In the case of the
$\Delta v$=-3 sequence, this happens to coincide with a trough in $H_2O$
absorption, and appears to be somewhat stronger than the general trend.
This band system may be a good candidate for detection in brown dwarfs.

The $B-X$ system covers a more restricted range at shorter wavelengths, but
with a narrow and very strong peak at about 19,000 cm$^{-1}$ or 0.53 $\mu$m
in the yellow part of the visible spectrum, as can be seen in Fig.
\ref{fig:8}, where $TiH$ has its highest absorption for both the band
systems.  As with the $A-X$ system, the various peaks are caused by the
overlapping sequences for a given value of $\Delta v$.  This is a general
characteristic for most electronic band systems with $|\Delta v|$ being
zero or a small integer.  Although as stated earlier the band systems
overlap, there is a substantial minimum between the maxima of the two
systems at about 14,300 cm$^{-1}$ or 0.7 $\mu$m in the red part of the
visible spectrum.

The partition function can be calculated from eq. (\ref{molpart}) using the
spectroscopic data of the 15 electronic states listed in Tables 11 and 12 of
Burrows et al. (2005), including the $X$, $A$, and $B$ states used to calculate
the opacity.  Since the energy levels of the four spin substates of the $X$
electronic state are available, these can be treated as four separate
states for calculating the partition function, so the sum can be performed
over 18 states.  The calculated values of the partition function at 200 K
intervals from 1200 to 4800 K are listed in Table 13 of Burrows et al. (2005).
As previously, only the partition function for the main isotopic version
need be calculated.  The variation of the abundance with temperature at
a gas pressure of 1 atmosphere is show in Fig. \ref{fig:18}, discussed in
more detail in \S\ref{atmosabun}.

$CrH$ - As discussed in Burrows et al. (2002), this molecule plays an
important role in L dwarfs and sunspots.  The calculations investigated here
are based on the $X^6\Sigma^+ - A^6\Sigma^+$ system using 12 bands with
$v'$ = 0, 1, 2, and $v''$ = 0, 1, 2, 3, with $\min(J',J'') \le 30\frac{1}{2}$,
giving rise to 6 $\Delta v$ sequencies in the range of -3 to 2.  For each
band, 6 strong main $P$- and $R$-branches are available, corresponding to
each of the 6 spin substates, together with 18 weaker satellite bands.
This yields a relatively modest number of about 13,800 lines between about
6200 cm$^{-1}$ and 14,500 cm$^{-1}$.  As both electronic states are $\Sigma$
states, i.e. have an orbital angular momentum quantum number $\Lambda$=0,
they are strictly Hund's case (b) (Herzberg 1950) coupling, so a rotational
quantum number apart from electron spin, $N$, can be defined.  However, the
net electron spin of 5/2 does interact, giving rise to 6 values of $J$ for
a given value of $N$ from $J=N+5/2$, $J=N+3/2$, $J=N+1/2$, $J=N-1/2$,
$J=N-3/2$, and $J=N-5/2$, which have slightly different energies.  The
effect of the splitting of these levels is to increase the contribution to
the opacity.

The main isotope is $^{52}CrH$ with an isotopic fraction of 0.83, with the
other isotopic versions of any significance being $^{50}CrH$,
$^{53}CrH$, and $^{54}CrH$, and having fractions of 0.044, 0.10, and 0.02,
respectively.  As with $TiH$, eq. (\ref{einsteina}) can be used to calculate
the line strengths in cm$^2$molecule$^{-1}$ using the Einstein $A$
coefficients given by eq. (\ref{avv}), with the line widths and isotopic
shifts being calculated similarly.

Unfortunately, only the $X$ and $A$ electronic states used to calculate the
opacity are readily available for calculating the partition function. 
The partition function for isotopically
substituted $CrH$ is the same as that derived under the assumption that
chromium is pure $^{52}Cr$ to within $\sim$0.3\%.  The monochromatic absorption of $CrH$
is replotted on a larger scale in Fig. \ref{fig:8}, together with $TiH$ and
$FeH$ at 1800 K and a total pressure of 10 atmospheres, and its abundance
as a function of temperature at a gas pressure of 1 atmosphere is shown
in Fig. \ref{fig:18}, together with the other metal hydrides.

$FeH$ - In Dulick et al. (2003), details were given for the calculation of
$FeH$ opacities produced by the $F^4\Delta_i - X^4\Delta_i$ (Wing-Ford)
system, which are briefly discussed here.  A total of 25 bands with both
$v'$ and $v''$ taking values of 0, 1, 2, 3, and 4, with 9 sequences of
$\Delta v$ in the range of -4 to 4, and with the rotational quantum
numbers limited by $\min(J',J'') \le 50\frac{1}{2}$ for the four spin
substates, were provided.  This yields a total of about 116,000 transitions
in the range from about 2000 cm$^{-1}$ in the infrared to 16,000 cm$^{-1}$
in the red part of the visible spectrum, i.e. 5 $\mu$m to 0.625 $\mu$m,
respectively.

Major problems with the $FeH$ Wing-Ford system are that $FeH$ itself is
not easily studied in the laboratory, and both the $F^4\Delta_i$ and
$X^4\Delta_i$ electronic states are intermediate between Hund's (a) and
(b) coupling cases (Herzberg 1950), with the coupling changing with $J$.
Moreover, the rotational levels are perturbed by unknown electronic states.
For Hund's case (b) coupling, the situation is like $CrH$, as discussed
above, with $N$ being a well-defined quantum number representing rotation
before electron spin is added, except that the net electron spin is 3/2,
yielding four values of $N$ for a given $J$, i.e., $J=N+3/2$, $J=N+1/2$,
$J=N-1/2$, and $J=N-3/2$; the selection rules are $\Delta N = 0,\pm1$ and
$\Delta J = 0,\pm1$.  Of the 28 branches corresponding to allowed
combinations of $\Delta J$ and $\Delta N$, there are four each of the
main $P$-, $Q$-, and $R$-branches with $\Delta J = \Delta N$, which are
the strongest branches, with the remainder being weaker satellite
branches.  Table 4 in Dulick et al. (2003) lists the H\"onl-London factors
of all 28 branches.  It is worth noting that for the four $P$- and
$R$-branches, the H\"onl-London factors scale as $O(J)$, but for the
others they scale as $O(J^n)$, where the integer $n =- 1$ or is greater
than 0.

For Hund's (a) coupling, the quantum number $\Omega = \Lambda + \Sigma$ is
defined, and is the sum of orbital angular momentum quantum number
($\Lambda$), where $\Lambda$ = 2 here for the $\Delta$ states, and the
projection of the net electron spin on the internuclear axis ($\Sigma$),
where $\Sigma$ = -3/2, -1/2, 1/2, and 3/2 for quartet states.  Hence,
$\Omega$ takes the values 1/2, 3/2, 5/2, and 7/2.  The selection rules
are such that $\Delta \Lambda = 0$ and $\Delta J = 0,\pm1$, so there are
12 branches, i.e., 4 $P$-, $Q$-, and $R$-branches for the different spin
substates, with the H\"onl-London factors for these being given
in Table 3 of Dulick et al. (2003).  Note that the $P$- and $R$-branches are
strong, and scale as $O(J)$, whereas the $Q$-branches are weak and scale
as $O(J^{-1})$, as is the case with the case (b) coupling.  The general
intermediate coupling between Hund's cases (a) and (b) is very complicated
for quartet states, and depends on spin-orbit and spin-spin
interactions.  Instead, intermediate coupling was determined by fitting to
experimental results, as discussed in Dulick et al. (2003), for both the
intensities and the term values for a number of selected vibrational levels.
The latter are listed in Tables 8a to 8j in Dulick et al. (2003) for the $X$
and $F$ states,
then used to calculate the opacity.  The use of the Einstein $A$ coefficients
together with the treatment of broadening are exactly the same as $TiH$ and
$CrH$ covered above.  The main isotope is $^{56}FeH$, with a fractional
abundance 0.917, the other isotopic forms are $^{54}FeH$, $^{57}FeH$, and
$^{58}FeH$, which have fractional abundances of 0.058, 0.022, and 0.028,
respectively, and, as previously, the treatment is the same.

The partition function can be calculated for the main isotopic version with
the data for the energies of the electronic states listed in Table 9a
of Dulick et al. (2003), and the vibrational and rotational constants of
these states ican be obtained from Table 9b of the reference.  A total of
10 states are listed in Dulick et al. (2003), including the $X$ and $F$ states,
and the individual energies of several spin substates in Table 9a are given,
yielding 29 states or substates included in the summation of eq.
(\ref{molpart}).  Table 9c in that reference compares analytic values
between 1000 and 3500 K from eq. (\ref{molpart}) with direct summation of
the levels.  It is found that the values from direct summation are always
less, ranging from 78\% of the values at 1000 K, to only about 30\% of the
value at 3500 K, by which time $FeH$ has for all practical purposes
completely dissociated.  As direct summations will always underestimate a
partition function when levels are omitted, whereas analytic solutions allow
for them, we have reason to believe that the analytic results are more
accurate.  As with $TiH$ and $CrH$, $FeH$ is replotted on a larger scale in
Fig. \ref{fig:8} at 1800 K and 10 atmospheres total gas pressure, and its
abundances as a function of temperature at a gas pressure of 1 atmosphere are
plotted on Fig. \ref{fig:18}, along with the other metal hydrides.
See \S\ref{atmosabun} for more details.

$MgH$ - Data for the $X^2\Sigma^+ - X^2\Sigma^+$ pure rotation and
vibration-rotation transitions in the infrared, the
$B'^2\Sigma^+ - X^2\Sigma^+$ electronic transitions in the visible,
and the near-ultraviolet spectrum are available from Skory et al. (2003),
and the $A^2\Pi - X^2\Sigma$ electronic transitions, also in the visible
and near ultraviolet, are available from Weck et al. (2003a).

Before the $X - X$ pure rotation and vibration-rotation bands can be
calculated, a file containing the energy levels and the corresponding
vibrational and rotational quantum numbers from Weck et al. (2003a) should be
combined with the file containing the transitional energies, Einstein $A$
coefficients, and the corresponding quantum numbers from Skory et al. (2003),
by matching up the quantum numbers.  The same procedure should be performed
for the $A - X$ and $B' - X$ systems.  However, an additional
complication is that the resultant line lists are not in general sorted
in order of ascending rotational quantum numbers for a given pair of
vibrational quantum numbers, which in turn are not in general ordered
in ascending values.  For the previous hydrides, this is the case, and
can result in the saving of a considerable amount of processing time when
calculating isotopic shifts; thus, the data should be resorted before
proceeding.

The data for the $X - X$ system contain a list of over 2500 lines from
the far infrared up to just above 10,000 cm$^{-1}$ or 1 $\mu$m.  The
small number of lines below 100 cm$^{-1}$ are completely swamped in the far IR by $H_2O$
and $CH_4$ (by more than three orders of magnitude) and can be ignored.
The data for the $A - X$ system contains over 10,000 lines from about
9,000 cm$^{-1}$, i.e. 1.1 $\mu$m, thus overlapping with the upper end
of the $X - X$ system, up to about 32,000 cm$^{-1}$ in the near
ultraviolet.  The dataset for the $B' - X$ system has about the same
number of lines, and covers almost exactly the same spectral region
as the $A - X$ system, except that it starts at a slightly shorter
wavelength and is missing an overlap with the $X - X$ system.  Thus,
there are no gaps in the spectral coverage between the three systems.

As with the other hydrides discussed, the same methods can be
used to calculate the line strength with the Einstein $A$ coefficients,
together with the line broadening and the isotopic shifts.  Because only
data for the most abundant isotope $^{24}MgH$ are readily available, the
shifted lines for the minor isotopic forms $^{25}MgH$ and $^{26}MgH$
should be calculated, taking account of the isotopic abundances, when
line strengths are calculated using the reordered data.

\begin{figure}
\epsscale{1.00}
\centerline{\includegraphics[angle=-90,width=19cm]{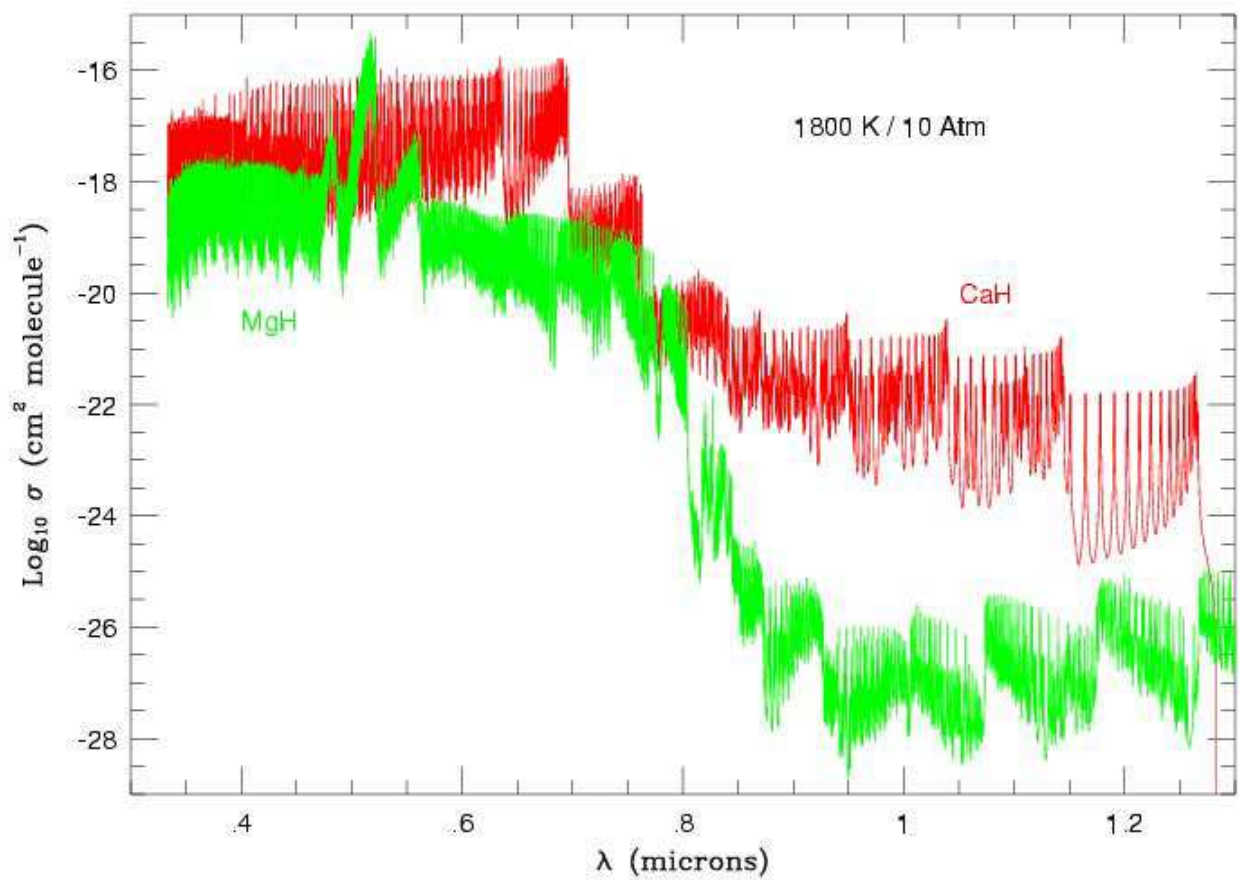}}
\caption{A subset of $CaH$ (red) and $MgH$ (green) taken from Fig.
\ref{fig:7} and plotted on larger scales.  The absorption produced by $CaH$
is calculated from the $A-X$, $B-X$, and $D-X$ system, and that due to
$MgH$ is calculated from the $X-X$ (vibration-rotation), $A-X$, and $B'-X$
systems, with the $X-X$ being the only source of absorption of $MgH$
longward of about 1 $\mu$m.}
\label{fig:9}
\end{figure}

The partition functions can be calculated using the spectroscopic data
for the $X$, $A$, and $B'$ electronic states.  Since the $A$ and $B'$
states are, respectively, about 19,200 cm$^{-1}$ and 22,400 cm$^{-1}$
above the $X$ state, their contributions, even at the highest
temperatures of interest, are small (less than 1\%), and higher states can be neglected.
The monochromatic absorption of $MgH$ is replotted on a larger scale in
Fig. \ref{fig:9}, together with that for $CaH$, at 1800 K and 10
atmospheres total gas pressure, and as discussed in more detail in
\S\ref{atmosabun}, its abundance as a function of temperature at 1
atmosphere is plotted alongside the other metal hydrides in Fig.
\ref{fig:18}.

%<<<
$CaH$ - Unlike the other metal hydrides where the opacities can be
calculated by reading in data from line lists available from the referenced
sources, in the case of $CaH$ we demonstrate that individual lines can be
calculated from band data.  Leininger \& Jeung (1995) and Weck et al. (2003b)
provide the required data.  In the case of the former, band oscillator
strengths are provided
%>>>
in their Table V for transitions in the $A^2\Pi - X^2\Sigma^+$,
$B^2\Sigma^+ - X^2\Sigma^+$, and $D^2\Sigma^+ - X^2\Sigma^+$ electronic
systems for the most abundant $^{40}CaH$ isotopic form, giving a total
of 154 values of $v''$ from 0 to 13 and of $v'$ from 0 to 11,
respectively, for each of the three band systems.  This is a total of
462 bands strengths, of which somewhat more than a half have
non-negligible values.  Given a band oscillator strength in the form of
$f_{v'v''}$ for a particular electronic system, a set of rotational lines
can be calculated for valid combinations of $J''$ and $J'$ that follow
the selection rules in order to calculate the strengths from eq.
(\ref{stren}) for each available combination of $v''$ and $v'$.  The
appropriate H\"onl-London factors for doublet transitions
(Kov\'{a}cs 1969) should be used.

The vibrational component of the vibrational-rotational energy levels can be
calculated from the standard anharmonic formulae given by Herzberg (1950),
and the rotational components can be calculated from the formulae for doublet
states, including spin-orbit splitting when available, from Kov\'{a}cs (1969).
The electronic, vibrational and rotational spectroscopic constants,
including vibrational and rotation coupling and spin-orbit coupling, can be
obtained from Huber and Herzberg (1979).  As no information on broadening
appears to be available, eq. (\ref{simplewid}) can be used to calculate the
line widths.

Only one other isotopic form of $CaH$ is of any significance, namely
$^{44}CaH$, with a fractional abundance of 0.02.  Exactly the same method
can be used to calculate its opacity using eq. (\ref{rho}) with
isotopically shifted constants.  The partition function of the main
isotopic form can be calculated using the spectrosopic constants of the
electronic states in eq. (\ref{partfun}), and, as previously, since the 
abundance of the second-most abundant isotope of Ca, $^{44}$Ca, is
$\sim$2\%, the isotopic shift correction to the partition function 
can be neglected.  As stated above for $MgH$, the monochromatic absorption of $CaH$ is replotted on a
large scale in Fig. \ref{fig:9} at 1800 K and 10 atmospheres gas pressure,
and its abundance as a function of temperature is shown in Fig. \ref{fig:18},
which is discussed in greater detail in \S\ref{atmosabun}.

\subsection{$H_2S$ and $PH_3$ Infrared Opacities}
\label{precomp}

Though the molecules $H_2S$ and $PH_3$ are not very abundant
in substellar atmsopheres, they are the primary reservoirs of
sulfur and phosphorus, respectively, and have distinctive spectral
features that can be used to diagnose their chemistry.
In this short section, we
summarize their spectroscopy and the databases where one can
obtain the relevant line lists, and provide a figure at a representative
temperature/pressure point depicting their absorption cross sections versus
wavelength.

$H_2S$ - The data on line broadening are available from Kissel et al (2002),
and HITRAN (Rothman et al. 2003, 2005) and GEISA (Jacquinet-Husson et
al. 1999, 2003, 2005) do provide some additional data.  The monochromatic
opacities are tabulated at 1 cm$^{-1}$ intervals from 200 cm$^{-1}$ to about
19570 cm$^{-1}$, so essentially all the infrared and a large fraction
of the visible light is covered.

$PH_3$ - Like $H_2S$, data can now be obtained from HITRAN
(Rothman et al. 2003, 2005) and GEISA (Jacquinet-Husson et al. 1999, 2003,
2005).  The data used for the precomputed tables are derived from Levy \&
Lacome (1993) for $H$ and $He$ broadening, and Levy, Lacome \&
Tarrago (1994) for the temperature dependence of collision broadening.
More recent data are available from Bouanich et al. (2004) for the
$H_2$-broadening coefficients in the $\nu_2$ and $\nu_4$ bands.

The resulting monochromatic opacities are tabulated at 1 cm$^{-1}$
intervals from 200 cm$^{-1}$, but unlike $H_2S$, reach only to about
2500 cm$^{-1}$, so only fractions of the mid- and far-infrared regions
are covered.

The monochromatic absorptions due to $H_2S$ and $PH_3$, along with $CO$, as
previously discussed, are shown in Fig. \ref{fig:3}, at 1600 K and 10
atmospheres, but as mentioned above, due to their low abundances, they are
only minor opacity sources.  Their abundances are shown in Fig.
\ref{fig:17}, for a total pressure of 1 atmosphere, together with the other
molecules discussed so far, except for the metal hydrides.

\subsection{Ultraviolet Opacities}
\label{uvopac}

For a substellar object sufficiently close to its parent star where 
irradiation from that star becomes important, a non-negligible fraction of
the incoming radiation will be at short wavelengths, depending on the
spectral type of the star.  Such radiation can play an important role in
the structure of the atmosphere, and ultraviolet and short wavelength visible
opacities need to be calculated.  The species discussed here are important
opacity sources at short wavelengths.

$Fe$ - When iron is in the gas phase, an appreciable fraction can be in
the monatomic form.  Kurucz (1995) gives a database of nearly 800,000
lines from about 100 to 0.055 $\mu$m, i.e. respectively, from 100 to
nearly 2 million cm$^{-1}$ from the far infrared to the far ultraviolet.
This covers an unrealistically large range for many practical calculations,
and opacities need only be calculated from 10,000 cm$^{-1}$ to
67,000 cm$^{-1}$, i.e. 1 $\mu$m to 0.15 $\mu$m, respectively.  Longer
wavelengths in the infrared can be neglected because the opacity due to
atomic iron is smaller than that of many of the other species,
and shorter wavelength ultraviolet opacities can also be neglected
as there is little flux from a parent star with a spectral type later
than $F8$.  Even with this restricted range, a wider grid spacing may be
needed to save memory and processing time.  As with the infrared
opacities, because $Fe$ is a refractory species, no opacities are required
below 1000 K.

The data from Kurucz (1995) are in the form of $\log_{10}$ of the $gf$
value for each transition, from which the integrated line strength can be
calculated using eq. (\ref{oscil}).  The Van der Waals, Stark, and
natural broadening are also available, which enables the broadening of
each line to be calculated using the Van der Waals theory:

\begin{equation}
\Delta\bar\nu = \frac{1}{2\pi c}\left[\gamma_W\left(N_H + C_{H_2}N_{H_2} +
C_{He}N_{He}\right) \left(\frac{T}{10,000}\right)^{0.3} + \gamma_N\right]\, ,
\label{widvdw}
\end{equation}

\noindent where $\gamma_W$ and $\gamma_N$ are the Van der Waals and
natural broadening coefficients, respectively,\footnote{Because the abundance of electrons
is negligible for most of the temperature range of interest, the 
Stark coefficient is not relevant for these calculations.} $N_H$, $N_{H_2}$,
and $N_{He}$ are the number densities of the corresponding species, and
$C_{H_2}$ and $C_{He}$ are coefficients for $H_2$ and $He$, respectively,
because their polarizabilities are different from $H$.  The partition
function can be calculated using eq. (\ref{partfun}) by direct summation
of the first 31 levels listed by Moore (1949).

\begin{figure}
\epsscale{1.00}
\centerline{\includegraphics[angle=-90,width=19cm]{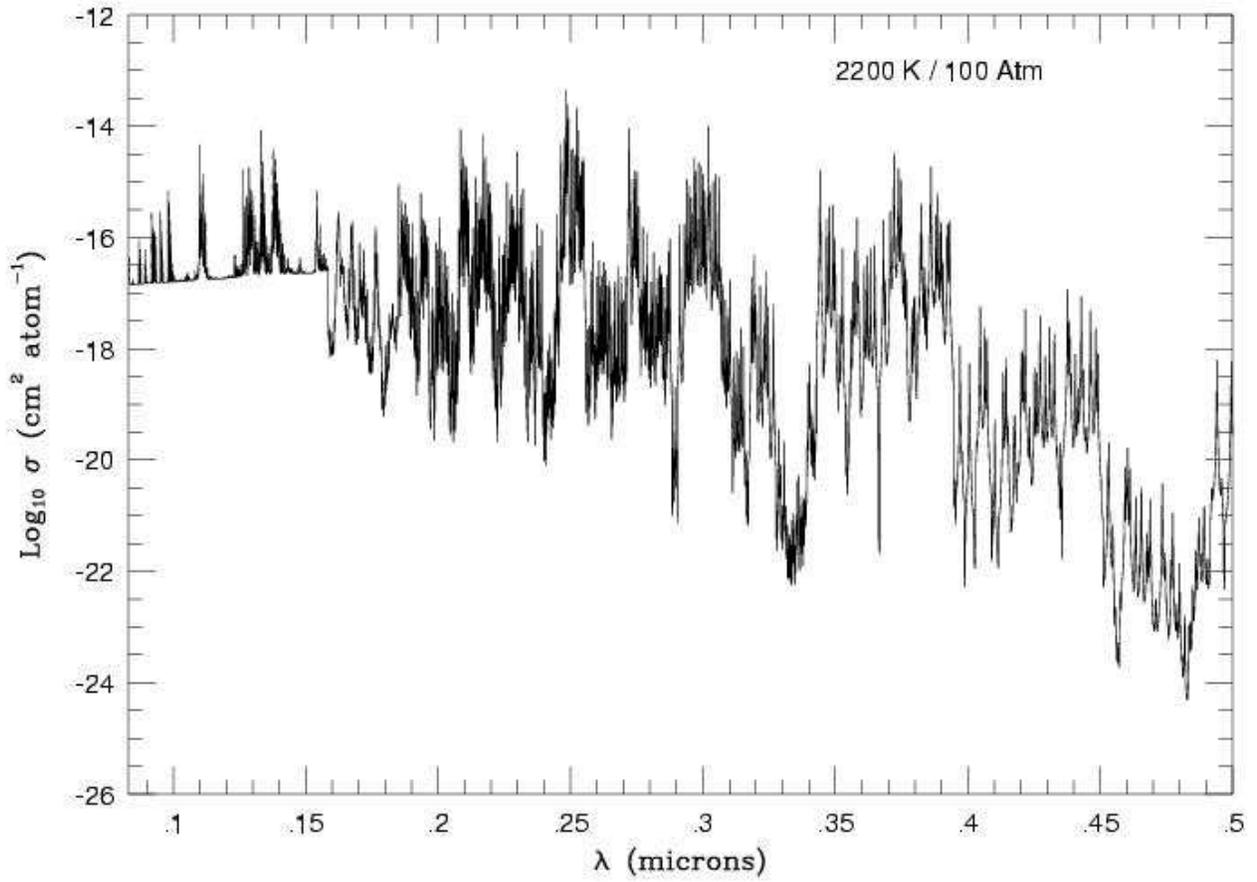}}
\caption{The log (base 10) of the sum of the monochromatic absorption
$\sigma$ in cm$^2$atom$^{-1}$ as a function wavelength $\lambda$
in $\mu$m due to the contributions of line and bound-free transitions
of atomic iron at 2200 K and 100 atmospheres in the visible and
ultraviolet parts of the spectrum.  A pressure of 100 atmosphere is
high, but as the absorption changes so rapidly over
very short wavelength intervals, this is adopted here so that the
general behavior of absorption can be represented graphically.  At the
shortest wavelengths, the minimum absorption is clearly governed by the
contribution due to the bound-free transition.}
\label{fig:10}
\end{figure}

\begin{figure}
\epsscale{1.00}
\centerline{\includegraphics[angle=-90,width=19cm]{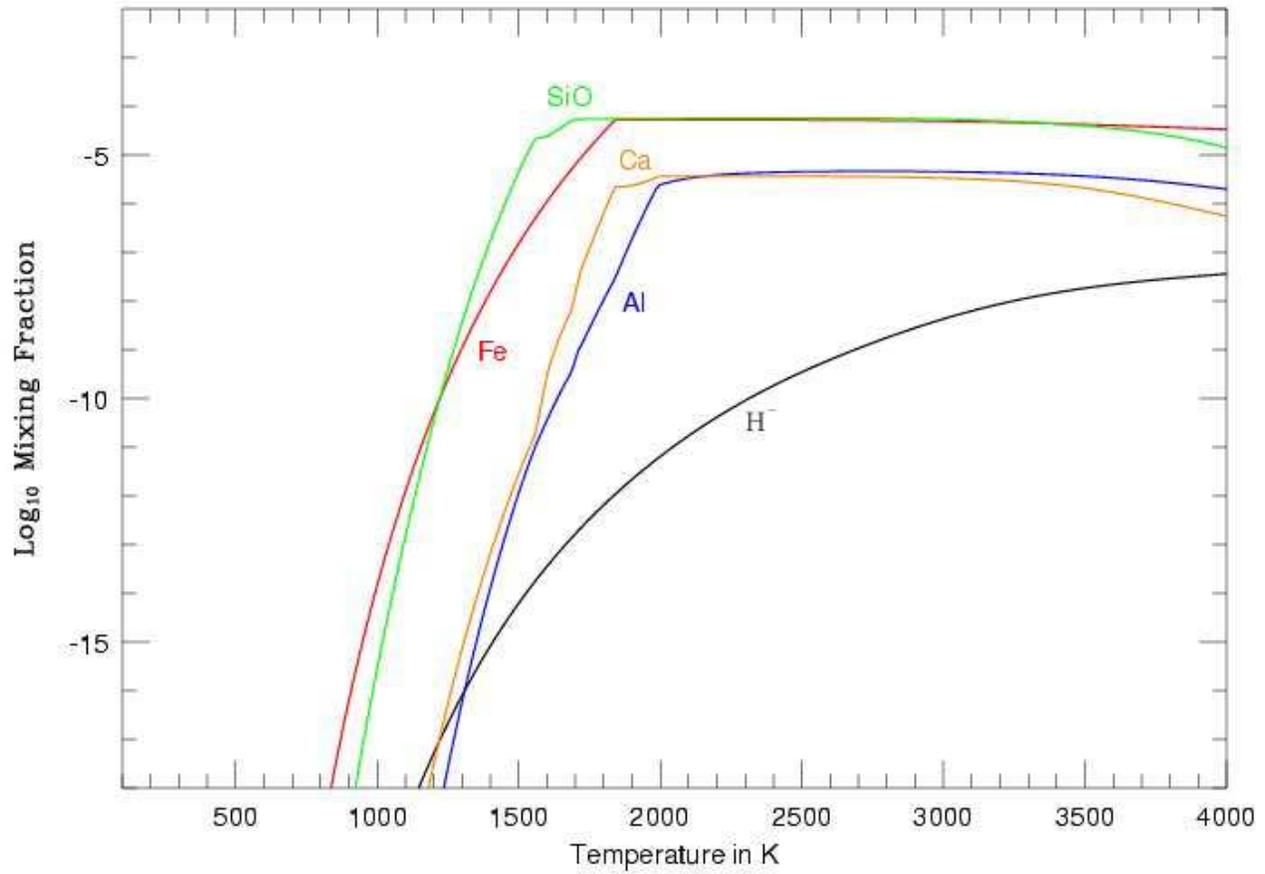}}
\caption{The log (base 10) of the mixing fraction as a function of
temperature at a total gas pressure of 1 atmosphere for the negative
hydrogen ion (black), $Fe$ (red), $SiO$ (green), $Al$ (blue), and $Ca$
(orange), plotted on the same scales as before.  These are some of the
atomic species together with $SiO$ which are important opacity sources in
the visible and ultraviolet, in addition to the other species plotted.
With the exception of $H^-$, they all start decreasing with decreasing
temperature between 2000 and 1500 K, due to the formation of condensates.}
\label{fig:20}
\end{figure}

Figure \ref{fig:10} shows the monochromatic absorption of $Fe$ in the
ultraviolet and visible parts of the spectrum due to the sum of the
contributions from the lines discussed here and the continuous bound-free
absorption discussed in \S\ref{boundfree}, at 2200 K and a pressure of
100 atmospheres.  A gas pressure of 100 atmospheres is high, 
and is only chosen so that the lines can be broadened
sufficiently to see the general trend.  At a realistically lower
pressure the general trend may be obscured by the fine details
with the opacity changing by many orders of magnitude in very small
spectral intervals.  The abundances of $Fe$, $Al$, $Ca$, $SiO$, and $H^-$,
are plotted in Fig. \ref{fig:20} for a total gas pressure of 1 atmosphere,
and are discussed in \S\ref{atmosabun}.

$CO$ - An ultraviolet line list is available from Kurucz (1993), that is
quite separate from the infrared data used in \S\ref{hitran}, and includes
transitions due the $A^1\Pi - X^1\Sigma^+$ (4$^{th}$ Positive) electronic
system.  Data on nearly 400,000 lines from 0.430 $\mu$m to 0.112 $\mu$m,
i.e. from about 23,000 cm$^{-1}$ to about 89,000 cm$^{-1}$, are provided.
With the $gf$ values and other data, the integrated lines strengths from
eq. (\ref{oscil}) can be calculated.  The broadening can be calculated
using eq. (\ref{widvdw}), except that the natural widths are not available
from Kurucz (1993).  

\begin{figure}
\epsscale{1.00}
\centerline{\includegraphics[angle=0,width=19cm]{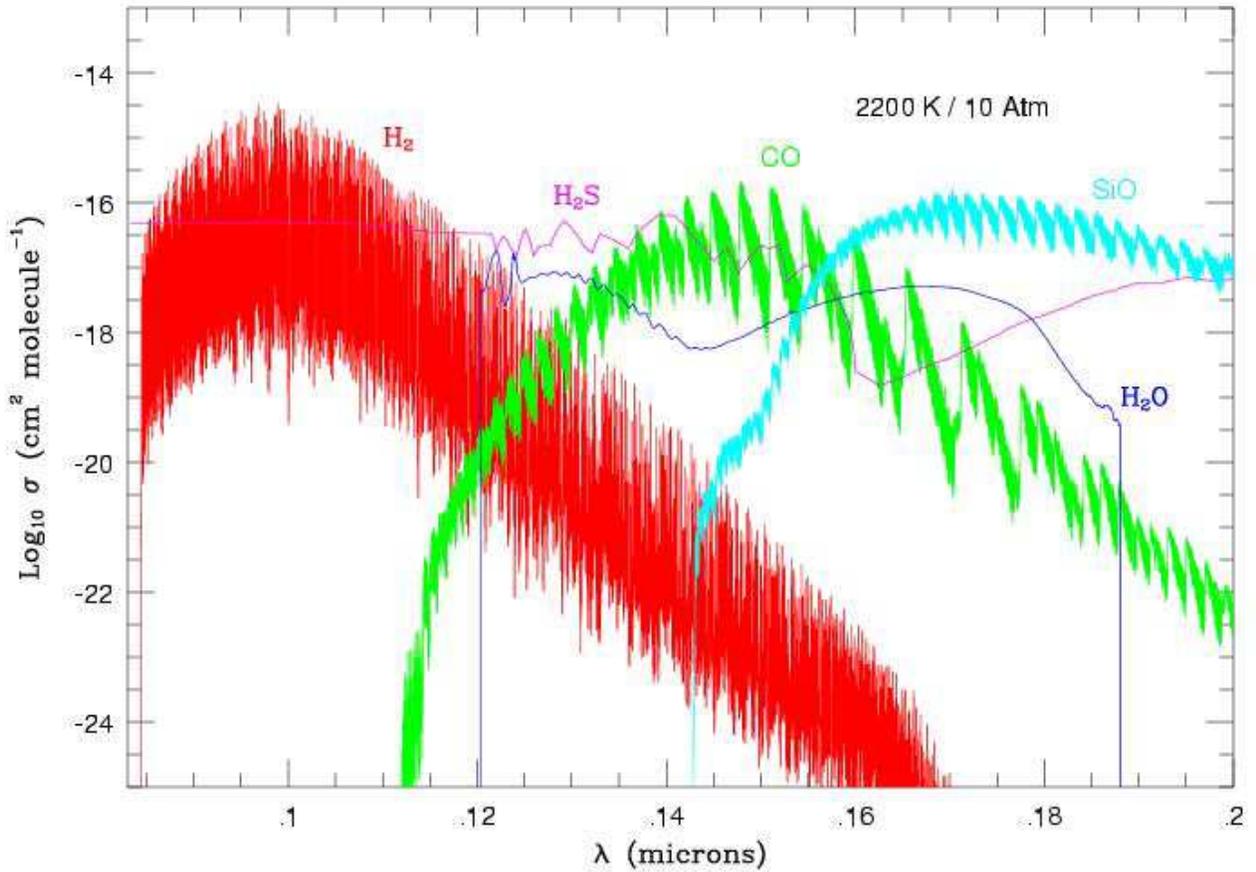}}
\caption{The log (base 10) of the monochromatic absorption $\sigma$ in
cm$^2$molecule$^{-1}$ as a function wavelength $\lambda$ in $\mu$m
at a temperature of 2200 K and a pressure of 10 atmospheres in the
ultraviolet for $H_2$ (red), $CO$ (green), $H_2O$ (blue), $H_2S$ (magenta),
and $SiO$ (cyan).  All the transitions are electronic, and in particular
the absorption for $H_2$ is calculated from the $B-X$ and $C-X$
(respectively, the Lyman and Werner) systems, and for $CO$ from the
$A-X$ system, as well as for $SiO$.  Since we have no individual line data for
$H_2O$ and $H_2S$, the temperature and pressure dependences are not
calculated, and only smoothed plots can be made.  As with the previous
figure, a high pressure is chosen to smooth out for graphical
representation several of the absorptions.}
\label{fig:11}
\end{figure}

\begin{figure}
\epsscale{1.00}
\centerline{\includegraphics[angle=-90,width=19cm]{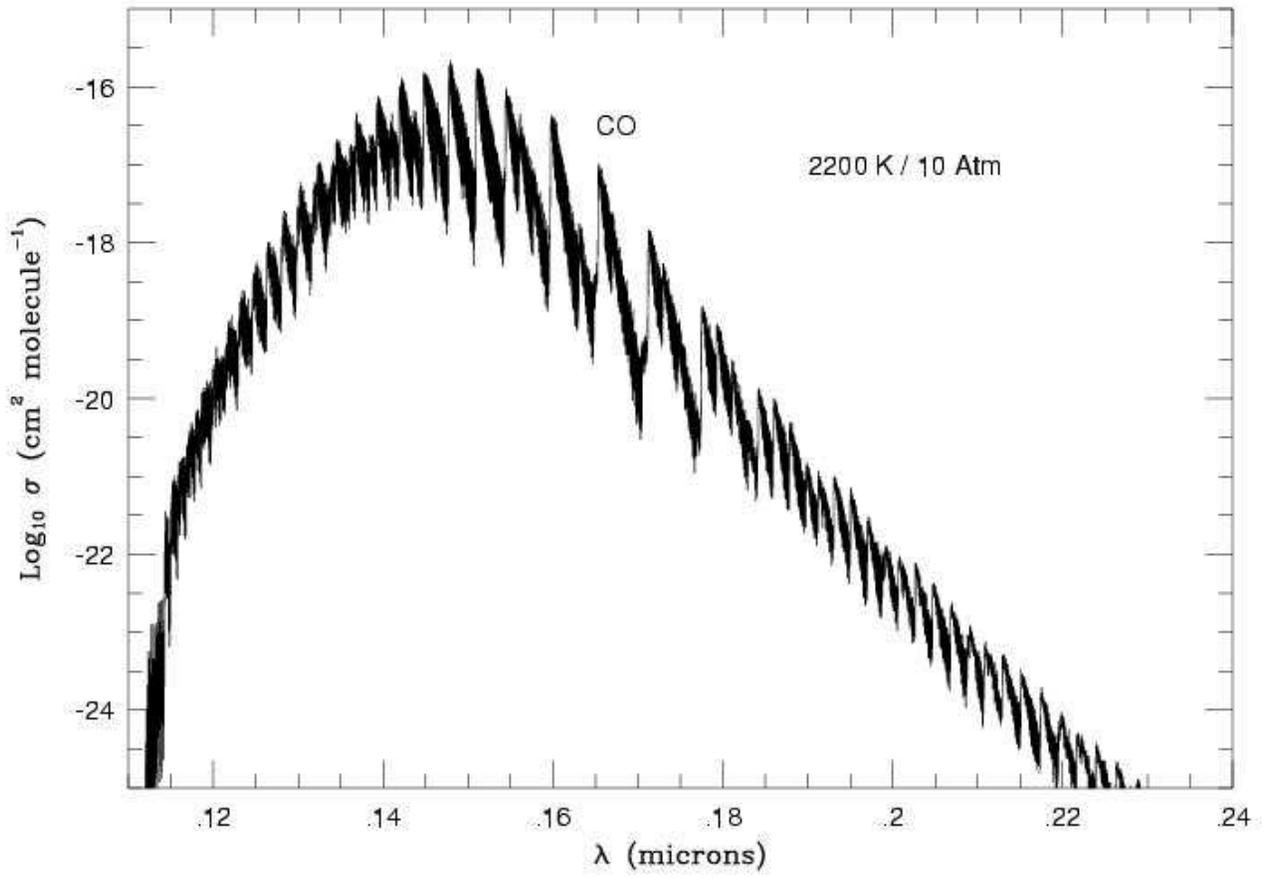}}
\caption{The absorption of $CO$ as given in the previous figure
for the same temperature and pressure, but extended to longer wavelengths
to give more complete coverage.}
\label{fig:12}
\end{figure}

The monochromatic absorption of $CO$ in the ultraviolet, together with 
those of $SiO$, $H_2$, $H_2O$, and $H_2S$, is shown in Fig. \ref{fig:11}
at 2200 K and total pressure of 10 atmospheres, and is repeated for $CO$,
only in greater detail, in Fig. \ref{fig:12}.  As with $Fe$, the pressure
is unrealistically high, but is chosen as a compromise to show the
various contributions to the absorption clearly.

$SiO$ - Although the most important molecular gas-phase species of silicon
is $SiO$, over most of the temperatures of interest most of the silicon
has condensed out into refractory silicates, which are the most abundant
condensates.  Nevertheless, at the upper temperature range of interest
$SiO$ will be present and can be a significant ultraviolet opacity
source, so should be considered.  As with the ultraviolet opacity of
$CO$, data from Kurucz (1993) are avialable with 1.7 million lines from
0.43 $\mu$m to 0.14 $\mu$m , i.e. about 23,000 cm$^{-1}$ to 70,000 cm$^{-1}$,
respectively.  The transitions are for the $A^1\Pi - X^1\Sigma^+$ electronic
system, which is exactly analogous to the 4$^{th}$ Positive system of $CO$. 

The partition function can be calculated using eq. (\ref{molpart}) with
data from Huber \& Herzberg (1979).  Only the ground electronic state makes
any significant contribution for the temperatures of interest. As with $CO$,
no information is available to enable isotopic variations to be considered.

\begin{figure}
\epsscale{1.00}
\centerline{\includegraphics[angle=0,width=19cm]{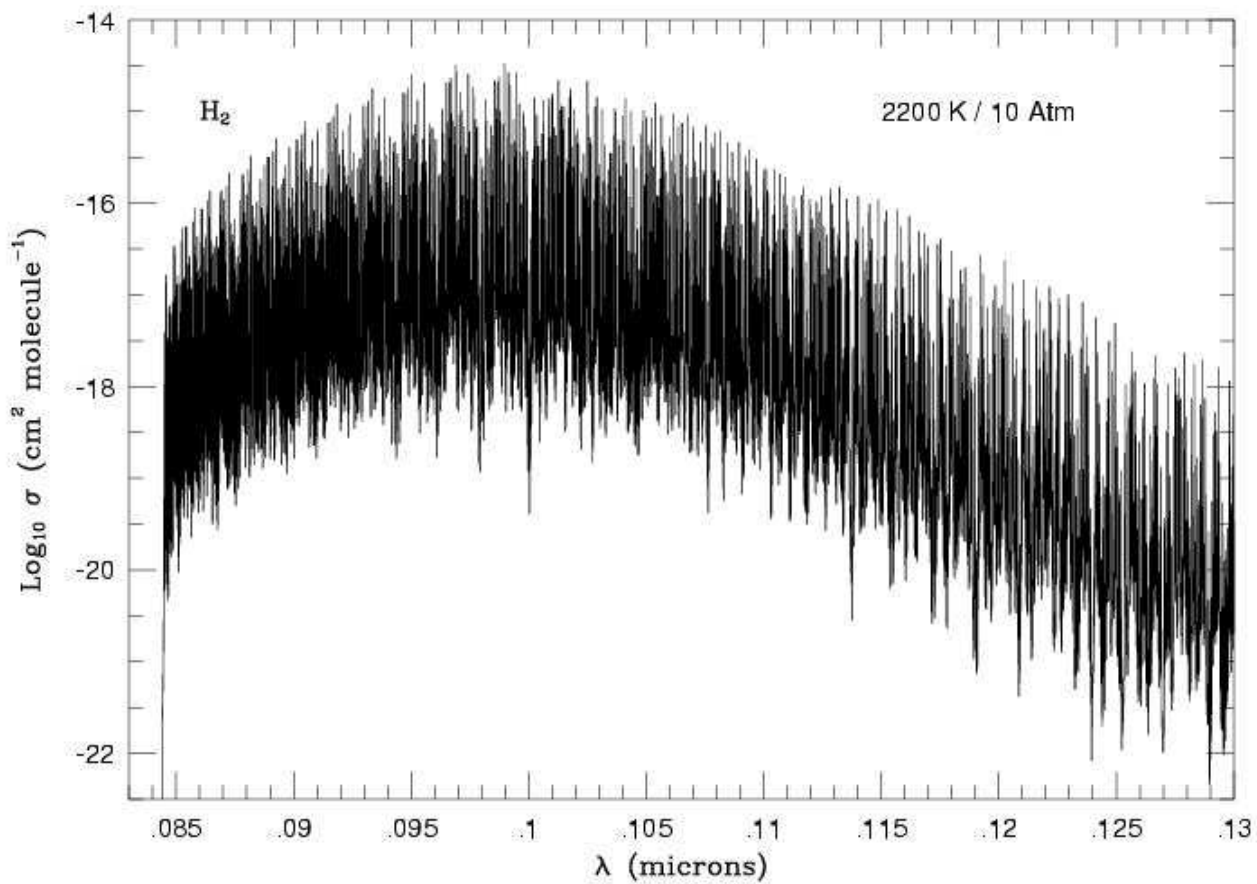}}
\caption{The short-wavelength absorption of $H_2$ as given in Fig. \ref{fig:11}
for the same temperature and pressure, but on a larger wavelength scale to
show more detail.  $H_2$ is by far the most abundant species,
absorbs strongly in the ultraviolet, and so completely dominates that part
of the spectrum.}
\label{fig:13}
\end{figure}

Figure \ref{fig:11} shows the absorption of $SiO$ compared with several
other species in the ultraviolet.  Because it's abundance is relatively low
compared with $CO$ or $H_2$, a more detailed plot like Fig. \ref{fig:12}
for $CO$ or Fig. \ref{fig:13} for $H_2$ (discussed immediately below),
is not included.  Figure \ref{fig:20} shows the abundance of $SiO$ along
with some other species (\S\ref{atmosabun}).

$H_2$ - This is by far the most abundant molecule, but with no
permanent dipole moment, its only contribution in the infrared is
collision-induced absorption (as discussed in \S\ref{ciaopac}).  Since
the first excited electronic state is at a very high energy above the
ground state, there is no absorption at shorter wavelengths until the
ultraviolet is reached, where the first and most important bands are due
to the $B^1\Sigma^+_u - X^1\Sigma^+_g$ (Lyman) and $C^1\Pi_u - X^1\Sigma^+_g$
(Werner) electronic systems.  About 28,500 transitions are available
from Kurucz (1993), from which the line strengths can be calculated
using eq. (\ref{oscil}).  Since the $gf$ values given do not include the
$2J''+1$ factor, this must be added in the calculations.  Also, no
information on line broadening is readily available, so eq.
(\ref{simplewid}) has to be used.  A set of precomputed tables can be
generated for later use.

The partition function can be calculated using eq. (\ref{molpart}) and
using data from Huber \& Herzberg (1979), and as with $CO$ and $SiO$,
only the ground electronic state makes any significant contribution for
the temperatures of interest.  Figure \ref{fig:11} includes $H_2$,
where it is seen that even per molecule, its strongest absorption exceeds
that of the other species shown, and a more detailed opacity is replotted
in Fig. \ref{fig:13}.

$H_2O$ - In addition to its infrared absorption which is discussed in
\S\ref{hitran}, separate data on its ultraviolet opacity can be obtained
from Kurucz (1993), which corresponds to bands associated with transitions
from the ground electronic state to an excited electronic state.
The data contain only a listing of the monochromatic absorption at a set
of tabulated points from 0.188 $\mu$m to 0.12 $\mu$m, corresponding to a
range of 53,000 cm$^{-1}$ to 83,000 cm$^{-1}$, respectively, with no
information on the lines producing this absorption.  Except for the
stimulated emission factor, no temperature or pressure dependence can be
calculated, and the opacity at required frequency points would simply be
calculated by linear interpolation from the data provided.
As explained in the caption to Fig. \ref{fig:11}, no data for individual
lines are available, and only a smooth opacity could be plotted in that
figure.

$H_2S$ - We consider $H_2S$ to be a significant ultraviolet opacity
source, but the availability of data are even
more limited than for $H_2O$.  Monochromatic opacities can be obtained
from Lee, Wang, \& Suto (1987) for the range 0.24 $\mu$m to 0.059 $\mu$m,
i.e. 42,000 cm$^{-1}$ to 204,000 cm$^{-1}$, respectively, (extending well
beyond the normal range of interest), in the form of a short table for
wavelengths below 0.16 $\mu$m, and a figure for the longer wavelength range.

As with the ultraviolet opacity of $H_2O$, it is not possible to consider
temperature and pressure dependencies, other than stimulated emission, and
the opacity at the required frequencies can only be calculated by linear
interpolation when required.  Like $H_2O$, only smooth data could be
plotted in Fig. \ref{fig:11}.

%<<<
$H$ - At the upper temperatures of interest molecular hydrogen will become
sufficiently dissociated that atomic hydrogen will be present, which
absorbs strongly in the ultraviolet in the Lyman continuum and lines.
There will be much weaker absorption in the Balmer, Paschen, Brackett, etc.
continua and lines, depending on the temperature, which generally will have
little importance.  However, because these are easy and quick to calculate,
the general line transitions of atomic hydrogen are discussed here, and
continuous processes in \S\ref{boundfree}.
%>>>

Given the lower ($n_i$) and upper ($n_j$) principle quantum numbers,
though all the line strengths are available in tabulated 
form (Weise, Smith, \& Glennon 1966) for modern computations,
the method of Carson (1988a) can be used to gain insight 
into the associated spectroscopy and for obtaining the line
strengths analytically when $n_i \leq 7$ and $n_j > n_i$.
The intrinsic line strengths
are obtained by solutions of hypergeometric functions listed
in that paper for this range of quantum numbers in terms of the square of the
Bohr radius for a general hydrogenic atom.  Since some very large intermediate
numbers are often involved in the calculations, care has to be taken to
avoid machine overflow or underflow.  The line strengths have to be
converted to the required strengths in cm$^2$atom$^{-1}$ with stimulated
emission included using

\begin{equation}
S = \frac{8\pi^3}{3h} a_o^2e^2\bar{\nu} S_{ij}
\frac{e^{-hcF_i/kT}}{Q(T)} \left[1-e^{-hc\bar{\nu}/kT}\right]\, ,
\label{hydrbb}
\end{equation}

\noindent where $a_o$ is the Bohr radius, $S_{ij}$ is the line strength in
units of $a_o^2$ and $F_i$ is the term value in cm$^{-1}$ of the lower state
given by $RZ^2(1-\frac{1}{n_i^2})$, where $R$ is the Rydberg constant and
$Z$ is the nuclear charge, which is unity here.  The term value for the
upper state, $F_j$, is likewise given by $RZ^2(1-\frac{1}{n_j^2})$, and the
line frequency in cm$^{-1}$ is just the difference.  Because the $l$ dependence 
of the energy levels of hydrogen is so weak, it is sufficient to
ignore the corresponding splittings, and the statistical weight of $n_i$
is already included in $S_{ij}$ in eq. (\ref{hydrbb}).  For the range of temperatures and
pressures of interest, only the ground state contributes significantly
to the partition function, so $Q(T)=2$ for all $T$.

For values of $n_i > 7$ for which $n_j-n_i \leq 6$, an approximate
method based on Menzel (1969) can be used, which is modified to obtain the
strength in $a_o^2$, then eq. (\ref{hydrbb}) is used to obtain the required
line strength in cm$^2$atom$^{-1}$.  This is given by

\begin{equation}
S_{ij} \simeq \frac{n_i^5n_j^2}{n_j^2-n_i^2} A_{\Delta n}
\left[1+\frac{3\Delta n}{2n_i} - \frac{B_{\Delta n}}{n_i^2}\right]\, ,
\label{menzel}
\end{equation}

\noindent where $\Delta n = n_j-n_i$, and $A_{\Delta n}$ and $B_{\Delta n}$
are two sets of six coefficients as given by Menzel.

Finally, for the most general case outside the restrictions above, the
method of Bethe \& Salpeter (1957) can be applied using

\begin{equation}
S_{ij} \simeq \frac{64}{3^{1/2}\pi}
\frac{n_i^5n_j^5}{\left[\Delta n(n_i+n_j)\right]^4}\, ,
\label{bethe}
\end{equation}

\noindent to obtain the strength in units of $a_o^2$.

Once the line positions and strengths have been calculated, the lines have
to be broadened.  
%
%A Lorentzian profile can be used according to the Van der
%Waals theory discussed in \S\ref{vald}.  Collisions with other hydrogen
%atoms (self broadening), as well as with $H_2$ and $He$, can be calculated
%using the available ionization potentials and polarizabilities.  Although
%the broadening is dependent on the angular momentum quantum numbers for
%each transition, since the individual $l$ substates are neglected in the
%discussion above, and in view of the considerable extra complexity in
%taking this into account, particularly the wings, the extra work is
%probably not justified.  Moreover, the opacity due to lines of atomic
%hydrogen are at best going to make a relatively minor contribution in
%brown dwarf and giant planet atmospheres.
%
Because the lines of hydrogen are so well known, and they are only
a minor opacity source for the temperatures and pressures of interest,
they have not been included in any figures here.

\subsection{Calculation of Alkali Atomic Opacities with Data from VALD}
\label{vald}

All the lines of the alkali elements $Li$, $Na$, $K$, $Rb$, and $Cs$ can be
calculated using the data from the Vienna Atomic Line Database (VALD -
Piskunov 1994).  Because of the relatively small number of lines, the
opacities can be calculated when required, rather than precomputed and
stored in tables.  The abundances as a function of temperature at a total
pressure of 1 atmosphere are plotted in Fig. \ref{fig:19}
(\S\ref{atmosabun}).

Lithium is an important indicator of youth for objects above about 65 $M_J$
where $M_J$, is the mass of Jupiter, which is about $10^{-3}M_\odot$.
Above this mass it is quickly destroyed by nuclear burning.  Sodium and
potassium are the most and second most abundant alkali elements,
%<<<
respectively, and
%>>>
because of their low ionization potentials, are responsible for most of
the free electrons at low temperatures, and thus, ultimately for the
negative hydrogen ion bound-free and free-free absorption.  In their own
rights, their resonance lines are extreemly broad and blanket a large
region of the visible and near infrared parts of the spectrum
(Burrows, Marly, \& Sharp 2000).  Because they tend to remain in the
atmosphere when more refractory species have rained out, and because their
resonance lines are in a part of the spectrum which is otherwise
relatively transparent, their absorption can play a surprisingly important
role, even though their abundances are relatively low.  Although rubidium
and cesium have low abundances, like sodium and potassium, their resonance
lines are in a part of the spectrum that is relatively transparent,
are observed in spectra, and so are included as minor species here.

The line strengths can be calculated from eq. (\ref{oscil}) using the
tabulated $gf$ values, excitation energies, and transition energies, together
with a factor of $2J''+1$ for the initial state.  The energy levels up to
about 20,000 cm$^{-1}$ and statistical weights can be used to calculate the
partition functions.

With the exception of the resonance lines of sodium and potassium, simple
Lorentzian profiles can be calculated based on Van der Waals broadening.
For a number of transitions, the Van der Waals broadening coefficients are
available together with the natural broadening and Stark broadening (for
highly ionized and dense atmospheres).  Given the Van der Waals broadening
coefficient $\gamma_W$, the full-width at half-maximum in cm$^{-1}$ can be
calculated by scaling the broadening due to collisions of a perturbing
species relative to collisions with atomic hydrogen using

\begin{equation}
\Delta\bar\nu = \frac{\gamma_W}{2 \pi c} \left[\frac{m_H(m+m_p)}{m_p(m+m_H)}
\frac{T}{10000}\right]^{\frac{3}{10}} \left[\frac{\alpha_p}{\alpha_H}\right]
^{\frac{2}{5}} N_p \, ,
\label{widscaled}
\end{equation}

\noindent where $m$ is the mass of the atom undergoing the transition,
$m_H$ is the mass of the hydrogen atom, $m_p$ is the mass of the perturbing
species, $\alpha_H$ and $\alpha_p$ are the polarizabilities of hydrogen and
the perturbing species, respectively, and $N_p$ is the number density of
the perturbing species $H$, $H_2$, or $He$.  The combined line width is
obtained by evaluating eq. (\ref{widscaled}) for each of the perturbing
species, then taking the square root of the sums of the squares of the
line widths, weighted by the abundances of the perturbing species.

When the Van der Waals broadening coefficient is not available, it can still
be calculated using the angular momentum quantum numbers of the participating
states, from the symbols given in VALD, and the theory in Lang (1999) and
Dimitrijevi\'{c} \& Peach (1990), in order to calculate the $C_6$
interaction coefficient.  From this the line width can be calculated using

\begin{equation}
\Delta\bar\nu = 1.664461 \left[\frac{kT}{N_A}\left(\frac{1}{m} +
\frac{1}{m_p}\right) \right]^{\frac{3}{10}} C_6^{2/5} N_p/c \, ,
\label{widtheory}
\end{equation}

\noindent where $N_A$ is Avogadro's number.
%
%and the numerical factor in
%the equation is obtained from a complicated combination of factors
%involving $\pi$, $\Gamma$ function, and trigonometric functions, as detailed
%in Dimitrijevi\'{c} \& Peach (1990).

The resonance lines of sodium and potassium are very strong, and at high
pressures, absorption continues out into the far wings across much of the
visible and near IR parts of the spectrum.  Either truncating the lines,
as discussed in \S\ref{lines}, or extrapolating a Lorentzian profile into
the far wings, is unrealistic, so a more appropriate treatment should be
used, such as the more detailed theories by Burrows \& Volobuyev (2003), 
Allard et al. (2003), or Zhu, Babb, \& Dalgarno (2006).  A
normal Lorentzian profile with an appropriate truncation is generally used for the
other lines of sodium and potassium, which are much weaker, and for which
no suitable theory exists.  This applies also to all the lines of the
other alkali elements, since their abundances are much lower (Lodders 1999).

\section{CONTINUOUS OPACITIES}
\label{continuous}

Although the major contribution to the total opacity in brown dwarf and giant
planet atmospheres is the large number of lines produced by several molecular
species, and in some cases also by monatomic species,
continuous sources of opacity must also be considered, because otherwise
Rosseland mean opacities would be divergent.
%
%are required, no part of the spectrum, no matter
%how narrow in frequency range, can have a zero opacity.  Otherwise the result
%is divergent.  In a real gas this can never occur, although in theory windows
%between the lines can have very small opacities which could mistakenly be
%neglected.  If no background continuum is calculated, since the wings of the
%lines are truncated, there is a risk of zero opacity windows between these
%lines.  In certain situations, particularly at high temperatures when
%there are many free electrons, and at low temperatures where there are dust
%grains in the atmosphere, continuous scattering and absorption opacities can
%make a significant contribution.
Because the various continuous opacity sources can be calculated much more
rapidly than some of the line sources, the opacities can be calculated on
the fly when required, for the particular temperature and pressure,
rather than being stored in precomputed tables.

The various continuous processes that can take place are as follows:

1. Thomson scattering by free electrons, which is not considered here due to
the fact that the temperature is  too low for a significant number of free
electrons to be present to make an important contribution.  It is
trivial to include if required.

2. Rayleigh scattering, which is also not discussed here, because at the
temperatures and pressures of interest, the opacity is totally dominated
by gas absorption or scattering due to grains.  Again, it would be simple to
include, and the polarizabilities for the most abundant species would be
required.

3. Free-free opacity, i.e. inverse bremstrahlung, which is due to a free
electron moving in the field of a neutral atom or molecule, or a positively
charged atomic or molecular ion, and produces absorption across the whole
spectrum.  By convention, we count the free electron as part of the
system, so the charge of the species, neutral or ionized, includes that
electron.  Therefore, $H^-_{ff}$ (negative hydrogen ion free-free) refers
to an electron moving in the field of a neutral hydrogen atom.  At the
temperatures of interest there will be few free electrons, and nearly all
the free electrons will be produced from the ionization of the alkali
elements, so free-free processes are relatively unimportant, except at the
highest temperatures of interest.  Consequently this is discussed only
briefly.

All free-free absorption coefficients are expressed in cm$^5$ particle$^{-1}$
electron$^{-1}$, where a particle is an atom, ion, or molecule, and the
absorption per unit volume in cm$^{-1}$ is obtained by multiplying this by
the abundances of both the particle in question and the electrons.  The
mass absorption coefficient in cm$^2$ g$^{-1}$ is obtained by dividing by the
mass density.

4. Bound-free opacity, namely the photoionization of an electron from a neutral
or positively charged atomic or molecular species, or photodetachment of an
electron from a negative ion.  Unlike free-free opacities, this absorption is
confined to specific regions of the spectrum, and is relatively strong.
The processes we consider are $H^-_{bf}$ (the
photodetachment of an electron from a bound $H^-$ ion), and photoionization
of neutral species.  Bound-free opacity also includes photodissociation of
molecules.

5. Collision-induced opacity due to induced dipole-dipole absorption by
collisions between $H_2$ with itself, and with other species.  This can be
a large opacity source in the infrared.

6. Mie scattering by grains, which is briefly discussed.

Table 1 lists the gas-phase species which are important sources of absorption,
together with the last 8 entries which give the species included in Mie
scattering calculations.

\subsection{Free-Free Absorption}
\label{freefree}

$H_{ff}$ - This absorption is produced by a free electron moving in the field
of a proton.  The convention used here is to count the net charge of the whole
system.  A suitable theory is one based on Carson (1988a) in the form of
a table giving the monochromatic cross section in units of
cm$^5$ proton$^{-1}$ electron$^{-1}$ as functions of temperature and
photon energy.  

For the low temperatures of interest here there will be very few free protons,
so this source of absorption is of little importance.  However, if a
hydrogenic approximation is made for ions of other atoms with low
ionization potentials, such as the alkali elements, this source of
absorption could play a minor role near the upper temperatures of interest.

$H^-_{ff}$ - This absorption is produced by a free electron moving in the
field of of a neutral hydrogen atom, and is rather more important than the
previous process.  Nevetheless, it will make only a minor contribution to the
total opacity.  The absorption can be calculated from fitted coefficients
from Wishart (1979).

$He^-_{ff}$ - In the case of free electrons moving in the field of a neutral
helium atom, Table 2 of Bell, Berrington, \& Croskery  (1982) can be used.
Pre-scaling the tabulated values of the cross sections by the
square of the photon energy and the square root of the temperature
reduces the changes in the values in the resultant new table, and
increase the accuracy of the interpolation.

$H_{2ff}^-$ - Because $H_2$ is the most abundant species, a free electron
moving in the field of a neutral $H_2$ molecule is likely to be the
most important negative ion free-free opacity source.  The data are
available from Table 1 of Bell (1980), and can be treated in the same way
as $He^-$, with the same pre-scaling used before bicubic spline interpolation
is used.

Other Species - If required, the opacity of the following species can
be included: $C^-_{ff}$ (Bell, Hibbert, \& Berrington 1988),
$N^-_{ff}$ (Ramsbottom, Bell, \& Berrington 1992), $Cl^-_{ff}$
(John \& Morgan 1975), $Ne^-_{ff}$, $Ar^-_{ff}$, $Kr^-_{ff}$, $Xe^-_{ff}$,
$Li^-_{ff}$, $Na^-_{ff}$, $Cs^-_{ff}$, $O^-_{ff}$, $Hg^-_{ff}$,
$N_{2ff}^-$, $O_{2ff}^-$, $CO^-_{ff}$, $CO_{2ff}^-$, and $H_2O^-_{ff}$
(John 1975).  However, because of their low abundances in substellar
atmospheres, they are unimportant for this particular work.

\subsection{Bound-Free Absorption}
\label{boundfree}

$H$ - At the upper temperature range of interest, free atomic hydrogen will
be present, which will contribute bound-free absorption to the total opacity.
If $\sigma_n(\bar{\nu})$ is the partial cross section as a function of
wavenumber for the atom in an initial state with the principle quantum
number $n$, with $n$=1, 2, 3,..., then for a general hydrogenic atom

\begin{equation}
\sigma_n(\bar{\nu}) = \frac{64\pi^2}{3^{3/2}} \left(\frac{e^2}{hc}\right)^3
\frac{RZ^4}{\bar{\nu{n}}^3} \frac{e^{-RZ^2\frac{hc}{kT}(1-1/n^2)}}{Q}
G({n,Z,\bar{\omega}}) \left(1-e^{-hc\bar{\nu}/kT}\right)\, ,
\label{hydrobf}
\end{equation}

\noindent in cm$^2$atom$^{-1}$ for a given $n$, where $Z$ is the nuclear
charge, which is 1 for hydrogen, $R$ is the Rydberg constant,
$Q$ is the partition function, $\bar{\omega}$ is the wavenumber offset
from an absorption threshold, $G({n,Z,\bar{\omega}})$ is the Gaunt factor,
and the other quantities have their usual meanings.  Equation
(\ref{hydrobf}) includes all factors, including the population of the
initial level with the quantum number $n$ and the stimulated emission factor.
At a given wavenumber, eq. (\ref{hydrobf}) is evaluated for all values of $n$
for which

\begin{equation}
\bar{\omega} = \bar{\nu} - \frac{RZ^2}{n^2} \geq 0 \, ,
\label{threshold}
\end{equation}

\noindent is valid, i.e., at and above the absorption threshold for a given
value of $n$.  Below the threshold, no contribution need be calculated.  We
neglect here the slight effects due to level dissolution and related 
pseudo-continuum opacities (Hummer \& Mihalas 1988).  The
total cross section at a given value of $\bar{\nu}$ is obtained by summing
the partial cross sections $\sigma_n(\bar{\nu})$ over $n$ for each frequency
grid point $\bar{\nu}$.  A value of 20 is a more than suitable
value for the upper limit of $n$, and should account for all the
excited levels present for the temperatures and pressures of interest.

The Gaunt factor $G({n,Z,\bar{\omega}})$ is a correction to the cross section,
and is obtained from an algorithm by Carson (1988b).  It is exact for the
range $1 \leq n \leq 7$, and approximations are used for larger values of $n$.
The strongest absorption is the Lyman continuum arising from the ground
electronic state in the ultraviolet, which is always present.  Absorption at
longer wavelengths is due to excited electronic states, which are only
important when these states are sufficiently populated at high enough
temperatures.

Although the partition function should be calculated by summing the levels,
in practice for the temperatures and pressures of interest in cool
atmopsheres, only the ground state contributes significantly, and so it has
the value of 2.

\begin{figure}
\epsscale{1.00}
\centerline{\includegraphics[angle=-90,width=19cm]{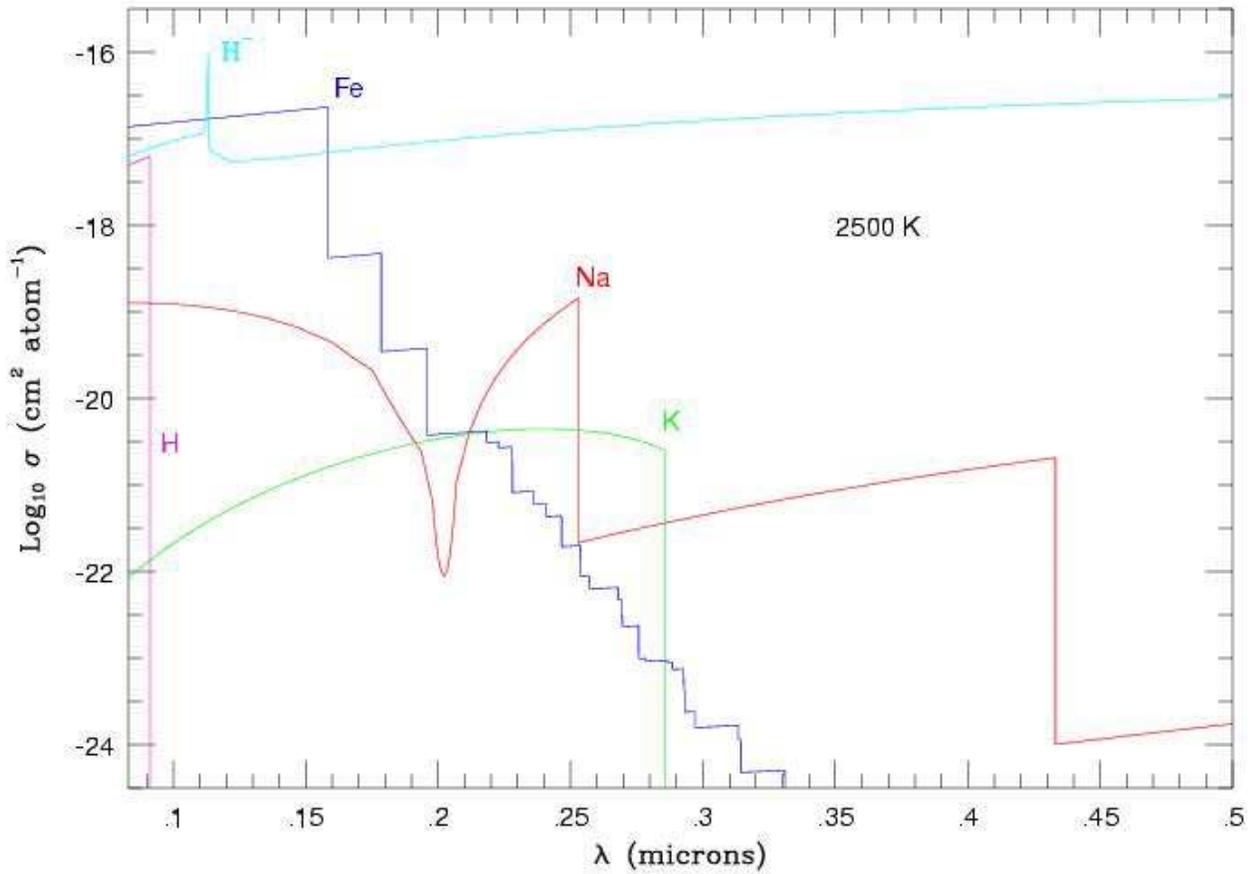}}
\caption{The log (base 10) of the monochromatic absorption $\sigma$ in
cm$^2$atom$^{-1}$ as a function wavelength $\lambda$ in $\mu$m at
a temperature of 2500 K due to bound-free transitions for the atomic
species $Na$ (red), $K$ (green), $Fe$ blue, and $H$ (magenta).
In addition the negative hydrogen ion $H^-$ (cyan) is
also plotted.  The effects of pressure have been neglected.  In the case
of $Fe$, this is plotted with the atomic lines in Fig. \ref{fig:9}.
The absorption shown here due to $H$ is caused by the Lyman
continuum, and the absorption due to $H^-$ is a smooth continuum
covering the whole ultraviolet and visible spectrum, as well as the near
infrared above the absorption threshold.  However, note the spike just
shortward of 0.11 $\mu$m.}
\label{fig:14}
\end{figure}

The monochromatic bound-free absorption per atom (or ion in the case of
$H^-$) is shown in Fig. \ref{fig:14} for $H$, $H^-$, $Na$, $K$, and $Fe$
for a temperature of 2500 K.  Because of the relatively low temperature,
the absorption due to the Balmer continuum is too weak to show, and is an
unimportant opacity source for most of the temperatures of interest;
however, the Lyman continuum is very important if there is a considerable
amount of irradiation at short ultraviolet wavelengths due to the parent
star.

$H^-$ - At high enough temperatures for some of the $H_2$ to be dissociated,
some of the alkali elements will be ionized, and
a portion of the freed electrons can be attached to the free hydrogen
atoms, resulting in the $H^-$ ion.  For photons with wavenumbers above
6083 cm$^{-1}$ (wavelengths shorter than 1.64 $\mu$m) in the infrared
this extra electron can be photodetached.

Data from from Bell \& Berrington (1987), Mathisen (1984), and Wishart
(1979) are available.  The data are in the form of a table of 85 cross
sections with corresponding wavenumbers from the absorption threshold to about
556,000 cm$^{-1}$ (0.018 $\mu$m) in the ultraviolet, from which the
absorption at the required wavenumbers can be obtained by linear interpolation.
As the data extend well beyond the short wavelengths of interest in the
ultraviolet, no extrapolation is required.  After a rapid rise above the
absorption threshold in the near infrared, there is very broad absorption
extending from the near infrared, across the entire visible, and into the
ultraviolet, gradually falling at shorter wavelengeths, except for a strong
and very sharp peak in the ultraviolet.  This peak corresponds to a
resonance with the $^2P$ state, and is in fact where $H^-$ has its
maximum absorption.

As can be seen in Fig. \ref{fig:14}, $H^-$ is the strongest opacity
source per species here, except for wavelengths shorter than 0.16 $\mu$m
where $Fe$ exceeds it most of the time.  As can also be seen from
Fig. \ref{fig:20}, the abundance of $H^-$ smoothly decreases with
decreasing temperature.

$Na$ - Data are available from the Strasbourg TOPBASE database
(Cunto \& Mendoza 1992; Cunto et al. 1993) for many ions and neutral
atoms of astrophysical interest up to calcium in 
the periodic table and for iron.  In the case of neutral
sodium, the excitation energies and statistical weights of the ground
state and 31 excited states are listed in the reference, together with
partial cross sections as a function of photon energy in the form of a
table for each state.  The total cross section $\sigma(\bar{\nu})$ as a
function of wavenumber can be obtained from the summation, with
stimulated emission included:

\begin{equation}
\sigma(\bar{\nu}) = \sum_{i=1} g_i \sigma_i(\bar{\nu})
\frac{e^{-E_ihc/kT}}{Q(T)} \left(1-e^{-hc\bar{\nu}/kT}\right)\, ,
\label{topbase}
\end{equation}

\noindent in cm$^2$atom$^{-1}$, where $i$ is now just a running index
rather than a quantum number, as in eq. (\ref{hydrobf}), with the ground
state being indexed by $i$=1, $g_i$, $E_i$, and $\sigma_i(\bar{\nu})$ are,
respectively, the statistical weight, excitation energy in cm$^{-1}$ above
the ground state, and the partial cross section as a function of wavenumber
for the $i^{th}$ state.

Simple linear interpolation can be used to obtain $\sigma_i(\bar{\nu})$ at the
required value of $\bar{\nu}$.  Summation is skipped over those indices for
which the photon energy is below the absorption threshold, and for a
photon energy above the upper range of the tabulated values the cross
section of the last tabulated cross section can be used.  This very simple
extrapolation avoids an artificially discontinuous drop in opacity.

\begin{figure}
\epsscale{1.00}
\centerline{\includegraphics[angle=-90,width=19cm]{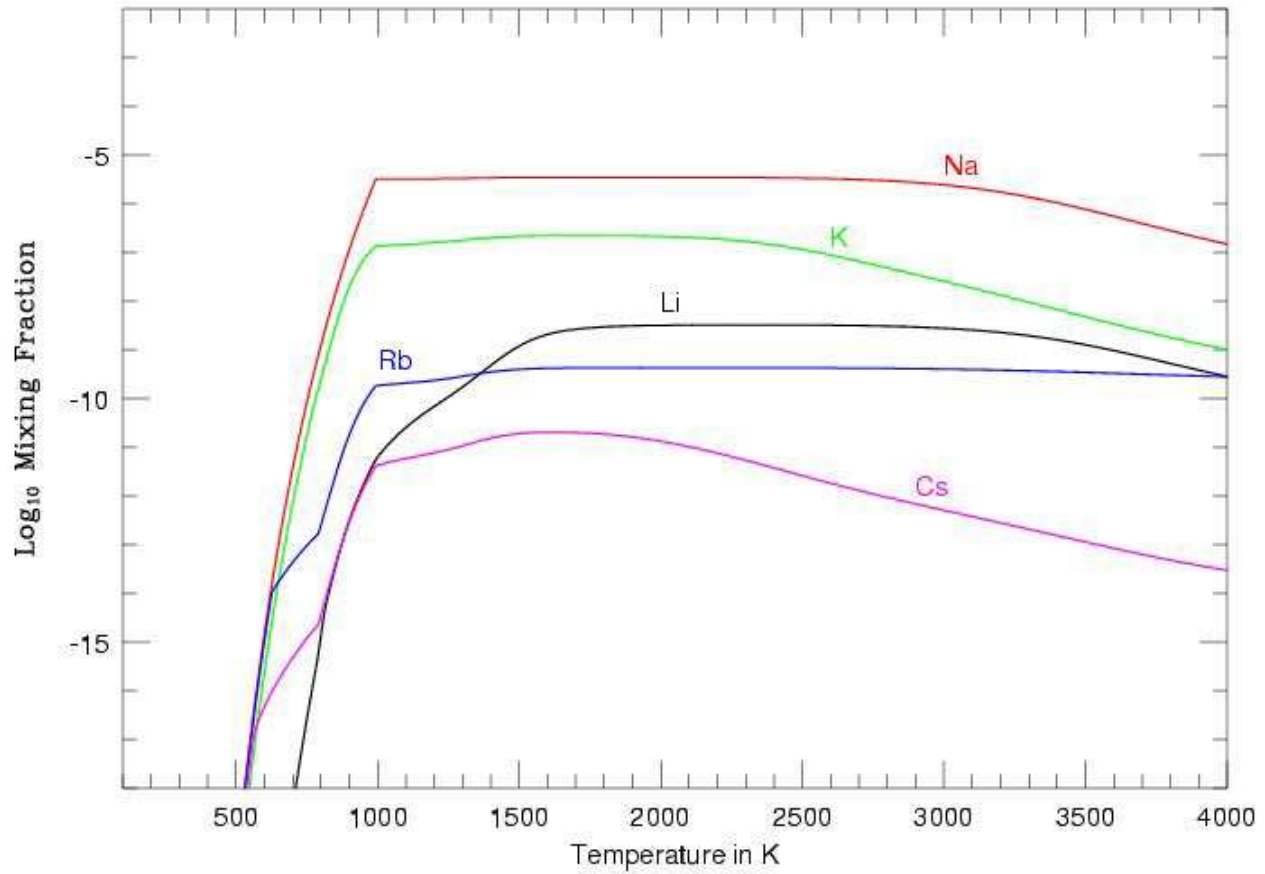}}
\caption{The log (base 10) of the mixing fraction as a function of
temperature at a total gas pressure of 1 atmosphere for the alkali
elements in their monatomic form: $Li$ (black), $Na$ (red), 
$K$ (green), $Rb$ (blue), and $Cs$ (magenta), shown on the same
scale as previously.  In all cases, they start decreasing in abundance
with decreasing temperature at about 1000 K, due to the formation of
condensates, and in the case of $Rb$ and $Cs$, kinks at
lower temperatures are due to the most abundant condensed species being
replaced by another one.  Unlike the other elements, $Rb$ has a
constant abundance above about 2000 K, but this is an artifact of
thermodynamic data missing for some species of $Rb$.}
\label{fig:19}
\end{figure}

%<<<
The partition function can be obtained by summing the ground state and the
first two excited states which have energies less than 20,000 cm$^{-1}$ above
the ground state, using data from Moore (1949).  The Boltzmann factors for
states with energies at and above 20,000 cm$^{-1}$ over the ground state make
a negligibly small contribution to the partition function, much less than 1
percent, even at the temperatures well above the highest of interest.
Consequently, such states are not considered, as previously stated.
%>>>
Because of its low
ionization and excitation energies, $Na$ has important bound-free
absorption at relatively long wavelengths, even at a temperature
of 2500 K, as plotted in Fig. \ref{fig:14}.  The sodium abundances, along
with those of the other alkali elements, are plotted in Fig. \ref{fig:19}
(see also Lodders 1999).

$K$ - Unfortunately, potassium is one of the elements not available in
the TOPBASE database, but an alternative though considerably less
extensive database given by Verner \& Yakovlev (1995) can be used
that does include it.  They cover, with fitted coefficients, all the neutral
atoms from hydrogen to zinc ($Z=30$), together with most of
their positive ions that contain at least one bound electron.

However, the only data of relevance for potassium are the fit coefficients
for absorption from the outermost electron in the $[Ar].4s^1$ ground state
configuration.  Unlike sodium, no data are available for excited states.  
Verner \& Yakovlev (1995) also give cross sections for inner electrons,
but these are at wavelengths in the far ultraviolet and X-ray that are of
little interest here.

From Verner \& Yakovlev (1995), the general expression for the partial cross
section is given by

\begin{equation}
\sigma_{nl}(\bar{\nu}) = 10^{-18}\sigma_o\left\{\left[(y-1)^2 + y_w^2\right]
y^{(P-10)/2-l}\left[1+\sqrt{y/y_a}\right]^{-P}\right\}\, ,
\label{verner}
\end{equation}

\noindent in cm$^2$absorber$^{-1}$, where $n$ is the principle quantum number,
$l$ is the angular momentum quantum number taking the values 0, 1, 2, ...
for $s$, $p$, $d$, etc., $\sigma_o$, $y_w$, $y_a$, and $P$ are fit parameters,
and $y$ is the photon energy relative to another fit parameter $E_o$.  Since
we are expressing photon energies in units of cm$^{-1}$, but $E_o$ is
given in eV in the reference, $y=\bar{\nu}/s_oE_o$, where the conversion
factor $s_o$=8065.539 cm$^{-1}$eV$^{-1}$ has to be applied.  Equation
(\ref{verner}) can be applied only for photon energies at and above the
absorption threshold energy.  Otherwise, there is no contribution.

In the case of potassium, only one partial cross section is available, and
since the absorption threshold energy is the ionization energy,
eq. (\ref{verner}) is evaluated only at and above the ionization energy
with $n$=4 and $l$=0.  Since the statistical weight is already included
in the cross section, and only the ground state is considered,
eq. (\ref{verner}) need only be divided by the partition function and
multiplied by the stimulated emission factor to obtain the total
required absorption in cm$^2$atom$^{-1}$.  As with sodium,
the first three energy levels from Moore (1949) are used to calculate the
partition function.  Because we have data only on a single level, the
absorption plotted in Fig. \ref{fig:14} contains only one absorption edge.
As discussed in \S\ref{atmosabun}, the abundances are plotted in Fig.
\ref{fig:19}.

$Fe$ - Like potassium, the TOPBASE database from Cunto \& Mendoza (1992)
and Cunto et al. (1993), does not include bound-free data for
iron, which has to be included, as it can be very important.
Such data, however, are available in Bautista (1997).
From Kurucz (1995), data are available for the bound-free absorption for 30
levels, starting from the ground state.  The data are in the form of an
energy in Hz above the ground state, together with various parameters,
from which the total absorption in cm$^2$atom$^{-1}$ can be found from
summing over the partial cross sections using

\begin{equation}
\sigma(\bar{\nu}) = 10^{-18}\sum_{i=1} g_i \alpha_i \sigma_i^x
\left[\beta_i+x(1-\beta_i)\right] \frac{e^{-E_ihc/kT}}{Q(T)}
\left(1-e^{-hc\bar{\nu}/kT}\right)\, ,
\label{febf}
\end{equation}

\noindent with stimulated emission included, where $\sigma_i$ is the
partial cross section at the threshold for the $i^{th}$ level, $\alpha_i$ and
$\beta_i$ are fitted coefficients for that level, $g_i$ is the statistical
weight, and $F_i$ is the corresponding term value in cm$^{-1}$ above the
ground state for that level, which we convert from the tabulated value in
Hz.  $F_i$ is also the photon energy of the absorption threshold, and is
used to define the dimensionless wavelength $x=F_i/\bar{\nu}$ used in
eq. (\ref{febf}).

As before, summation is required only for those levels for which
$\bar{\nu} \ge F_i$, and the partition function is obtained by summing over
the levels listed in Moore (1949).  Summing over the first 30 levels up
to just above 20,000 cm$^{-1}$ from the ground state should be adequate.
Previously, in Fig. \ref{fig:10}, the combined bound-bound and bound-free
absorption were plotted.  Because $Fe$ is so important, it is also included
in Fig. \ref{fig:14} for bound-free absorption alone.  (See Fig.
\ref{fig:20} for its abundance as a function of temperature.)

$CH$ - Bound-free data are available for the $CH$ molecule from
Kurucz, Dishoeck, \& Tarafdar (1987) in the form of a table of cross
sections for the range of temperatures and photon energies
of interest.  Bicubic spines can be used for interpolation,
and the partition function is not required, as it is implied in the
table.  However, for a solar composition with [$C$]/[$O$] $<$ 1, this
species is of little importance at any temperature.

$OH$ - Bound-free data are available from Kurucz, Dishoeck, \&
Tarafdar (1987) for the $OH$ molecule in a form very similar to $CH$, and
the same interpolation scheme can be used.  For a solar composition
at sufficiently high temperatures when $H_2O$ has mostly dissociated,
this molecule can play a minor role, but at the temperatures of interest
here where $H_2O$ is mostly associated, it is unlikely to be important.
Neither $CH$ nor $OH$ are shown in our plots.

\subsection{Collision-Induced Absorption}
\label{ciaopac}

The last major gas-phase contribution to the opacity that is of interest
in cool atmospheres
is collision-induced absorption (CIA).  During a collision a molecule
can have a transient induced dipole moment, which can allow a rotational
or vibrational-rotational transition to take place in the infrared, that
would otherwise not be possible.  Symmetric diatomic molecules that have no
permanent dipole moment, as well as symmetric polyatomic molecules that
have a number of vibrational modes that are inactive in the infrared are
of particular interest.  Because the absorption is due to the collision
of gas-phase species, it will scale as the square of the gas pressure, and
so is most important for high pressures and low temperatures relevant to
giant planets and brown dwarfs.

In a gas with a typical astrophysical composition, by far the most important
sources of CIA are due to $H_2-H_2$ and $H_2-He$ collisions, together with
collisions between $H_2$ and other common gas-phase species.  In the absence
of CIA (and discounting small Rayleigh scattering), these two most abundant
gases are very transparent in the infrared, with $H_2$ having only very weak
quadrupole absorption, and $He$ having essentially no absorption at all.
Thus, the treatment of CIA is particularly important.

CIA cross sections can be found mostly in Borysow \&
Frommhold (1989), Borysow \& Frommhold (1990), Borysow, Frommhold, \&
Moraldi (1989), Zheng \& Borysow (1995), Zheng \& Borysow (1995b),
Borysow \& Frommhold (1988), and Borysow, Frommhold,
\& Birnbaum (1985), as detailed below, for $H_2-H_2$, $H_2-He$ and
$H_2-CH_4$ collisions, the last involving collisions with a
third common species.  Data on $H_2-H_2O$ and other collision pairs
of relevance are not readily available.

The data for $H_2-H_2$, $H_2-He$, and $H_2-CH_4$ are similar, consisting
of cross sections in $\log_{10}($cm$^{-1}$amagat$^{-2})$ per collision pair
tabulated in wavenumbers from 10 cm$^{-1}$ in the far infrared
to 18,000 cm$^{-1}$ in the visible, for 500 values, and for temperatures
ranging from 50 K to 5000 K, for 25 values, where an amagat is a unit of
density in terms of Loschmidt's number, $L_o$.
($L_o = 2.68676\times 10^{19}$ cm$^{-3}$ is defined such
that $L_o = N_A/V_o$, where $N_A$ and $V_o$ are, respectively,
Avogadro's number and the volume occupied by 1 mole of gas at standard
temperature and pressure.)

At a given temperature and wavenumber, the monochromatic absorption for a
collision pair in cm$^{-1}$amagat$^{-2}$ can be obtained by bilinear
interpolation in wavenumber and temperature, and linear interpolation in
the tabulated log of the absorption, yielding a value of
$\sigma_{H_2-X}(\bar{\nu})$, where $X$ is the second species colliding with
the $H_2$ molecule, and can be another $H_2$ molecule or another species.
The required monochromatic volume opacity $\kappa_v(\bar{\nu})$ in
cm$^2$cm$^{-3}$, i.e. in cm$^{-1}$, can then be obtained from

\begin{equation}
\kappa_v(\bar{\nu}) = \sigma_{H_2-X}(\bar{\nu})N_{H_2}N_X/L_o^2\, ,
\label{cia}
\end{equation}

\noindent where $N_{H_2}$ and $N_X$ are, respectively, the number densities
in cm$^{-3}$ of $H_2$ and the species colliding with it.
Equation (\ref{cia}) can be evaluated for each of the three collisions,
with the results summed and added to the total opacity, after correction
for stimulated emission.  At the end of all the calculations, the final
monochromatic mass opacity in cm$^2$g$^{-1}$ is obtained by dividing by
the gas mass density.

Data are available from the following references:

$H_2-H_2$ - Borysow et al. (1985), Zheng, \& Borysow (1995a) and (1995b),
Borysow \& Frommhold (1990), Lenzuni, Chernoff, \& Salpeter (1991), and
Guillot et al. (1992),

$H_2-He$ - Borysow, Frommhold, \& Birnbaum (1988), Borysow, Frommhold,
\& Moraldi (1989), and Borysow \& Frommhold (1989), and

$H_2-CH_4$ - Courtin (2004).

\begin{figure}
\epsscale{1.00}
\centerline{\includegraphics[angle=-90,width=19cm]{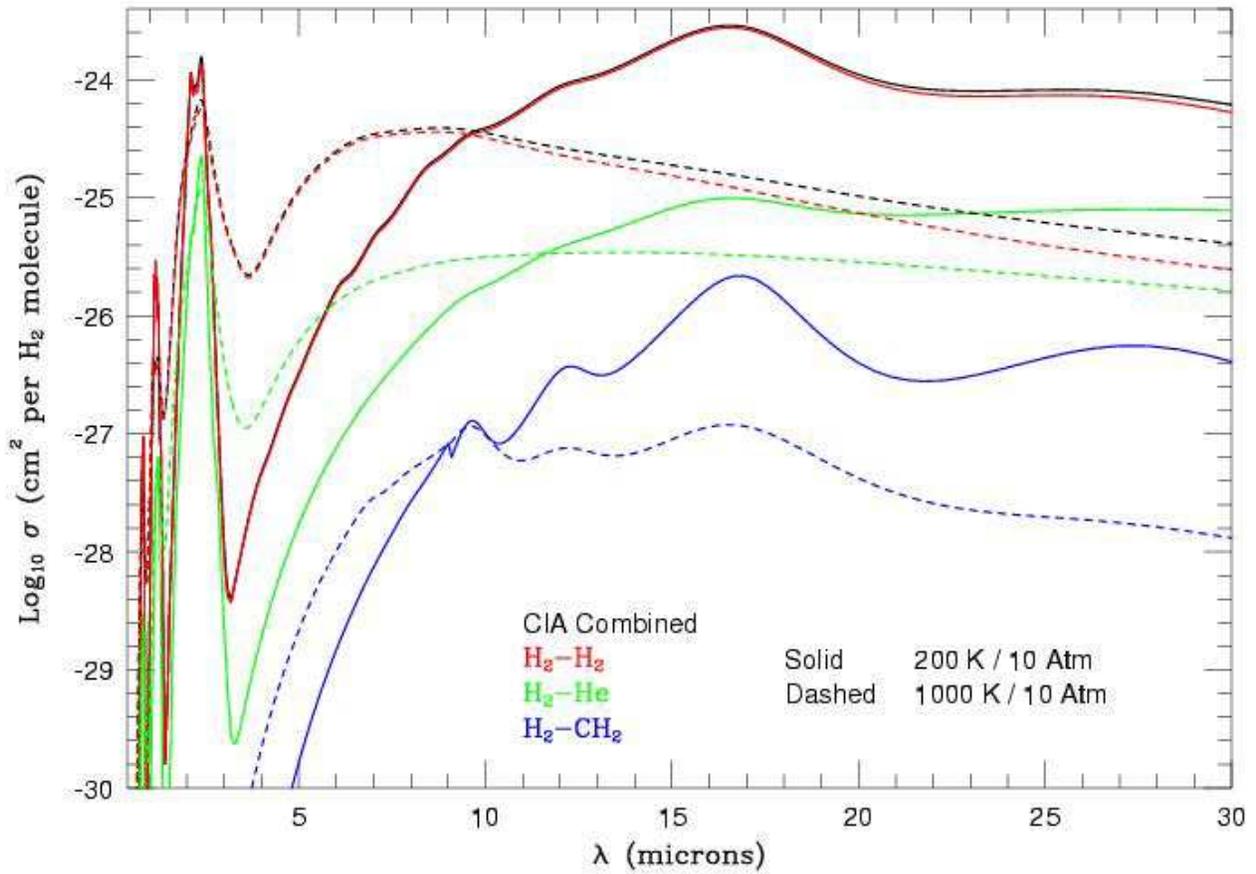}}
\caption{The log (base 10) of the monochromatic collision induced
absorption $\sigma$ in cm$^2$ per $H_2$ molecule as a function
wavelength $\lambda$ in $\mu$m at 200 K and 10 atmospheres, shown by
solid curves, and 1000 K and 10 atmospheres show by dashed curves.
The red curves are due to $H_2-H_2$ CIA expressed as the
absorption per $H_2$ molecule in the gas, regardless of whether
any particular $H_2$ molecule is undergoing a collision; the
green curves are due to $H_2-He$ CIA, but weighted by the
abundance of helium, and likewise the blue curves are due to
$H_2-CH_4$ CIA, weighted by the abundance of methane.  The black
curves are the combinations of these contributions at the two temperatures,
i.e. the CIA due to $H_2$ colliding with any of $H_2$,
$He$, and $CH_4$.  Note that the CIA scales as the square of
the pressure.}
\label{fig:15}
\end{figure}

Figure \ref{fig:15} shows the monochromatic absorption due to
$H_2-H_2$, $H_2-He$, $H_2-CH_4$, and the combined CIA, i.e. the sum of
the contributions due to $H_2$ colliding with another $H_2$, $He$, and
$CH_4$ weighted by their respective abundances at 200 and 1000 K at a
pressure of 10 atmospheres.  The absorption is expressed
in cm$^2$ per $H_2$ molecule, regardless of the fraction of $H_2$
molecules participating in a collision at any given instant.  As CIA is
proportional to the square of the pressure, plots at 1 and 100
atmospheres would be exactly the same, except that the absorption would be
scaled by factors of 0.01 or 100, respectively.

\subsection{Calculation of Grain Opacities}
\label{grainopac}

The final source of opacities discussed here is due to scattering and
absorption caused by solid and liquid particles, for which the treatment is
completely different than the gas phase.  This obviously requires liquid or
solid particles to be suspended in the atmosphere in the form of clouds, which
for equlibrium chemistry requires the temperature to be below the condensation
temperature of the species considered.  This can take place in the atmospheres
of very late M dwarfs, and all T and L dwarfs.

In the limiting case when particles are very small compared with the
wavelength of the radiation, Rayleigh scattering takes place; in the
opposite case when the particules are very large compared with
wavelength, geometric scattering is applicable.  In the most general (and
most complicated) case when the wavelength of the radiation is of
approximately the same order as the particle size, Mie scattering
theory should be applied.

The theory we present here is based on the work of Van de Hulst (1957),
where we start with a set of uniform spheres of the same substance in
suspension (liquid or solid), all having the same radius.  The
scattering and absorption of the radiation depends on only four
parameters: the real and imaginary parts of the refractive index, the
radius of the particles, and the wavelength of the photon being
scattered or absorbed.

Specifically, the scattering and extinction coefficients are

\begin{equation}
Q_{sca} = {\sigma_{scat}\over {\pi a^2}} =
{2\over x^2}\sum_{n=1}^\infty (2n+1)[\mid a_n \mid^2 + \mid b_n \mid^2]\, ,
\label{sca}
\end{equation}

\noindent and

\begin{equation}
Q_{ext} = {\sigma_{ext}\over {\pi a^2}} =
{2\over x^2}\sum_{n=1}^\infty (2n+1)Re(a_n+b_n)\, ,
\label{ext}
\end{equation}

\noindent respectively, where $a$ is the particle radius, $x$ is the size
parameter ($=2\pi a/\lambda$), and $a_n$ and $b_n$ are the Mie coefficients.
Following Deirmendjian (1969), for numerical computation these coefficients
can be expressed in terms of the complex index of refraction,
$m=n_{real}-in_{imag}$, and Bessel functions of fractional order:

\begin{equation}
a_n = {{{\left[{A_n(mx)\over m} + {n\over x}\right]}
J_{n+{1\over 2}}(x)-J_{n-{1\over 2}}
(x)}\over{{\left[{A_n(mx)\over m} + {n\over x}\right]}\left[J_{n+{1\over 2}}(x)
+(-1)^niJ_{-n-{1\over 2}}(x)\right] -
J_{n-{1\over 2}}(x)+(-1)^niJ_{-n+{1\over 2}}(x)}}\, ,
\label{coefan}
\end{equation}

\noindent and

\begin{equation}
b_n = {{{\left[mA_n(mx) + {n\over x}\right]}
J_{n+{1\over 2}}(x)-J_{n-{1\over 2}}
(x)}\over{{\left[mA_n(mx) + {n\over x}\right]}\left[J_{n+{1\over 2}}(x)
+(-1)^niJ_{-n-{1\over 2}}(x)\right] - J_{n-{1\over 2}}(x)+
(-1)^niJ_{-n+{1\over 2}}(x)}}\, ,
\label{coefbn}
\end{equation}

\noindent where

\begin{equation}
A_n(mx) = {{J_{n-{1\over 2}}(mx)\over{J_{n+{1\over 2}}(mx)}} - {n\over{mx}}}\, .
\label{bigan}
\end{equation}

Equations (\ref{sca}) and (\ref{ext}) are evaluated by summing up the first
50 to 200 terms, depending on the value of the size parameter, with large
size parameters requiring more terms in order to produce accurate scattering
and extinction coefficients.  For size parameters larger than $\sim$75,
variations in $Q_{ext}$ and $Q_{sca}$ are much reduced.  An asymptotic form
of the Mie equations for large $x$ outlined fully by Irvine (1964) is used
to find the limits of the scattering coefficients.  In the large $x$ limit,
the extinction coefficient, $Q_{ext}$, approaches 2.0.  Interpolation
between the full Mie theory results and these asymptotic limits yields
the coefficients for large size parameters.

In practice, there will be a range of particle sizes for the grains or drops
suspended in a cloud.  Sudarsky (2002) assumed a range of
particle sizes given by

\begin{equation}
n(a) \propto \left({a\over a_0}\right)^6 \exp\left[-6\left({a\over a_0}\right)
\right]\, ,
\label{cloud}
\end{equation}

\noindent where $n(a)$ is proportional to the number of particles in the
size range $a$ to $a+da$ and $a_0$ is the size at the peak of the
distribution.  Equation (\ref{cloud}) reproduces the distributions in
cumulus water clouds in Earth's atmosphere fairly well if
$a_0 \simeq 4~\mu$m (Deirmendjian 1964).  However, stratospheric
aerosols in Earth's stratosphere can be represented by the ``haze''
distribution given by

\begin{equation}
n(a) \propto {a\over a_0}\exp\left[-2\left({a\over a_0}\right)^{1/2}\right]\, ,
\label{haze}
\end{equation}

\noindent which is also obtained from Deirmendjian (1964).

Given a size distribution, eqs. (\ref{sca}) and (\ref{ext}) are evaluated
for a range of sizes for every wavelength in a grid, and weighted mean
scattering and extinctions are calculated at each
wavelength for a particular condensed species, using the refractive index
of that species at each wavelength in turn.  In general, the real and
imaginary parts of the refractive index vary with wavelength; in particular,
the imaginary part can vary by many orders of magnitude.  These calculations
are repeated for any additional species present in the cloud, using the
refractive indices for that species.  Thus, in a parcel of gas inside
a cloud, the total opacity is given by the contribution of the condensates,
in addition to that of the gas-phase species previously discussed.

\begin{figure}
\epsscale{1.00}
\plotone{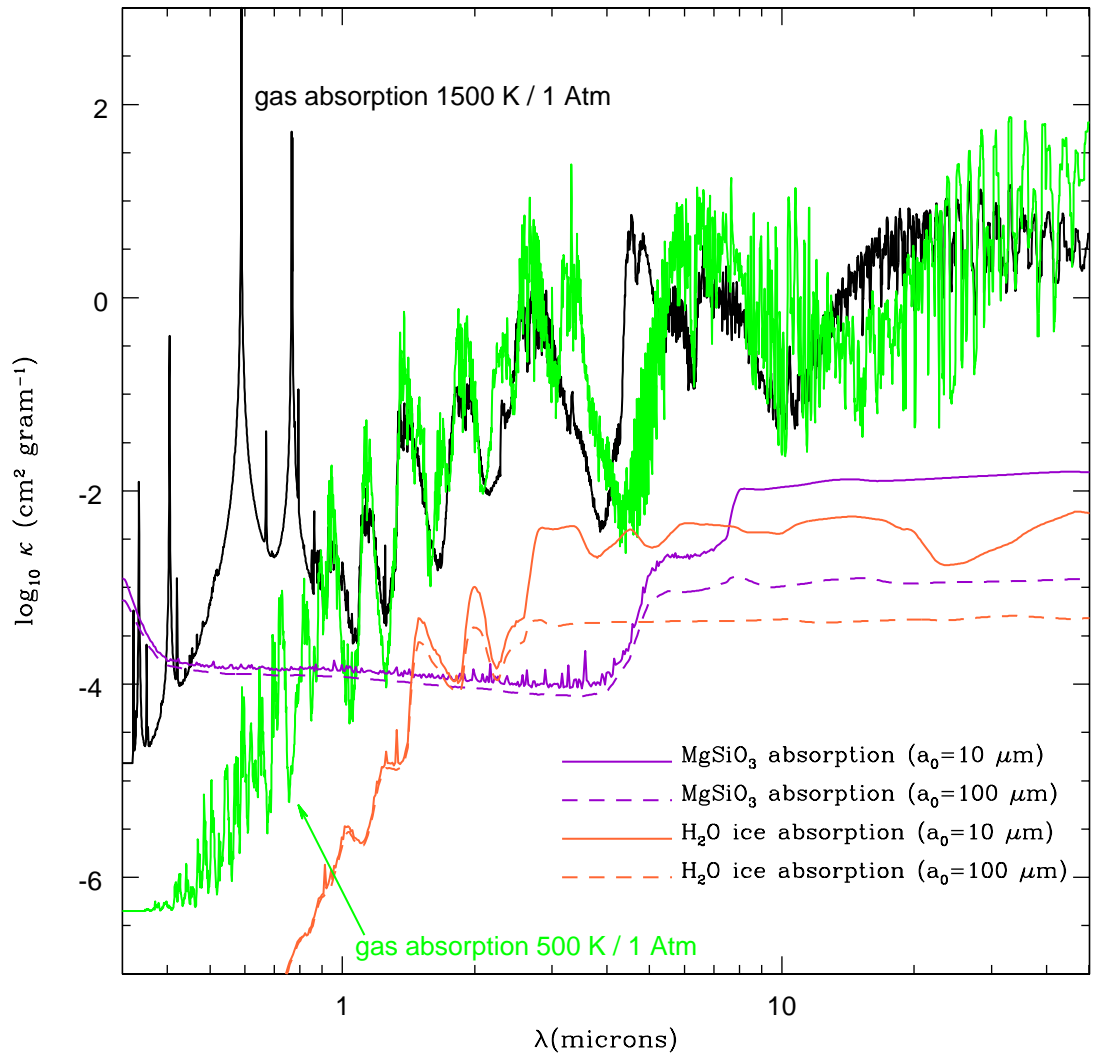}
\caption{The log (base 10) of the monochromatic opacity $\kappa$ in
cm$^2$g$^{-1}$ as a function of wavelength $\lambda$ in $\mu$m for
separate gas and grain absorption.  The green and black curves represent
at a pressure of 1 atmosphere the monochromatic gas opacity at
temperatures of 500 K and 1500 K, respectively, with the contributions
of the individual species weighted by their abundances.  The magenta
and orange curves represent the absorption due to grains of $MgSiO_3$
and $H_2O$ ice, respectively, with the solid and dashed curves showing
the absorption due grains or ice particles with radii of 10 $\mu$m and
100 $\mu$m, respectively.  The opacity is weighted by the abundances;
in the case of $MgSiO_3$ it is given by the least abundant element,
which is $Si$, and in the case of ice the abundance is based on the
equilibrium with $H_2O$ in the vapor phase.  In both cases, the
slight effects of temperature and pressure are neglected, except that the
temperature has to be low enough for the phase to be stable, which in
the case of $H_2O$ ice is at or below 273 K.}
\label{fig:16}
\end{figure}

Figure \ref{fig:16} compares the monochromatic absorption of the gas at
500 K and 1500 K at a pressure of 1 atmosphere, and the absorption due to
$MgSiO_3$ grains and water ice for two grain (or ice) sizes.  The gas
opacity is calculated from the contributions due to many of the
individual species that have been previously discussed.

\subsection{Abundance-Weighted Total Opacities}
\label{total}

In order to calculate the total opacity of a gas, or a gas plus condensed
species, the monochromatic opacity as a function of wavelength or wavenumber
for each species has to be calculated first in the form of an absorption
in cm$^2$ per atom, ion, or molecule, for the required temperature and
pressure.  Once the absorption for each species has been obtained, they
are combined together.  This requires calculating the abundance of
each species, then weighing their contribution by these abundances.

Because all intermediate calculations are based on number densities of the
species, it is convenient to perform all opacity calculations as volume
opacities, i.e in cm$^2$cm$^{-3}$ = cm$^{-1}$, which are obtained by
multiplying the opacities of each species by their number densities
in cm$^{-3}$, then at the end to convert to the total mass opacity by
dividing by the mass density.  This can be sumarized by writing

\begin{equation}
\kappa_{mass} = \frac{1}{\rho}\sum_iN_i\kappa_i =
\frac{N_A}{\mu N_t}\sum_i N_i\kappa_i\, ,
\label{totopac}
\end{equation}

\noindent where $\kappa_{mass}$ is the monochromatic mass opacity in
cm$^2$g$^{-1}$, $N_i$ and $\kappa_i$ are, respectively, the number density
in cm$^{-3}$ of species $i$ and the monochromatic opacity of that species in
cm$^2$species$^{-1}$, $\rho$ is the density (possibly including suspended
condensates) in g cm$^{-3}$, $N_A$ is Avogadro's number, $\mu$ is the
mean molecular weight, and $N_t$ is the total number density in cm$^{-3}$ of
all particles, gas and condensates.

In order to calculate the total opacity, the quantities $N_i$ in
eq. (\ref{totopac}), the number densities for each species, have to be
calculated.  This is discussed in \S\ref{atmosabun}.  However, one of the
dilemmas in calculating total opacities is that \textit{ab initio} it is
not always possible to know which species are likely to be important
opacity sources without investing the time in the first place to determine
whether they are indeed abundant enough.  In many cases, this
is obvious, but in others this can only be determined after it has been
included in chemical equilibrium calculations.  However, even then, a
species with a low abundance can turn out to be a very important opacity
source.  This can take place when its absorption features fall in a part
of the spectrum that is otherwise quite transparent for the particular
temperature and pressure being investigated.  Good examples are $TiO$ in
M dwarfs, and the lines of the alkali elements in T and L dwarfs.  Moreover,
even species that do not contribute a significant opacity, or for which no
relevant data are available, can indirectly affect the opacity by altering
the abundances of those species that do.

\section{ATMOSPHERIC ABUNDANCES}
\label{atmosabun}

Before any opacities can be calculated, the abundances of a mixture of
a large number of species have to be determined for the given temperature
and pressure.  For the results shown here, nearly 500 species were
followed, including over 150 condensates, containing 27 elements, using
the Allende Prieto, Lambert, \& Asplund (2001), 
Asplund et al. (2005), Allende Prieto \& Lambert (2005) composition, which
replaces the Anders \& Grevesse (1989) composition previously used.  The
27 elements used are the following:
$H$, $He$, $Li$, $C$, $N$, $O$, $F$, $Ne$, $Na$, $Mg$, $Al$, $Si$, $P$,
$S$, $Cl$, $Ar$, $K$, $Ca$, $Ti$, $V$, $Cr$, $Mn$, $Fe$, $Co$, $Ni$, $Rb$,
and $Cs$.  Other than $He$, which should be included for
collision-induced opacities, the noble gases neon and argon play no role
in the chemistry or opacity, but should still be included to ensure that
for a given pressure, the gas has the correct density.  In spite of their
very low abundances, rubidium and caesium should be included because they
have strong resonance lines that can be relatively important in the near
infrared, and together with the other alkali elements contribute free
electrons at relatively low temperatures because of their low ionization
energies.  Lithium should also be included because below a mass of about 65
M$_J$ it is not destroyed by nuclear burning, and it is present in more
massive brown dwarfs and stars that are too young to have destroyed their
lithium.

When calculations are performed with different values of metallicity ($Z$),
the abundances of all the metals should be scaled with $Z$, but their
fractions within $Z$ remained constant.  At temperatures high enough for
ions and electrons to be important, electrons should be added as an
additional ``element" with a zero total abundance, and the stoichiometric
coefficient of an ionized atom, including the electron itself, is equal
in magnitude, but opposite in sign, to the charge.

In all cases the abundances can be calculated assuming chemical
equilibrium between the different species in the gas phase, together with
any condensates that may be present, using the methods described in detail
in Sharp \& Huebner (1990), Burrows \& Sharp (1999), and Fegley \& Lodders (2001).  For a given
temperature, pressure, and composition, the equilibrium abundances of the
various species can be determined by minimizing the total Gibbs free energy
of the system.  This requires a knowledge of the free energy of each species
as a function of temperature, which is normally obtained from thermodynamic
data.  However, suitable spectroscopic data for an important gas-phase
species can be used to calculate the free energy when no thermodynamic data
are available, as summarized in Appendix A of Burrows et al. (2005).   Since
the work of Sharp \& Huebner (1990), a number of substantial improvements
to the databases have been made, and were used in Burrows \& Sharp (1999).
Extensive updates to these thermodynamic databases have been made by 
Lodders \& Fegley (2002).  Data from Barin (1995) for a number of species
replace the earlier data from the JANAF tables from Chase (1982) and
Chase et al. (1985).  In the earlier work, Burrows \& Sharp (1999) used
Tsuji (1973) and Turkdogan (1980) for most of the species for which data
in the JANAF tables were not available, and most of these data are now
replaced by the far better data from Barin (1995).  The most important
gas-phase species from Barin (1995) are:
$CaH$, $CrO$, $TiO$, and $VO$, and the most important condensates
are: $Al_2O_3$, $CaTiO_3$, $FeSiO_3$, $MgTiO_3$, $Mg_2TiO_4$, $MgTi_2O_5$,
various oxides and nitrides of chromium, titanium, and vanadium, and a
number of other compounds involving combinations of at least two elements
out of aluminum, calcium, magnesium, manganese, and silicon, combined with
oxygen.

Most thermodynamic tabulations of free energy of a species are given in
terms of the formation energy of the species from their elements in their
reference states at the temperature being considered.  The problem with this
is that there is a discontinuity in the derivative of the free energy when
one of the elements in its reference state changes phase.  One way to
circumvent this problem is to redefine the free energy of formation of a
species in terms of its constituent elements in their monatomic neutral
gaseous phase, even if that phase is not stable at the temperature being
considered.  Thus, the element in that state is defined as having a free
energy of zero.  Moreover, if charged species are considered so electrons
are included, free electrons should also be defined to have a zero free
energy of formation.  Using this method, polynomials can be fitted to the
free energy to represent a smooth function with temperature.

Our equilibrium code was originally based on the SOLGASMIX code from Besmann
(1977), which used calories as the measurement of heat and 1 atmosphere as
the reference pressure (as used also by the JANAF tables).  The free energies
of formation of a species from its elements in their standard reference state
are used, then these are converted to the free energy of formation
from the elements in their neutral monatomic gaseous state using

\begin{equation}
\Delta G_{pi} = \Delta G^o_{pi} - \sum_{j=1}^n\nu_{pij}\Delta G^o_{1j}\, ,
\label{convertg}
\end{equation}

\noindent where $\Delta G_{pi}$ is the required Gibbs free energy of
formation of substance $i$ in phase $p$ in cal mol$^{-1}$,
$\Delta G^o_{pi}$ is the tabulated free energy in terms of the elements
in their standard reference states, $\nu_{pij}$ is the stoichiometric
coefficient of species $i$ in terms of each of its constituent
elements $j$, and $\Delta G^o_{1j}$ is the tabulated free energy of
formation from its reference state of element $j$ in its monatomic
gaseous phase, where the gas phase is always indexed by 1.  If the
reference state of the element is the monatomic gaseous phase, then
this term is zero.  The summation is performed over the $n$ different
elements contained in the species, which include electrons for
charged species as an additional ``element."

The data from Barin (1995) are expressed in terms of Joules, with
the reference pressure being 1 bar.  Moreover, it is convenient
to use the Gibbs energy functions as tabulated, then convert
these to the free energies in cal mole$^{-1}$.  Given $g'_{pi}$,
the Gibbs energy function in J mol$^{-1}$ of species $i$ in phase $p$
with a reference pressure of 1 bar, the required Gibbs energy
function, $g_{pi}$ in cal mol$^{-1}$ with a reference pressure of
1 atmosphere,  can be obtained from

\begin{equation}
g_{pi} = [g'_{pi} - P_c]/J_c\, ,
\label{convertjc}
\end{equation}

\noindent where for a gas-phase species, $P_c $ is the conversion factor
in cal mol$^{-1}$, given by $P_c = R\ln(1.01325)$, with
$R$ = 1.986 being the gas constant in cal mol$^{-1}$, and $J_c$ = 4.184
J cal$^{-1}$.  For a condensed phase, since the typical bulk modulus 
is in the megabar range, the compressibility of the
species is neglected, so $P_c$ is replaced by zero.  The free
energy is then obtained from

\begin{equation}
\Delta G_{pi} = \Delta H_{pi}(298) - Tg_{pi} + \sum_{j=1}^n
\left[\nu_{ij}\left\{Tg_{1j} - \Delta H_{1j}(298)\right\}\right]\, ,
\label{gef}
\end{equation}

\noindent where $\Delta H_{pi}(298)$ is the enthalpy of formation
of species $i$ in phase $p$ from its elements in their standard
reference states at 298.15 K, and as with eq. (\ref{convertg}),
the summation is over the constituent elements, with the indices
referring to element $j$ in phase 1, the gas phase.  In the case
of data used from Barin (1995), the value of $\Delta H_{pi}(298)$
is often given only for the phase that is stable at 298.15 K.
However, for phases stable at elevated temperatures, the enthalpy
change is given at each phase boundary with increasing temperature,
so the value of $\Delta H_{pi}(298)$ for a high temperature phase
could be obtained by taking the enthalpy of formation of the phase
stable at 298.15 K, and adding to it the enthalpy change for the
high temperature phase, summing over any intermediate phases.

For a number of phases at intermediate temperatures that have too
few points to fit a smooth function using the data from Barin (1995),
the relationship between the Gibbs energy function, enthalpy, entropy
and specific heat could be used.  Dropping for convienience the
indices $p$ and $i$ and the primes because heat is measured in Joules,
but making it clear that it is a function of $T$, the Gibbs energy
function can be obtained from

\begin{equation}
g(T) = \frac{H(298) - H(T_o)}{T} + S -
\frac{1}{T}\int_{T_o}^T C_p dT + \int_{T_o}^T\frac{C_p}{T}dT\, ,
\label{intgef}
\end{equation}

\noindent where $H(298)$ is the enthalpy at 298.15 K, $H(T_o)$
is the enthalpy at some tabulated temperature $T_o$ in the range where
the phase is stable, $S$ is the corresponding entropy at $T_o$, and
$C_p$ is the specific heat at constant pressure.  All these quantities
are given in Barin (1995).  If $C_p$ is constant over the temperature
range being considered, which is the case for all liquid phases
tabulated, then eq. (\ref{intgef}) can be integrated to give

\begin{equation}
g(T) = \frac{H(298) - H(T_o)}{T} + S +
C_p\left[T_o/T - \ln(T_o/T) - 1\right]\, .
\label{solvegef}
\end{equation}

Using eq. (\ref{solvegef}), together with eqs. (\ref{convertjc})
and (\ref{gef}), it is possible to generate additional points,
including those outside the range of stability of the phase being
studied, and make a good fit.  If extrapolation outside the
range of stability is included for the purposes of making a fit,
such as below the freezing point of a liquid phase, one should ensure
that the free energy outside the range of stability is always
greater than the phase that is stable in that region, in order
to guarantee that the phase being extrapolated never forms in preference
to the more stable phase.  If the specific heat varies with temperature,
it should be possible to replace $C_p$ by a fitted polynomial and to
integrate eq. (\ref{intgef}).

In addition to the above methods, for some gas-phase species,
such as $TiH$, no suitable thermodynamic data appear to be available,
but it is possible to use the spectroscopic data for this purpose.
The free energy can be calculated from the data using eq. (A1) in
Burrows et al. (2005), and is directly in the form required.

At the temperatures for which data are tabulated using the various methods
already discussed, least-square fits can be made for a set of polynomials
whose highest order is given by

\begin{equation}
\Delta G_{pi}(T) = aT^{-1} + b + cT + dT^2 + eT^3\, ,
\label{fitgibbs}
\end{equation}
\noindent

\noindent where $a$, $b$, $c$, $d$, and $e$ are fitted coefficients,
and $\Delta G_{pi}(T)$ is the fitted Gibbs free energy of formation
at temperature $T$ of species $i$ in phase $p$.  The polynomials are
evaluated at the tabulated points and the deviations from the tabulated
values are obtained.  The polynomial with the best fit over the
temperature range of particular interest, namely where the phase is
stable, should be selected, but in the event of a near tie between two
or more polynomials, the lowest order polynomial should be selected,
with the unused coefficients being set to zero.  For monatomic neutral
species in the gas phase and for free electrons, $\Delta G_{pi}(T)$ is zero
by definition for all $T$.  Because discontinuities associated with
phase changes of the elements in their standard reference states have
been removed, $\Delta G_{pi}(T)$ varies smoothly as a function of $T$,
and eq. (\ref{fitgibbs}) should represent a good fit over the
temperature range required.  Once a good fit has been obtained, the
data can be added to a thermodynamic database.

In performing the calculation for a particular temperature, pressure,
and composition, the Gibbs free energy for each species of interest
is obtained from the database using the fitted coefficients at the
temperature required, then the total free energy of the system is
minimized to obtain the abundances of the gas-phase species, together
with any condensates.  The total dimensionless free energy is given by

\begin{equation}
\frac{G(T)}{RT} = \sum_{i=1}^m \left[n_{1i}
\left\{ \frac{\Delta G_{1i}(T)}{RT} + \ln P +
\ln\left(\frac{n_{1i}}{N}\right)
\right\} \right] + \frac{1}{RT} \sum_{p=2}^{s+1} 
\Big[n_{p1} \Delta G_{p1}(T)\Big]\, ,
\label{eqmin}
\end{equation}

\noindent where $R$ is the gas constant, and for the first sum for
the gas phase with $p=1$ , $P$ is the total pressure in atmospheres,
$N$ is the number of moles, $m$ is the number of species,
$n_{1i}$ is the number of moles of species $i$, and $\Delta G_{1i}(T)$
is the corresponding free energy of that species.  The second sum is
over the $s$ condensed phases, which may include multiple phases
of the same species, but except at a phase boundary, only one phase
of a particular species in a condensed form is present at any
time, since we have not considered solid or liquid solutions.
Consequently, $n_{p1}$ is the number of moles of a condensed
species and $\Delta G_{p1}(T)$ is the corresponding free energy of
that species.  Since there is only one species per phase, for
convenience we have set $i$ equal to 1.

Equation (\ref{eqmin}) then has to be minimized, subject to the
constraint given by the mass balance for each element $j$

\begin{equation}
\sum_{i=1}^m \nu_{1ij} n_{1i} + \sum_{p=2}^{s+1} \nu_{p1j} n_{p1}
= b_j\, ,
\label{balance}
\end{equation}

\noindent where $\nu_{1ij}$ and
$\nu_{p1j}$ are the stoichiometric coefficients of the species
containing the element, and $b_j$ is the total number of gram-atoms
of that element in all forms.  When ionized species are considered,
$j$ includes the electron as an ``element," which can have positive
or negative stoichiometric coefficients, but it must sum to zero
for net charge neutrality; thus, the corresponding value of $b_j$ is
zero.  The solution is found by the method of Lagrange
undetermined multipliers and Taylor expansions about an arbitrary
point after starting with trial abundances, as described by
Eriksson (1971) and Eriksson \& Ros\'{e}n (1973).

The most efficient method is to perform a set of calculations at a
given pressure and $Z$ for a sequence of decreasing temperatures at one
degree intervals, starting at a temperture above the range of interest,
in particular above the condensation temperature of the
most refractory species, using arbitrary starting values for the
first iteration.  For each new temperature down in the sequence, the
abundances from the previous temperature can be used as starting values.
This can be very efficient, and not much would be gained in computing
time by using larger temperature intervals at the expense of fewer
temperature points covered.  As the temperature decreases, the first
appearance or disappearance of a condensed phase should be noted, in
order to follow the condensation sequence with decreasing temperature.
When a phase disappears with decreasing temperature, it always changes
into one or more other condensed phases.  The only exception for a
realistic astrophysical mixture might be graphite, which can convert
into $CH_4$ in a carbon-rich ($[C]/[O] > 1$) mixture at low temperatures.

At higher temperatures, convergence is generally very
rapid, but as the temperature decreases and more condensed phases
appear, the iterations slow down, and there is a greater
likelihood of convergence failing at some low temperatures.  This
problem can usually be cured by removing species that have
negligible abundances, but still adversely affects the processing time.
In some cases at low temperatures it is possible for the code to
converge to solutions where the element abundances or net charge
neutrality are not valid, i.e. eq. (\ref{balance}) does not balance.
This can be checked by calculating residuals using

\begin{equation}
r_j = \frac{b_j - s_j}{b_j}\, ,
\label{residual}
\end{equation}

\noindent where $r_j$ is the residual fraction for element $j$,
$b_j$ is the specified abundance, as in eq. (\ref{balance}),
and $s_j$ is the sum of the abundances multiplied by the stoichiometric
coefficients of all species containing element $j$ on the left
hand side of eq. (\ref{balance}).  In the case of charge
conservation for $j$ corresponding to electrons as an ``element,"
$r_j$ is just the sums on the left hand side of eq. (\ref{balance}),
i.e. $r_j = s_j$.

Working in double precision with an accuracy of about 15 significant
figures, if $r_j$ is substantially larger in magnitude than
$O(10^{-15})$, the calculations should be discarded, and if necessary
repeated with some of the very low abundance species removed.

As discussed previously, because of the expense in calculating some of
the opacities, tables should be precomputed at selected temperatures
and pressures, then when required opacities can be calculated by
interpolation.  Additionally, for those opacities that do not require
extensive precomputations and are not already in precomputed tabular
form, they can be calculated on the fly when required.  In all cases,
the final mass opacities in cm$^2$g$^{-1}$ require the abundances of
the absorbing species to be known for the given temperature, pressure,
and chemical composition at which the calculations are being performed.

In the equilibrium calculations performed for the plots presented here,
a subset of 30 gas-phase species out of nearly 350 gas-phase
species were selected for detailed treatment.
The species are the neutral atoms: $H$, $He$, $Li$, $Na$, $K$, $Rb$,
$Cs$, $Al$, $Ca$, and $Fe$, the ions: $e^-$, $H^+$, and $H^-$,
the metal hydrides: $MgH$, $CaH$, $FeH$, $CrH$ and $TiH$, with the
remaining molecules being $H_2$, $N_2$, $CO$, $SiO$, $TiO$, $VO$,
$CaOH$, $H_2O$, $H_2S$, $NH_3$, $PH_3$, and $CH_4$.  Out of these,
$Al$, $Ca$, $N_2$, and $CaOH$ are not discussed here, but their
abundances can be important.  Note that like $H_2$, $N_2$ is a
symmetric molecule with no net dipole moment, and in
planetary/satellite atmospheres like of the Earth or Titan is a
source of CIA opacities in the infrared, as well as collisional
broadening of lines.  Although the temperature is too low for
scattering due to $e^-$ to be of any importance, the abundance of free
electrons indicates the level of ionization, and thus indirectly the
contribution due to the absorption produced by $H^-$.

Although we are discussing calculations involving only a solar composition,
a large range of stellar metallicities exist, even for nominally
Population I compositions. It is thus expected that the metallicities
of substellar objects will also have a large range.  To cover the
large ranges in temperature, pressure, and metallicity expected,
a recommended set of values could be as follows:

\smallskip

\begin{tabular}{ll}
50 K $\le T \le$ 5000 K &
in steps of $\Delta \log T$ = 0.0025,\\
8$\times 10^{-8}$ Atm $\le P \le$ 400 Atm &
in steps of $\Delta \log P$ = 0.1,\\
0.01$Z_{\bigodot} \le Z \le$ 3.16$Z_{\bigodot}$ &
in steps $\Delta \log Z$ = 0.5,
\end{tabular}

\noindent and at each value of $T$, $P$, and $Z$ the abundances
of each of a selected subset of species, say 30, could be stored for
later use in opacity calculations.

In order to see how the abundances of a number of important species vary
with temperature, Figs. \ref{fig:17}, \ref{fig:18}, \ref{fig:19}, and
\ref{fig:20}, show the $\log_{10}$ of the mixing fractions as a function
of temperature between 100 and 4000 K at a total gas pressure of 1
atmosphere for a number of species.  All the figures are plotted with
the same scales to make comparisons easier, with the $\log_{10}$ of the
mixing fractions plotted from -18 to -2.  The molecule $H_2$ is the most
abundant species over most of the temperature range, so is not plotted.
At the highest temperatures here atomic hydrogen replaces $H_2$ as the most
abundant species, and is likewise omitted.

Figure \ref{fig:17} is a plot of the abundances of the molecules
discussed in \S\ref{hitran}, namely $H_2O$, $NH_3$, $CH_4$, and $CO$,
the abundances discussed in \S\ref{oxides}, $TiO$ and $VO$, and $H_2S$
and $PH_3$ from \S\ref{precomp}.  In addition, because of the
importance of the equilibrium between $N_2$ and $NH_3$, $N_2$ is
also included.  Since $CO$ has a very high dissociation energy, and
since the abundance of oxygen likely exceeds that of carbon,
virtually all carbon is tied up in $CO$, with the surplus oxygen
being bound in $H_2O$, excepting some partial dissociation of $H_2O$
at the very highest temperatures and the formation of other
oxides and condensdates containing oxygen.  With decreasing
temperature a point is reached where $CO$ reacts with $H_2$ to form
$CH_4$, which replaces $CO$ as the dominant carbon-bearing species.
In the process, the oxygen tied up in $CO$ becomes available to form
additional $H_2O$, which more than compensates for some of the oxgyen
that has in the mean time been removed from $H_2O$ to form various
condensates, mostly those of silicon and magnesium.

Like $CO$ with oxygen, $N_2$ is the most abundant nitrogen bearing
species at high temperatures, because $N_2$ also has a very high
dissociation potential, and like $CO$, with decreasing temperature it
reacts with $H_2$ to form $NH_3$, which replaces it as the most abundant
nitrogen-bearing species at low temperatures.  However, nitrogen
is only weakly coupled to the rest of the system, because other
compounds containing nitrogen have low abundances.  Both $TiO$ and
$VO$ are influenced by the formation of condensates, that cause
them to be very rapidly removed from the gas phase when the condensates
form (Lodders 2002).  If rainout did not take place, $Fe$ would
react with $H_2$, causing $FeS$ to form and the abundance of $H_2S$
would rapidly fall like $TiO$ and $VO$, but because $Fe$ has been
removed from system, the abundance of $H_2S$ does not decrease
with decreasing temperature (Fegley \& Prinn 1983;
Visscher, Lodders \& Fegley 2006).

Figure \ref{fig:18} shows the mixing fractions of the metal hydrides,
whose opacities are discussed in \S\ref{hydrides}, Fig. \ref{fig:19}
shows the corresponding fractions of the alkali elements discussed in
\S\ref{vald}, and Fig. \ref{fig:20} shows the corresponding fractions
of a number of species discussed in \S\ref{boundfree} and \S\ref{uvopac}.
With the exception of the $H^-$ ion, all the abundances start dropping
rapidly at some specific temperature due to the formation of
condensates, which remove elements forming condensates from the gas
phase.

\begin{figure}
\epsscale{1.00}
\centerline{\includegraphics[angle=-90,width=19cm]{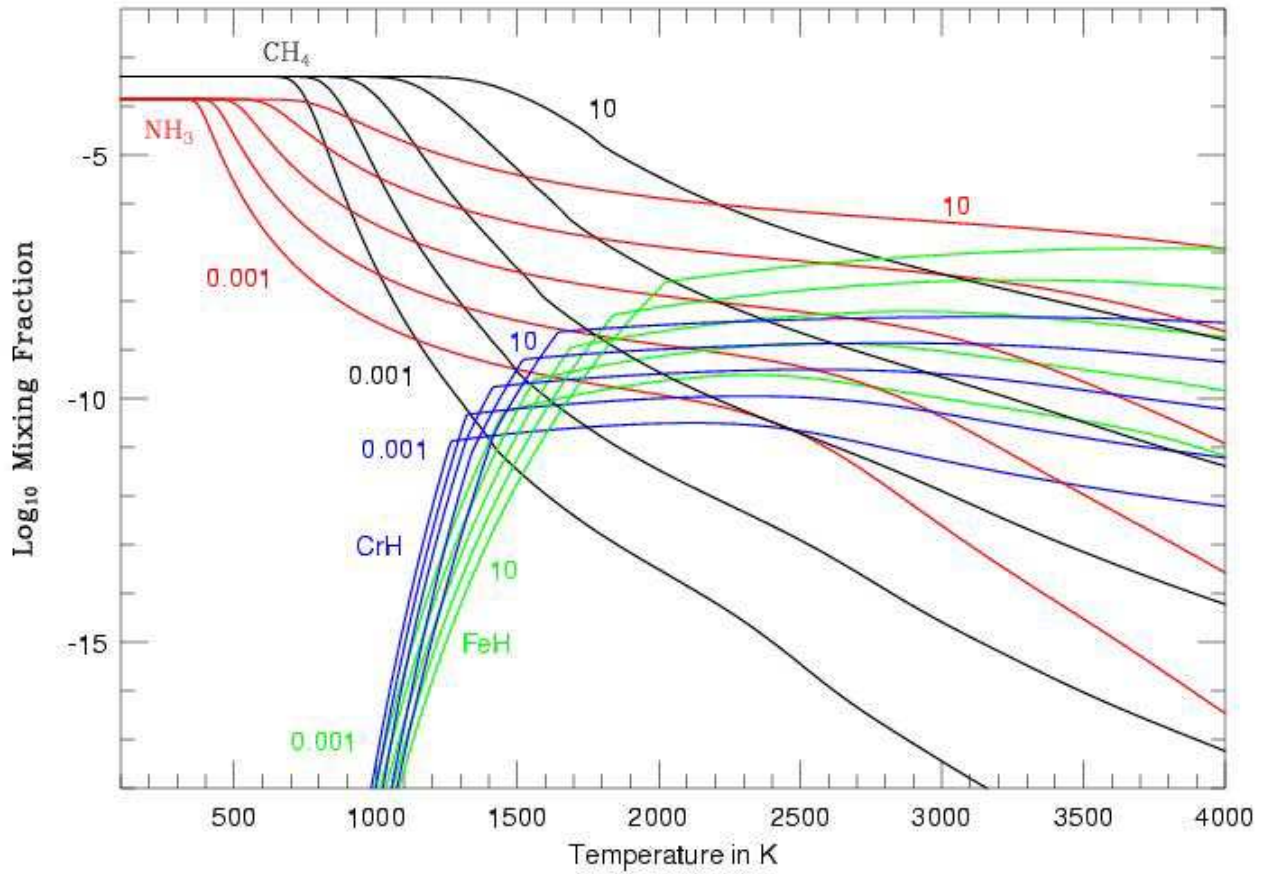}}
\caption{The log (base 10) of the mixing fraction as a function of
temperature for pressures of 0.001, 0.01, 0.1, 1, and 10 atmospheres
for $CH_4$ (black curves), $NH_3$ (red curves), $FeH$
(green curves), and $CrH$ (blue curves), with the same scales as
before.  For each species, only the curves for 0.001 and 10 atmospheres
are labeled (to avoid clutter), with the curves for the three intermediate
pressures lying between them.  In the cases of $CH_4$ and $NH_3$,
the temperatures at which these molecules are fully associated increase
with increasing pressure, and above these temperatures when these molecules
are less abundant the abundances increase with pressure, which is in full
accord with the principle of mass action.  Both $FeH$ and
$CrH$ are influenced by the formation of condensates between about
1000 and 2000 K, such that with increasing pressure the temperature at
which condensation takes place and reduces the abundance in the gas phase
increases.  Moreover, when no condensation is present, the abundance
increases with increasing pressure, such that with decreasing temperature,
the curve for a given pressure crosses over those due to higher pressures
that are rapidly falling at a higher temperature when condensation takes
place.}
\label{fig:21}
\end{figure}

Figure \ref{fig:21}, which is plotted on the same scales as the previous
four figures, shows the mixing fractions of $CH_4$, $FeH$, and $CrH$ for
pressures of 0.001, 0.01, 0.1, 1, and 10 atmospheres, and
Fig. \ref{fig:22} is the same for the abundances of $TiO$, $VO$, and $K$.
With the exception of $CH_4$, all the abundances fall rapidly at some point
due to the formation of condensates.

\section{CONCLUSION}
\label{conclusions}

In this work, we have discussed and detailed the main sources of opacity
in the cool atmospheres of brown dwarfs and extrasolar giant planets.
Since these objects have lower atmospheric temperatures than stars, a
number of diatomic and polyatomic molecules are present, which are not
found in abundance in most stellar types.  Such molecules can have a very complicated
spectrum with a large number of lines.  Because of the complex spectrum
of molecules, and because they are such important sources of absorption,
a considerable amount of effort is required to calculate their
contribution to the opacity.  Detailed calculations for many millions of
lines over an extensive frequency grid are involved.

Most other papers dealing with the opacities in the atmospheres of substellar
objects, consider only one or a few specific molecules, often in a
restricted wavelength region.  This is one of the first papers to provide
in one place a comprehensive discussion of the important molecular and
atomic opacity sources, covering a broad range of wavelengths from the
near ultraviolet through to the far infrared.

In practical opacity calculations, once the monochromatic absorption
of each species has been considered, their abundances have to be
calculated before they can be combined to obtain total opacities.
The minimization of the free energy of the system can be used to determine
equilibrium abundances, and we have provided a comprehensive summary of
the techniques needed to derive such abundances and results of such
calculations.  Once the complete monochromatic opacities are obtained,
mean opacities weighted by abundances can be calculated.

In order to model substellar objects correctly, a knowledge of how
the atmosphere absorbs radiation is necessary.  This paper is meant
to provide a compendium of approaches, references, and results to
aid the student interested in substellar dense atmospheres and the
associated spectroscopy and chemistry.

\acknowledgements

This work was supported in part by NASA under grants
NAG5-10760 and NNG04GL22G.  The authors would like to thank Ivan Hubeny
for reading the manuscript and providing detailed comments in early drafts.
They would also like to thank Richard Freedman for providing 
guidance with line broadening parameters.
Finally, AB acknowledges support through the Cooperative Agreement
\#{NNA04CC07A} between the University of Arizona/NOAO LAPLACE node and
NASA's Astrobiology Institute.

% References -------------------------------------------------------------------

% Table ------------------------------------------------------------------------

\clearpage
\begin{deluxetable}{cll}
%\tablenum{111}
\tablewidth{17cm}
\tablecaption{Summary List of Species, Section Numbers, \& Band Systems}
\tablehead{
\colhead{Species}  & \colhead{Section} & \colhead{Comments}}
\startdata
$H$ & \ref{uvopac} and \ref{boundfree} & Lines and continuum \\
$\,\,\, H^-$ & \ref{boundfree} & Continuum present when free electrons are also
                                 present \\
$He$ & \ref{ciaopac} & Important for CIA and line broadening \\
$Li$ & \ref{vald} & Alkali element - lines \\
$Na$ & \ref{vald} and \ref{boundfree} & Alkali element - lines and continuum \\
$K$ & \ref{vald} and \ref{boundfree} & Alkali element - lines and continuum \\
$Fe$ & \ref{uvopac} and \ref{boundfree} & Lines and continuum \\
$Rb$ & \ref{vald} & Alkali element - lines \\
$Cs$ & \ref{vald} & Alkali element - lines \\
$H_2$ & \ref{uvopac} and \ref{ciaopac} & CIA, and $B-X$ and $C-X$ UV bands \\
$CO$ & \ref{hitran} and \ref{uvopac} & Vib-rot bands in IR and $A-X$ band
                                       system in UV \\ 
$SiO$ & \ref{uvopac} & $A-X$ band system in UV \\
$H_2O$ & \ref{hitran} and \ref{uvopac} & Rot and vib-rot bands in IR and
                                         absorption in UV \\
$NH_3$ & \ref{hitran} & Vib-rot bands in IR \\
$CH_4$ & \ref{hitran} and \ref{ciaopac} & Vib-rot bands in IR \\
$H_2S$ & \ref{precomp} and \ref{uvopac} & Pre-computed vib-rot bands in IR
                                          and absorption in UV \\
$PH_3$ & \ref{precomp} & Pre-computed vib-rot bands in IR \\
$TiO$ & \ref{hitran} & $\alpha$, $\beta$, $\gamma$, $\gamma'$, $\delta$,
                       $\epsilon$, and $\phi$ band systems \\
$VO$ & \ref{hitran} & $A-X$, $B-X$ and $C-X$ band systems \\
$TiH$ & \ref{hydrides} & $A-X$ and $B-X$ band systems \\
$CrH$ & \ref{hydrides} & $A-X$ band system \\
$FeH$ & \ref{hydrides} & $F-X$ band system \\
$MgH$ & \ref{hydrides} & $X-X$, $A-X$ and $B'-X$ systems \\
$CaH$ & \ref{hydrides} & $A-X$, $B-X$ and $D-X$ systems \\
$MgSiO_3$ & \ref{grainopac} & Enstatite in Mie theory \\
%$Mg_2SiO_4$ & \ref{grainopac} & Forterite in Mie theory \\
%$MgAl_2O_4$ & \ref{grainopac} & Magnesium-aluminum spinel in Mie theory \\
%$Al_2O_3$ & \ref{grainopac} & Corundum in Mie theory \\
%$Ca_2Al_2SiO_7$ & \ref{grainopac} & Gehlenite in Mie theory \\
$H_2O$ & \ref{grainopac} & Water in Mie theory \\
%$NH_3$ & \ref{grainopac} & Ammonia in Mie theory \\
%$Fe$ & \ref{grainopac} & Iron grains in Mie theory \\
\enddata
\label{table:X1}
\end{deluxetable}
\clearpage

% Figures ----------------------------------------------------------------------

% figure 1
\clearpage

% figure 2
\clearpage

% figure 3
\clearpage

% figure 4
\clearpage

% figure 5
\clearpage

% figure 6
\clearpage

% figure 7
\clearpage

% figure 8
\clearpage

% figure 9
\clearpage

% figure 10
\clearpage

% figure 11
\clearpage

% figure 12
\clearpage

% figure 13
\clearpage

% figure 14
\clearpage

% figure 15
\clearpage

% figure 16
\clearpage

%Start section on abundance plots

% figure 17
\clearpage

% figure 18
\clearpage

% figure 19
\clearpage

% figure 20
\clearpage

% figure 21
\clearpage

% figure 22
\clearpage

\end{document}